\documentclass[pre,aps,showpacs,nofootinbib,twocolumn]{revtex4}
\usepackage{amssymb}
\usepackage{graphicx}

\input{tcilatex}

\begin{document}

\title{Importance of the Electron-Phonon Interaction with the Forward
Scattering Peak \\
for Superconducting Pairing in Cuprates}
\author{Miodrag L. Kuli\'{c}}

\address{J. W. Goethe-Universit\"{a}t Frankfurt am Main, Theoretische Physik,
Max-von-Laue-Strasse 1, 60438 Frankfurt am Main, Germany}

\begin{abstract}
We discuss first some basic experimental facts related to ARPES, tunnelling,
optics ad neutron scattering measurements. They give evidence for the
relevance of the electron-phonon interaction (EPI) in pairing mechanism of
HTSC cuprates. A controllable and very efficient theory for strong
correlations and their effects on EPI is discussed. The theory is based on
the 1/N expansion method in the t-J model without using slave bosons (or
fermions). The remarkable prediction is that strong correlations renormalize
EPI and other charge-fluctuation properties (by including nonmagnetic
impurity scattering) in such a way that the forward scattering peak (FSP)
appears in the corresponding effective interactions. The existence of FSP in
EPI is additionally supported by the weakly screened Madelung interaction in
the ionic-metallic structure of layered cuprates. Pronounced FSP in EPI of
HTSC cuprates reconciles many puzzling results, which could not be explained
by the old theory with the momentum independent EPI. For instance, EPI with
FSP gives that the couplings in the s- and d-wave pairing channel are of the
same magnitude near and below the optimal hole doping. It is shown that FSP
in the impurity scattering potential is responsible for robustness of d-wave
pairing in cuprates with nonmagnetic impurities and other defects. The ARPES
kink and the isotope effect in the nodal and anti-nodal points, as well as
the collapse of the elastic nonmagnetic impurity scattering in the
superconducting state, are explained by this theory in a consistent way. The
proposed theory also explains why the nodal kink is not-shifted in the
superconducting state while the anti-nodal kink is shifted by the maximal
superconducting gap. It turns out that in systems with FSP in EPI besides
the classical phase fluctuations there are also internal fluctuations of
Cooper pairs. The latter effect is pronounced in systems with long-ranged
pairing forces, thus giving rise to an additional contribution to the
pseudogap behavior.
\end{abstract}

\date{\today }
\maketitle

\section{Introduction}

In this year the physics community celebrate twenty years from the
remarkable discovery of high-temperature superconductors (HTSC) by G.
Bednorz and K. A. M\"{u}ller. However, all of us are aware of the fact that
our friend and teacher Vitalii Lazarevich Ginzburg \cite{VL} is the pioneer
in this fascinating field. He was not only the first who raised this
question but he also actively advocated the possibility of the
room-temperature superconductors (RTSC) \cite{VL-Book}. Moreover, he has
proposed the well known excitonic mechanism of pairing in quasi 2-D
structures, which, in fact, was a precursor for other studies of this
important subject. Finally, HTSC were discovered in 1986 by Bednorz and M%
\"{u}ller in rather exotic quasi-2D HTSC materials - cuprates \cite{BeM}. A
number of interesting approaches for reaching high critical temperature T$%
_{c}$ were elaborated in the famous VL superconductivity-group in the I. E.
Tamm department at the P. N. Lebedev Institute in Moscow. (VL is the popular
name of V. L. Ginzburg - see \cite{VL}.) One of the most important results,
which came from the "theoretical kitchen" of the VL group, was related to a
possible{\ limit of }T$_{c}${\ due to the electron-phonon interaction} (EPI).

Regrettably, study of the pairing mechanism in cuprate superconductors was
significantly influenced by prejudices related to EPI. One prejudice in the
search of HTSC\ materials is related to the question of the upper limit of T$%
_{c}$ within the phonon mechanism of pairing \cite{Cohen}. By assuming that
the \textit{total macroscopic longitudinal dielectric function} $\varepsilon
_{tot}(\mathbf{k},\omega =0)$ is positive for all $\mathbf{k}$ ($\varepsilon
_{tot}(\mathbf{k},\omega =0)>0$ , $\mathbf{k}$ is the momentum and $\omega $
is the frequency) the authors in Ref. \cite{Cohen} came to a pessimistic
conclusion that EPI is able to produce only a moderate $T_{c}<10$ $K$, since
as they predicted the EPI coupling constant is limited to its maximal value $%
\lambda _{EPI}^{\max }<0.5$. However, a number of experiments contradicted
this statement, since the large coupling constant $\lambda _{EPI}>1$ is
realized in a number of materials. For instance $\lambda _{EPI}\approx 2.5$
in the $PbBi$ alloy (although $T_{c}$ is not large), and $\lambda
_{EPI}\approx 1$ with $T_{c}\approx 40$ $K$ in $MgB_{2}$. There are a number
of other systems with $\lambda _{EPI}>0.5$, such as $A_{3}C_{60}$ with $%
\lambda _{EPI}\approx 1$ and $T_{c}\approx 40$ $K$, etc. The correct answer
to this question was given by D. A. Kirzhnitz \cite{Kirzhnitz}, a remarkable
man and physicist from the Ginzburg group, who has shown that thermodynamic
and lattice stability do not exclude the possibility that $\varepsilon
_{tot}(\mathbf{k},\omega =0)<0$. In fact a negative value of $\varepsilon
_{tot}(\mathbf{k},\omega =0)$ is realized in a large portion of the
Brillouin zone in most superconductors \cite{DKM}. This important result
means that the dielectric function $\varepsilon _{tot}$ \textit{does not
limit }$T_{c}$\textit{\ in the phonon mechanism of pairing} - see more in
\cite{VL-Book}, \cite{DKM}, \cite{Kulic-Review} and \cite{MaksimovReview}.

Concerning HTSC in cuprates one of the central questions is - what is the
pairing mechanism in these materials? In the last twenty years, the
scientific community was overwhelmed by all kinds of (im)possible proposed
pairing mechanisms, most of which are hardly verifiable, if at all. This
trend is still continuing nowadays (although with smaller slope), in spite
of the fact that the accumulated experimental results eliminate all but few.
A somehow similar situation happened also in experimental physics of HTSC
cuprates, where a whole spectrum of possible and impossible values of some
(for pairing mechanism) relevant quantities was reported. As illustration
the reader can look at (critical) analysis of experimental situation in
optics done by Ivan Bo\v{z}ovi\'{c} \cite{Bozovic-Plasma} - especially of
the measurements of the plasma frequency $\omega _{pl}$. The dispersion in
the reported value of $\omega _{pl}$, from $0.06$ to $25$ $eV$ - almost
three orders of magnitude - tells us that in some periods the science
suffers from the lack of rigorousness and objectiveness.

By the end of the twenty-years era in studying HTSC materials, more or less
three main directions in search for the pairing mechanism have crystallized.
The \textit{first approach}, based on \textit{strong correlations, was}
first proposed by P. W. Anderson in 1987 and followed by many others. The
starting point is that the physics of cuprates is the physics of doped Mott
insulator \cite{Patrick-Lee}, which is supposed to be well described by the
single-band Hubbard model (or its extensions). The promoters of this
approach believe that strong electronic correlations alone are the driving
mechanism behind the whole phase diagram of these materials - the
antiferromagnetic phase, metallic and superconducting one. Since strong
correlations belong to the strong coupling problem it is understandable that
even nowadays, exactly 20 years after the discovery of HTSC, there is still
no reliable theory of strong correlations in HTSC materials which is capable
to describe both the normal and the superconducting state.

The \textit{second approach} is based on pronounced \textit{%
antiferromagnetic spin fluctuations} which are present in cuprates
especially at very low doping - we call it the spin-fluctuation interaction
approach (SFI) \cite{Pines}. Later on we will argue that this
phenomenological approach - which can be theoretically justified in the very
weak coupling limit only - is based on two ingredients which are more in
conflict than in agreement with the existing experimental results. First,
the spin-fluctuation spectrum taken from the (low-frequency) NMR\
measurements differs from the one obtained in neutron scattering
measurements - see \cite{Kulic-Review}, \cite{Kulic-AIP}. The latter is
shifted to much higher energies. Second, in order to explain large $T_{c}$
within the SFI approach, one has to assume an unrealistically large
spin-fluctuation coupling constant $g_{sf}\approx 0.6$ $eV$, which implies
that, the pairing coupling constant is also unrealistically large $\lambda
_{sf}(\sim g_{sf}^{2})>2$. Our statement is confirmed by a number arguments.
Let us mention here some of them, while others are given in Sec. II. A. For
instance, such a large $\lambda _{sf}$ gives much larger resistivity than
the measured one. Also, $g_{sf}$ extracted from ARPES\ and neutron
scattering measurements is much smaller, i.e. $g_{sf}\approx 0.2$ $eV$ what
gives rather small pairing constant $\lambda _{sf}<0.3$, see Sec. II A.

Let us stress that both of the above approaches are based either on the
Hubbard model or on its (more popular) derivative, the t-J model. Here, the
central question is - do these two models show superconductivity at high
temperature? The Monte-Carlo calculations on the 2D repulsive ($U>0$)
Hubbard model give \textit{no evidence for superconductivity with a large
critical temperature} \cite{D-Ds}. On the other hand, there is a strong
tendency for superconductivity (either BCS like or
Berezinskii-Kosterlitz-Touless like in 2D systems) in the attractive ($U<0$)
Hubbard model for the same intensity of $U$. Recent numerical calculations
in the t-J model \cite{Pryadko} have shown that \textit{there is no
superconductivity in the t-J model} at temperatures characteristic for
cuprates. If it exists (at all!) $T_{c}$ must be very low $\sim 1$ $K$.
These numerical results tells us that the lack of HTSC\ in the repulsive ($%
U>0$)\ single-band Hubbard model, as well as in the t-J model, is not due to
2D-fluctuations (which at finite T suppress superconductivity with phase
coherence) but it is due to the inherent ineffectiveness of strong
correlations and spin fluctuations to produce HTSC in cuprates. This means
also that the simple single-band Hubbard and the t-J model are insufficient
to explain the pairing mechanism in cuprates, and some other ingredients
must be included. Since EPI is strong in oxides it is physically plausible
and justified that it should be accounted for. As we shall argue later on,
the experimental support for the importance of EPI comes from optics,
tunnelling, and especially from recent ARPES measurements.

Regarding the EPI one can ask whether it contributes significantly to d-wave
pairing in cuprates? Surprisingly, in spite of a number of experiments in
favor of EPI many of researchers believe that EPI is irrelevant and that the
pairing mechanism is due to SFI and strong correlations alone - see \cite%
{Pines}. This belief is mainly based on: (i) an incorrect stability
criterion, discussed briefly above, and (ii) the fact that in cuprates
d-wave pairing ($\Delta (\mathbf{k},\omega )\approx \Delta _{s}(k,\omega
)+\Delta _{d}(\omega )(\cos k_{x}-\cos k_{y})$ with $\Delta _{s}<0.1\Delta
_{d}$) is realized with gapless regions on the Fermi surface \cite{Scal},
which is incompatible with EPI. So, the EPI\ mechanism of pairing in
cuprates has been abandoned mainly because of these two (incorrect)
prejudices.

In fact, there is rich experimental evidence that EPI is sufficiently large
in cuprates in order to produce HTSC, i.e. the EPI coupling constant is $%
\lambda \gtrsim 1$. Let us quote some of them: $\mathbf{(1)}$ The
superconductivity-induced \textit{phonon renormalization }\cite{Kulic-Review}
is much larger in cuprates than in $LTSC$ superconductors. This is partially
due to the larger value of $\Delta /E_{F}$ in HTSC than in $LTSC$. $\mathbf{%
(2)}$ The \textit{line-shape} in the phonon Raman scattering is very
asymmetric (Fano line), which points to a substantial interaction of lattice
vibrations with some quasiparticle continuum (electronic liquid). For
instance, the phonon Raman measurements \cite{Kulic-Review} on $%
HgBa_{2}Ca_{3}Cu_{4}O_{10+x}$ at $T<T_{c}$ give very large softening
(self-energy effects) of the $A_{1g}$ phonons with frequencies $240$ and $%
390 $ $cm^{-1}$ by $6$ $\%$ and $18$ $\%$, respectively. At the same time
there is a dramatic increase of the line-width immediately below T$_{c}$,
while above T$_{c}$ the line-shape is strongly asymmetric. A substantial
phonon renormalization was obtained in ($Cu,C)Ba_{2}Ca_{3}Cu_{4}O_{10+x}$
\cite{Kulic-Review}. $\mathbf{(3)}$ The \textit{large isotope coefficients} (%
$\alpha _{O}>0.4$) in $YBCO$ away from the optimal doping \cite{Franck} and $%
\alpha _{O}\approx 0.15-0.2$ in the optimally doped $%
La_{1.85}Sr_{0.15}CuO_{4}$. At the same time one has $\alpha _{O}\approx
\alpha _{Cu}$ making $\alpha \approx 0.25-0.3$. This result tell us that
other, besides O, ionic vibrations participate in pairing. $\mathbf{(4)}$
Very important evidence that EPI plays an important role in pairing comes
from \textit{tunnelling spectra} in cuprates, where the phonon-related
features have been clearly seen in the $I-V$ characteristics \cite%
{Tunnelling-HTSC}. $(\mathbf{5})$ The \textit{penetration depth} in the a-b
plane of YBCO is significantly increased by the substitution $%
O^{16}\rightarrow O^{18}$, i.e. $(\Delta \lambda _{ab}/$ $\lambda
_{ab})=(^{18}\lambda _{ab}-^{16}\lambda _{ab})/^{16}\lambda _{ab}=2.8$ $\%$
at $4$ $K$ \cite{lambda-isotope}. Since $\lambda _{ab}\sim m^{\ast }$, the
latter result, if confirmed, might be due to a nonadiabatic increase of the
effective mass $m^{\ast }$. ($\mathbf{6}$) The \textit{breakthrough} in the
physics of cuprates came from the recent ARPES measurements \cite{Lanzara},
\cite{Shen}, \cite{Cuk}. The latter show a number of features typical for
systems with pronounced EPI and peculiar Coulomb interaction. For instance,
there is a \textit{kink} in the quasiparticle spectrum in the \textit{nodal
point} at characteristic (oxygen) phonon frequencies in the normal and
superconducting state. The ARPES measurements indicate that the EPI coupling
constant is of the order one, $\lambda \approx \lambda _{EPI}\sim 1$ in the
nodal direction, and $\lambda _{EPI}\sim 2$ near the antinodal point. The
authors of Ref. \cite{Cuk} even claim that they are able to resolve the fine
structure in $Re\Sigma (\mathbf{k},\omega )$ in ARPES, what results in the
Eliashberg spectral function with a number of peaks at the characteristic
phonon frequencies. The imaginary part of the ARPES self energy, $Im\Sigma
(k,\omega )$, has a \textit{knee-like structure} around $\omega _{ph}\sim 70$
$meV$, which is typical for systems with pronounced EPI \cite{Lanzara}, \cite%
{Shen}, \cite{Cuk}. Here, $\omega _{ph}$ is the characteristic maximal
phonon energy. For $\omega >\omega _{ph}$ one has $Im\Sigma (k,\omega )\sim
-\lambda _{c}\omega $, which is certainly due to Coulomb interaction (by
including SFI too) with rather moderate $\lambda _{c}\lesssim 0.4$. Recent
ARPES\ spectra in the anti-nodal point give evidence for the kink at 40 meV,
while the spectral function shows a peak-dip-hump structure which is
characteristic for systems with strong EPI ($\lambda _{EPI}\gtrsim 1$). Very
recent ARPES spectra \cite{Lanzara2} show a \textit{pronounced isotope effect%
} in the real part of the quasiparticle self-energy. These results give
clear evidence that the EPI is appreciate and strongly involved in pairing
and quasiparticle scattering.

On the \textit{theoretical side} there are self-consistent $LDA$
band-structure calculations which (in spite of their shortcomings) give a
rather large bare EPI coupling constant $\lambda \sim 1.5$ in $%
La_{1.85}Sr_{0.15}CuO_{4}$ \cite{Krakauer}. The tight-binding calculations
of EPI in YBCO gave $\lambda \approx 2$ and $T_{c}$ $\approx 90$ $K$ \cite%
{Zhao-tc}. The \textit{nonadiabatic effects} due to poor metallic screening
along the $c$-axis, not accounted for in \cite{Krakauer}, may additionally
increase $\lambda $ as pointed first in \cite{Falter}. All these facts are
in favor of a substantial EPI in cuprates. However, if the properties of the
normal and superconducting state in cuprates are interpreted in terms of the
standard EPI theory - which is well established in low-temperature
superconductors ($LTSC$) - some puzzles arise. One of them is related to the
normal-state resistivity in optimally doped cuprates, which indicates that
the transport coupling constant $\lambda _{tr}$ is rather small, i.e. $%
\lambda _{tr}<\lambda /3$ according to ARPES\ and tunnelling measurements
\cite{Kulic-Review}, \cite{Kulic-AIP}. For instance, the combined
resistivity and low frequency conductivity (Drude part) measurements give $%
\lambda _{tr}\approx 0.3$ if the plasma frequency takes the value $\omega
_{pl}\sim 1$ $eV$ - see more below. So, if one assumes that $\lambda
_{tr}\approx \lambda $, which is the case in most $LTSC$, such a small $%
\lambda $ can not give large $T_{c}$ $(\approx 100$ $K)$. This implies two
possibilities: $\mathbf{(a)}$ $\lambda _{tr}\ll \lambda $ and the pairing is
due to EPI, or $(\mathbf{b})$ $\lambda _{tr}\simeq \lambda $ but EPI is
responsible for pairing by virtue of some peculiarities in the equations
describing superconductivity - for instance because of non-Migdal vertex
corrections. Related to this dilemma, it is worth mentioning that a similar
puzzling situation ($\lambda _{tr}\ll \lambda $) is realized in $%
Ba_{x}K_{1-x}BiO_{3}$ (with T$_{c}\simeq 30$ $K$), where optical
measurements give $\lambda _{tr}\approx 0.1-0.3$ \cite{Salje}, while
tunnelling measurements \cite{Jensen} give $\lambda \sim 1$. Note, in $%
Ba_{x}K_{1-x}BiO_{3}$ there are no magnetic fluctuations (or magnetic order)
and no signs of strong electronic correlations. Therefore, EPI is favored as
the pairing mechanism in $Ba_{x}K_{1-x}BiO_{3}$. It seems that in this
compound long-range forces in conjunction with some nesting effects, may be
responsible for this discrepancy.

According to the above discussion, the microscopic theory of EPI and pairing
in cuprates must be able to explain the following three important facts: (1)
why $\lambda _{tr}\ll \lambda $; (2) why d-wave pairing is realized in the
presence of strong EPI; $(3)$ why is d-wave pairing robust in presence of
non-magnetic impurities and defects? Here we shall argue that the theory
which is based on the existence of the \textit{forward scattering peak}
(FSP) in EPI can account for both facts. We point out that d-wave pairing in
cuprates is very \textit{robust in the presence of nonmagnetic impurities}
contrary to the prediction of the standard theory. We shall argue in what
follows that this robustness is again due to strong correlations which give
rise also to FSP in the impurity scattering \cite{Kulic-Review}, \cite%
{Kulic-AIP}. The FSP in EPI and in non-magnetic impurity scattering can be
due to: (\textbf{i}) strong electronic correlations; (\textbf{ii}) the
presence of long-range Madelung potential in EPI, which is poorly screened
for phonons vibrating along the c-axis; (\textbf{iii}) out of plain
impurities and defects which are poorly screened. The concept of FSP in EPI
and nonmagnetic impurity scattering potential due to strong correlations was
for the first time put forward in \cite{Kulic1}, \cite{Kulic2}, \cite{Kulic3}%
. This approach, which was elaborated further by the present author and his
collaborators, see more in \cite{Kulic-Review}, \cite{Kulic-AIP}, is
essential to explain ARPES in cuprates.

This paper is organized as follows. In Sec. II we discuss direct and
indirect experimental evidence for the importance of the EPI in cuprates.
The recent ARPES\ results are discussed in more detail. Section III is
devoted to the \textit{systematic and controllable theory} of strong
correlations without using the slave boson (or fermion) technique. This
theory treats strong correlations in terms of Hubbard operators $X^{\alpha
\beta }$ - we call it the \textit{X-method} - which describe the motion of
the composite electrons. The latter takes care that the double occupancy at
a given lattice site is excluded. The central question in strongly
correlated systems is - \textit{how to calculate the coherent part in the
quasiparticle dynamics}? This was achieved within the \textit{X-method }by
using a systematic and controllable 1/N expansion. In Sec. IV we discuss the
renormalization of EPI by strong correlations. The 1/N expansion indicates
that strong correlations lead to FSP in EPI. We stress that this result is
straightforward in the X-method, while in the slave-boson method until now
there have been no reliable calculations of EPI. We point out that the
X-method predicts in the t-J model, that the EPI coupling becomes
long-ranged in the presence of strong correlations. As the result of this
specific renormalization, the EPI coupling constant in the d-channel reaches
the value of that in the s-channel for some optimal doping and below it.
Furthermore, in conjunction with the residual Coulomb interaction this gives
rise to d-wave pairing, since the residual Coulomb interaction is much
stronger in the s-channel than in the d-channel. It is also shown that FSP
appears in the nonmagnetic impurity scattering giving rise to robustness of
d-wave pairing in cuprates. Section V applies the proposed theory to
explanation of ARPES experiments, of the d-wave robustness in presence of
impurities, of transport properties, etc. The problem of the pseudogap is
discussed in Sec. VI from the point of view of the proposed theory. It turns
out that the long-range EPI force gives rise to internal fluctuations of
Cooper pairs, which together with phase fluctuations reduce strongly the
mean-field critical temperatures.

There are numerous experimental indications on the nanoscale inhomogeneities
of HTSC oxides, for instance recent STM experiments \cite{Davis}. In that
case the superconducting order parameter is inhomogeneous, i.e. $\Delta (%
\mathbf{k},\mathbf{R})$ where $\mathbf{R}$ is the center of mass
of Cooper pairs. In this review we shall not discuss this effects,
which are in our opinion play secondary role in pairing mechanism.
By understanding what is going on in homogeneous systems we have a
starting basis to study effects of inhomogeneities.

\section{ Experimental results in favor of the EPI}

The experimental situation which is related to the quasiparticle scattering
in the normal and superconducting state of cuprates is extensively analyzed
in a number of reviews \cite{Kulic-Review}, \cite{Kulic-AIP}. Here we
summarize briefly some experiments which are important for the microscopic
theory of pairing in cuprates and which indicates importance of EPI in the
quasiparticle scattering. Before this discussion let us analyze briefly
interplay of EPI and SFI in the pairing mechanism and their relative
strength by taking into account three important experimental facts: (1) T$%
_{c}$ in cuprates is high, i.e. $T_{c}^{\max }\sim 160$ $K$; (2)
the pairing in cuprates is d-wave like, i.e. $\Delta
(\mathbf{k},\omega )\approx
\Delta _{s}(k,\omega )+\Delta _{d}(\omega )(\cos k_{x}-\cos k_{y})$ with $%
\Delta _{s}<0.1\Delta _{d}$, and (3) the EPI coupling constant is rather
large, $\lambda \approx 1-2$. Let us assume that d-wave pairing in cuprates
is due to some non-phononic mechanism, for instance due to SFI. If EPI in
cuprates would isotropic like in LTSC materials then it would be very
detrimental for d-wave pairing as it is shown in \cite{Licht} by numerical
studies of interplay of SFI and EPI by the Eliashberg theory. Here we shall
elucidate this important problem in a physically plausible and
semi-quantitative way. In the case of dominating \textit{isotropic} EPI in
normal state, then near $T_{c}$ the linearized Eliashberg equations have an
approximative form
\[
Z(\omega )\Delta (\mathbf{k},\omega )=\sum_{\mathbf{q}}\dint\limits_{0}^{%
\omega _{\max }}\frac{d\Omega }{\Omega }V(\mathbf{k},\mathbf{q},\Omega
)\Delta (\mathbf{q},\omega )\tanh \frac{\xi (\mathbf{q})}{2T_{c}}
\]%
\begin{equation}
Z(\omega )\approx 1+i\Gamma _{ep}.  \label{LinElia}
\end{equation}%
For pure d-wave pairing one has $V(\mathbf{k},\mathbf{q},\omega )=V(\omega
)(\cos k_{x}-\cos k_{y})(\cos q_{x}-\cos q_{y})$ and $\Delta (\mathbf{k}%
,\omega )=\Delta (\omega )(\cos k_{x}-\cos k_{y})$ what gives the equation
for T$_{c}$
\begin{equation}
\ln \frac{T_{c}}{T_{c0}}=\Psi (\frac{1}{2})-\Psi (\frac{1}{2}+\frac{\Gamma
_{ep}}{2\pi T_{c}}).  \label{gama}
\end{equation}%
Here, $T_{c0}$ is the bare critical temperature due to the SFI mechanism and
$\Psi $ is the di-gamma function. At temperatures near $T_{c}$ one has $%
\Gamma _{ep}\approx 2\pi \lambda _{ep}T$ $\ $and the solution of Eq. (\ref%
{gama}) is
\[
T_{c}\approx T_{c0}e^{-\lambda _{ep}}.
\]

The latter result means that when $\lambda _{ep}=1-2$ and $T_{c}^{\max }\sim
160$ $K$ the bare $T_{c0}$ due to SFI must be very large, i.e. $%
T_{c0}=(400-1100)$ $K$, what is highly unrealistic. However, this result,
together with d-wave pairing in cuprates, imply that EPI in these materials
must be \textit{strongly dependent on the momentum transfer}, as it was
first argued in \cite{Kulic1}. Only in that case is EPI conform with d-wave
pairing, either as its main cause or as a supporter of non-phononic
mechanism such as SFI. In the following Sections we shall argue that the
strongly momentum dependent EPI is the main player in cuprates providing
strength of the pairing mechanism, while the residual Coulomb interaction
and SFI, although weaker, trigger it to d-wave pairing.

\subsection{Magnetic neutron scattering and spin fluctuation spectral
function}

In a number of papers the experimental results for the pronounced imaginary
part of the susceptibility $Im\chi (\mathbf{k},k_{z},\omega )$, at the AF
wave vector $\mathbf{k}=\mathbf{Q}=(\pi ,\pi )$, are interpreted as a
support for SFI mechanism of pairing \cite{Pines}. We briefly explain the
\textit{inadequacy} of such an interpretation.

(\textbf{1}) The breakthrough in understanding the role of spin-fluctuations
came from magnetic neutron scattering experiments on $YBa_{2}Cu_{3}O_{6+x}$
by Bourges group \cite{Bourges}.
\begin{figure}[tbp]
\resizebox{.3\textwidth}{!}
{\includegraphics*[width=8cm]{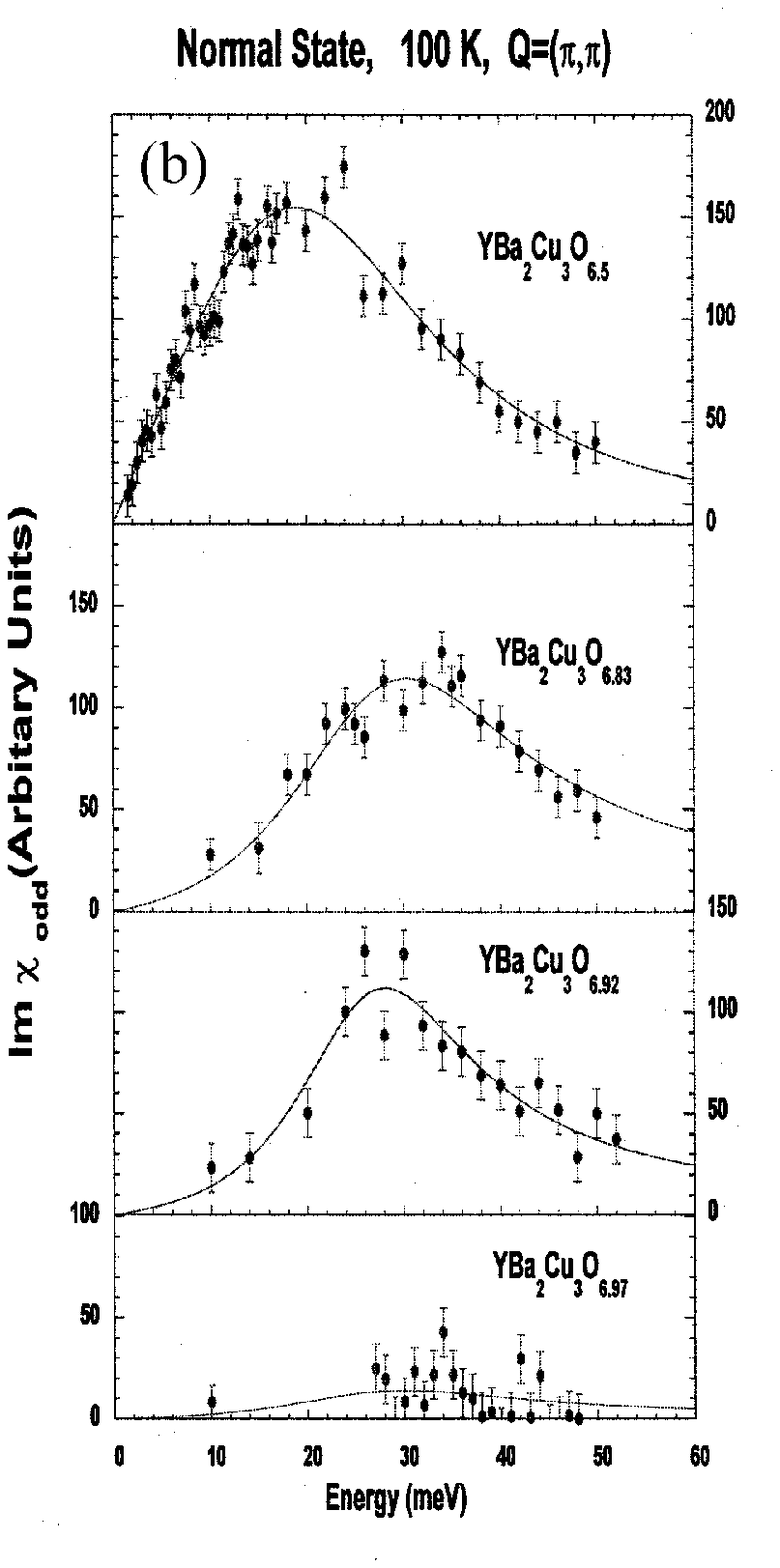}}
{\includegraphics*[width=6cm]{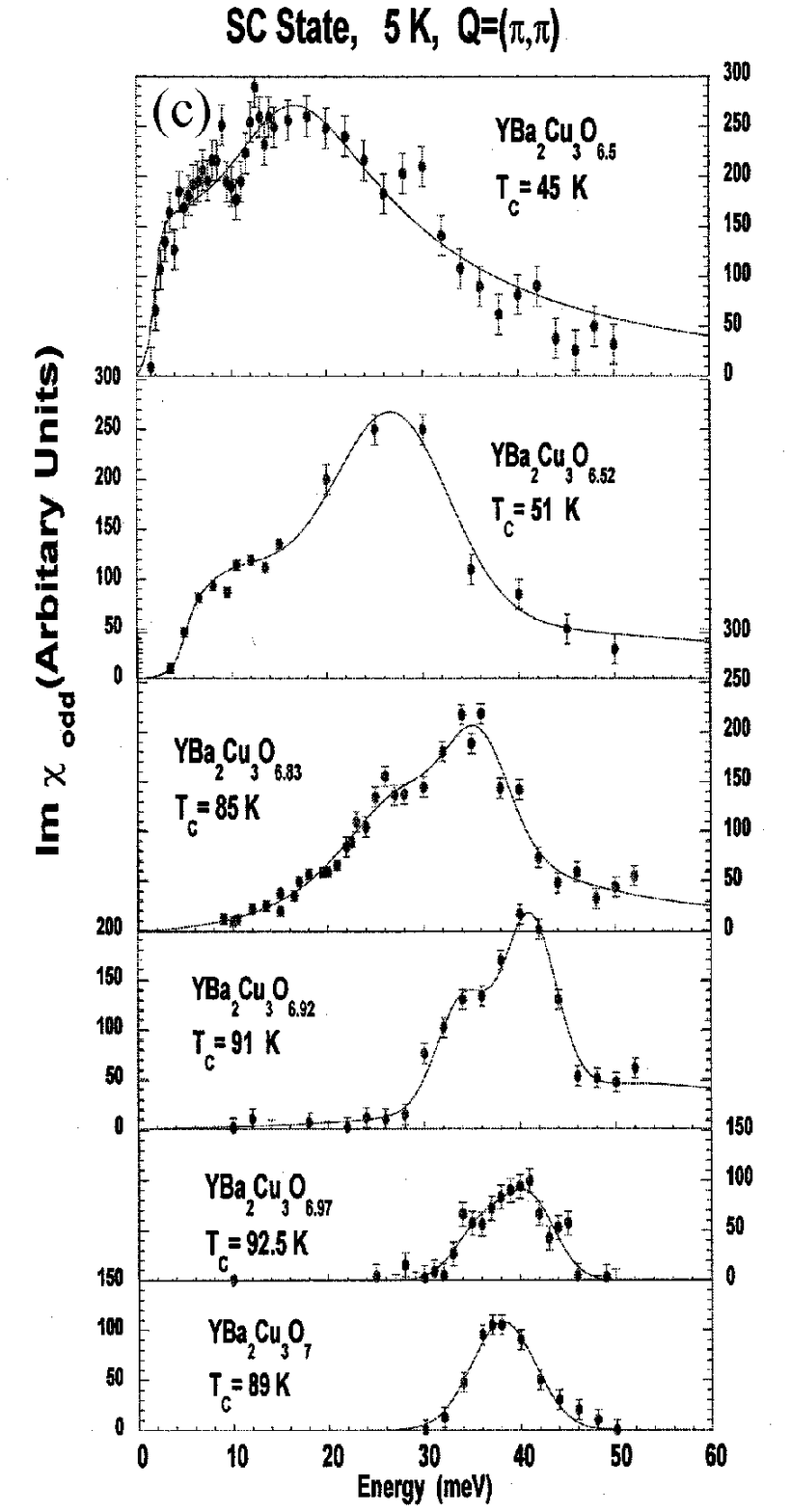}}
\caption{Magnetic spectral function $Im\protect\chi ^{(-)}(\mathbf{k},%
\protect\omega )$: (b) for $YBa_{2}Cu_{3}O_{6+x}$ in the normal state at $%
T=100$ $K$ and at $Q=(\protect\pi ,\protect\pi )$. $100$ counts in the
vertical scale corresponds to $\protect\chi _{max}^{(-)}\approx 350\protect%
\mu _{B}^{2}/eV$; (c) for $YBa_{2}Cu_{3}O_{6+x}$ in the superconducting
state at $T=5$ $K$ and at $Q=(\protect\pi ,\protect\pi )$. From Ref.
\protect\cite{Bourges}.}
\label{SuscFig}
\end{figure}
They showed that the spectral function $Im\chi ^{(odd)}(\mathbf{k},\omega )$
(the odd part in the bilayer system) is strongly dependent on hole doping.
By varying doping there is a huge reconstruction of $Im\chi ^{(odd)}(\mathbf{%
Q},\omega )$ in the frequency interval which is important for
superconducting pairing, say $5$ $meV<\omega <(60-70)$ $meV$. On the other
hand, there is only a small variation of the $T_{c}$. For instance, in
underdoped $YBa_{2}Cu_{3}O_{6.92}$ \ one has that $Im\chi ^{(odd)}(\mathbf{Q}%
,\omega )$ (and $S(\mathbf{Q})=\int_{0}^{70}d\omega \func{Im}\chi ^{(odd)}(%
\mathbf{Q},\omega )\sim \lambda _{SFI}\left\langle \omega \right\rangle $)
is much larger than that in the near optimally doped $YBa_{2}Cu_{3}O_{6.97}$%
, i.e. $S_{6.92}(\mathbf{Q})\gg S_{6.97}(\mathbf{Q})$, although the
difference in their critical temperatures $T_{c}$ is very small, i.e. $%
T_{c}=91$ $K$ in $YBa_{2}Cu_{3}O_{6.92}$ and $T_{c}=92.5$ $K$ in $%
YBa_{2}Cu_{3}O_{6.97}$. The reader can convince himself directly from the
Fig~\ref{SuscFig}(b). Such a large reconstruction of $\func{Im}\chi ^{(odd)}(%
\mathbf{Q},\omega )$ but a negligible change in $T_{c}$ in YBCO is a strong
argument against the $SFI$ mechanism of pairing in cuprates. Furthermore,
the \textit{anti-correlation} between the NMR spectral function $I_{\mathbf{Q%
}}=\lim_{\omega \rightarrow 0}Im\chi (\mathbf{Q},\omega )/\omega $ and $%
T_{c} $ (where $T_{c}\sim 1/I_{\mathbf{Q}}^{\alpha }$, $\alpha >0$) see \cite%
{Kulic-Review} additionally disfavors SFI models of pairing \cite{Pines},
i.e. the strength of pairing interaction is little affected by SFI. These
results also mean that in spite of pronounced spin fluctuations around $%
\mathbf{Q}$, the pairing quasiparticle self-energy is little affected by SFI.

(\textbf{2}) The next very serious argument against SFI pairing mechanism is
the \textit{smallness of coupling constant} $g_{sf}$. Namely, the real
spin-fluctuation coupling constant extracted from experiments is rather
small $g_{sf}\leq 0.2$ $eV$, in contrast to the large value ($%
g_{sf}^{(MMP)}\sim 0.6$ $eV$) assumed in the SFI theory by Pines and
collaborators in order to get large $T_{c}\sim 100$ $K$. (Note that the
pairing coupling in the SFI theory is $\lambda _{sf}\sim g_{sf}^{2}$, and
for the realistic value of $g_{sf}\leq 0.2$ $eV$ it would produce $\lambda
_{sf}\sim 0.2$ and very small $T_{c}\sim 1$ $K$ - see more in \cite%
{Kulic-Review}.) The large value for $g_{sf}^{(MMP)}$ gives resistivity much
larger than what is observed \cite{Kulic-Review}. The upper limit for $%
g_{sf}(\leq 0.2$ $eV)$ is extracted: (\textbf{i}) from the width of the $41$
$meV$ magnetic resonance peak in the superconducting state \cite{Kivelson2};
(\textbf{ii}) from the small magnetic moment ($\mu <0.1$ $\mu _{B}$) found
in the antiferromagnetic state of LSCO and YBCO. based on this fact in \cite%
{KulicKulic} it was estimated $\lambda _{sf}\sim 0.2$; (\textbf{iii}) from
ARPES experiments \cite{Shen} that give $\lambda _{sf}\sim 0.3$ - see also
\cite{Kulic-Dolgov}.

(\textbf{3}) The $SFI$ model is in some sense a phenomenological derivative
of the t-J Hamiltonian. However, recently it was shown that \textit{there is
no superconductivity in the t-J model} at temperatures characteristic for
cuprates \cite{Pryadko}. If it exists (at all!) T$_{c}$ must be very low.

(\textbf{4}) In the \textit{superconducting state} the magnetic fluctuations
are drastically changed, as expected for the singlet pairing state, since it
induces a spin gap in magnetic excitation spectrum of s-wave superconductors
- see Fig.~\ref{SuscFig}(c). However, the spectrum in superconducting state
of cuprates is more complex due to d-wave pairing and due to some
specificities of the band structure. For instance, at $T<T_{c}$ the sharp
peak in $Im\chi ^{(odd)}(\mathbf{k},\omega )$ is observed at $\omega
_{reson}=41$ $meV$ and $\mathbf{k}_{2D}=(\pi /a,\pi /a)$ - the \textit{%
resonance mode} - of the fully oxygenated (optimally doped) $%
YBa_{2}Cu_{3}O_{6+x}$ ($x\sim 1,$ $T_{c}\approx 92$ $K$) - see the bottom of
Fig.~\ref{SuscFig}c. With increased doping, the resonance mode becomes
sharper and moves to higher frequencies (scaling with $T_{c}$), while its
height decreases \cite{Bourges} - see Fig.~\ref{SuscFig}(c). In optimally
doped cuprates the spectral weight of the resonance peak is only 2-5 \% of
the total weight. Recently there were speculations in the literature that
d-wave superconductivity in cuprates might be due to the resonance mode!
This claim is untenable, since this mode can not cause superconductivity
simply because its intensity near $T_{c}$ is vanishingly small, thus not
affecting T$_{c}$ at all. If the magnetic resonance would be the origin for
superconductivity (and high T$_{c}$) the phase transition at T$_{c}$ would
be of the \textit{first order}, contrary to the experiments showing that it
is of the second order. In fact, the resonance mode is a \textit{affected by
superconductivity }but not its cause.

However, in spite of the fact that the strength of SFI is small, it can,
together with other contributions of the residual Coulomb interaction,
\textit{trigger} d-wave pairing. However, the \textit{strength of pairing is
due to EPI with FSP} - see more \ below and in \cite{Kulic-Review}, \cite%
{Kulic-AIP}.

\subsection{Dynamic conductivity and resistivity}

Optics measurements play very important role in studying the quasiparticle
dynamics since this method probes bulk sample contrary to ARPES\ and
tunnelling which probe tiny regions near the sample surface.

A diversity of the results on the reflectivity measurements in cuprates were
reported in the past. The dynamic conductivity $\sigma (\omega )$, which is
not a directly measured quantity but is derived from $R(\omega )$, was
analyzed in terms of the quasiparticle self-energy $\Sigma (k,\omega )$
instead of the transport self-energy $\Sigma _{tr}(k,\omega )$. In fact $%
\Sigma (k,\omega )\neq \Sigma _{tr}(k,\omega )$ and this was the main reason
for a number of erroneous conclusions. Here we discuss briefly the normal
state $\sigma (\omega )$ in the low frequency region $\omega <1$ $eV$ where
the \textit{intra-band} effects dominate the quasiparticle scattering. In
the low $\omega $ regime, the processing of data in the metallic state of
cuprates is usually done by using the generalized Drude formula for the
in-plane conductivity $\sigma (\omega )=\sigma _{1}(\omega )+i\sigma
_{2}(\omega )$ \cite{Dolgov}, \cite{Shulga}
\begin{equation}
\sigma _{ii}(\omega )=\frac{\omega _{p,ii}^{2}}{4\pi }\frac{1}{\Gamma
_{tr}(\omega ,T)-i\omega m_{tr}(\omega )/m_{\infty }}.  \label{Eq4}
\end{equation}%
$i=a,b$ enumerates the plane axis, $\omega _{p,ii}$, $\Gamma _{tr}(\omega
,T) $ and $m_{tr}(\omega )$ are the plasma frequency, transport scattering
rate and optical effective mass, respectively. For optimally doped HTSC
systems the best fit for $\Gamma _{tr}(\omega ,T)$ is given by $\Gamma
_{tr}(\omega ,T)\approx \max \{\alpha T,\beta \omega \}$ in a broad
temperature ($100$ $K<T<2000$ $K$) and frequency range ($10$ $meV<\omega
<200 $ $meV$), where $\alpha ,\beta $ are of the order one. These results
tell us that the quasiparticle liquid - which is responsible for transport
properties in HTSC - is not a simple weakly interacting Fermi liquid. In
that respect it is a well known fact that quasiparticles interacting with
phonons at finite $T$ are not described by the standard Landau-Fermi liquid,
in particular at higher temperatures $T>\Theta _{D}/5$, since the scattering
rate is of the order of quasiparticle energy, i.e. one has $\Gamma \sim \max
(\omega ,T)$. Such a system is well described by the \textit{%
Migdal-Eliashberg theory} whenever $\omega _{D}\ll E_{F}$ is fulfilled,
which in fact treats quasiparticles beyond the original Landau quasiparticle
concept. Note that even when the original Landau quasiparticle concept
fails, the transport properties may be described by the Boltzmann equation,
which provides a broader definition of the Landau-Fermi liquid.

As we said, in a number of articles it was \textit{incorrectly} assumed that
$\Gamma (\omega ,T)\approx \Gamma _{tr}(\omega ,T)$ holds in cuprates. Since
the theory of EPI gives that $\Gamma (\omega ,T)=const$ for $\omega >\omega
_{ph,\max }$ (the maximal phonon frequency) and experiments give $\Gamma
_{tr}(\omega ,T)\sim \omega $ for $\omega >\omega _{ph,\max }$, it was
(incorrectly) concluded that EPI in cuprates is small and can not be
responsible for superconductivity and transport properties in the normal
state. However, the thorough theoretical calculations of $\sigma (\omega )$
in a number of materials (including HTSC and LTSC systems) by including
realistic EPI spectral function $\alpha _{tr,EP}^{2}F(\omega )$ \cite{Dolgov}%
, \cite{Shulga}, \cite{MaksSav} showed that as a rule one has: (1) $\Gamma
_{tr}(\omega ,T)\neq \Gamma (\omega ,T)$ and (2) $\Gamma _{tr}(\omega
,T)\sim \omega $ is linear in the broad region up to $\omega _{sat}\simeq
-Im\Sigma _{tr}(\omega _{sat})\gg \omega _{ph,\max }$. Such a behavior of $%
\Gamma _{tr}(\omega ,T)$ and $\Gamma (\omega ,T)$ is also realized in HTSC
materials and in other metallic non-superconducting oxides; an extensive
discussion is given in \cite{Kulic-Review} and \cite{MaksimovReview}.

The main conclusion from the optics measurements is that EPI can explain the
optical spectra of cuprates for $\omega <1$ $eV$. Moreover, this analysis
shows that the EPI coupling constant is rather large, i.e. $1<\lambda
_{EPI}\lesssim 2$. The transport spectral function $\alpha _{tr}^{2}(\omega
)F(\omega )$ can be in principle extracted from the transport scattering
rate
\begin{equation}
\Gamma _{tr}(\omega ,T=0)=\frac{2\pi }{\omega }\int_{0}^{\omega }d\Omega
(\omega -\Omega )\alpha _{tr}^{2}(\Omega )F(\Omega )  \label{gamma-trans}
\end{equation}
- see \cite{Kulic-Review}, \cite{MaksimovReview}. However, real measurements
are performed at finite $T(>T_{c}$) where $\alpha _{tr}^{2}(\omega )F(\omega
)$ is the solution of the inhomogeneous Fredholm integral equation of the
first kind. This inverse problem at finite temperatures in cuprates was
studied in \cite{Dolgov} (see also \cite{Shulga}), where the smeared
structure of $\alpha _{tr}^{2}(\omega )F(\omega )$ in $YBa_{2}Cu_{3}O_{7-x}$%
\ was obtained, in a qualitative agreement with the shape of the phonon
density of states $F(\omega )$. At finite $T$ the problem is more complex
because the fine structure of $\alpha _{tr}^{2}(\omega )F(\omega )$ gets
blurred, as the calculations in \cite{Kaufmann} confirm. The latter showed
that $\alpha _{tr}^{2}(\omega )F(\omega )$ ends up at $\omega _{\max
}\approx 70-80$ $meV$, which is the maximal phonon frequency in cuprates.
This result indicates strongly that the EPI in cuprates is dominant in the
IR optics. Strictly speaking, this is still not a definitive proof for the
EPI since the solution of the Fredholm equation of the first kind is
extremely sensitive to the input, because the unknown function ($\alpha
_{tr}^{2}(\omega )F(\omega )$ which is a positive quantity) is under
integral and smoothed by some kernel. It is known that this smoothing can
miss some fine details of the solution.

We point out that if $R(\omega )$ (and $\sigma (\omega )$) are due to some
other bosonic process with large frequency cutoff $\omega _{c}$ in the
spectrum, as it is the case with SFI scattering where $\alpha
_{tr}^{2}(\omega )F(\omega )\sim g_{sf}^{2}Im\chi _{s}(\omega )$ and $\omega
_{c}\approx 400$ $meV$, the extracted $\alpha _{tr}^{2}(\omega )F(\omega )$
should end up at this high $\omega _{c}$. This was done in \cite{Carbotte}
for $T>T_{c}$, where it was found that $\alpha _{tr}^{2}(\omega )F(\omega )$
reaches negative values in a broad energy region $70$ $meV<\omega <\omega
_{c}$. These results rule out SFI as a dominant scattering mechanism in
cuprates and favor EPI. Recent

We stress that the extraction of $\Gamma _{tr}$ from $R(\omega )$ is a
rather subtle procedure, since it depends on the assumed value of dielectric
constant $\varepsilon _{\infty }$. For instance, if one takes $\varepsilon
_{\infty }=1$ then $\Gamma _{tr}^{EP}$ is linear up to very high $\omega $,
while for $\varepsilon _{\infty }>1$ the linearity of $\Gamma _{tr}^{EP}$
saturates at lower $\omega $. For instance, the extracted $\Gamma
_{tr}^{EP}(\omega ,T)$ in \cite{Puchkov} is linear up to very high $\omega $%
. Since in these papers there is no information (!)\ on $\varepsilon
_{\infty }$, the too linear behavior might mean that the ion background and
interband transitions (contained in $\varepsilon _{\infty }$) are not
properly taken into account in these papers. We stress again, that the
behavior of $\Gamma _{tr}(\omega )$ is linear up to much higher frequencies
for $\varepsilon _{\infty }=1$ than for $\varepsilon _{\infty }\approx 4-5$
- the characteristic value for HTSC, giving a lot of room for inadequate
interpretations of results. Note that recent elipsometric measurements on
YBCO \cite{BorisMPI} confirm the earlier results \cite{Bozovic} that $%
\varepsilon _{\infty }\geq 4$ and that $\Gamma _{tr}^{EP}$ saturates at
lower frequency than it was the case in Ref. \cite{Puchkov}. We stress again
that a reliable estimate of the value and of the $\omega ,T$ dependence of $%
\Gamma _{tr}(\omega )$ and $m(\omega )$ can be done, not from the
reflectivity measurements \cite{Puchkov}, but from elipsometric ones only
\cite{Bozovic}, \cite{BorisMPI}.

In concluding this part we stress two facts: (\textbf{1}) The large
difference in the $\omega ,T$ behavior of $\Gamma _{tr}^{EPI}(\omega ,T)$
and $\Gamma ^{EPI}(\omega ,T)$ is not a specificity of cuprates but it is
realized also in a number of $LTSC$ materials. In fact this is a common
behavior even in simple metals, such as Al or Pb where $\Gamma ^{EPI}(\omega
,T)$ saturates at much lower (Debay) frequency than $\Gamma
_{tr}^{EPI}(\omega ,T)$ and $\Gamma _{tr}^{\ast ,EPI}(\omega ,T)$ do - see
more in \cite{Kulic-Review}, \cite{MaksimovReview} and references therein.
In that respect the difference between simple metals and cuprates is in the
scale of phonon frequencies, i.e. $\omega _{\max }^{ph}\sim 100$ $K$ in
simple metals, while $\omega _{\max }^{ph}\sim 1000$ $K$ in cuprates. Having
in mind these well established and well understood facts, it is very
surprising that even nowadays, twenty years after the discovery of cuprates,
the essential and quantitative difference between $\Gamma $ and $\Gamma
_{tr} $ is neglected in the analysis of experimental data. For instance, by
neglecting the pronounced (qualitative and quantitative) difference between $%
\Gamma _{tr}(\omega ,T)$ and $\Gamma (\omega ,T)$, in the recent papers \cite%
{Puchkov} were made far reaching, but unjustified, conclusions that the
magnetic pairing mechanism prevails; (\textbf{2}) It is worth mentioning
that quite similar (to cuprates) properties, of $\sigma (\omega )$, $%
R(\omega )$ and $\rho (T)$ were observed in experiments \cite{Bozovic} on
isotropic metallic oxides $La_{0.5}Sr_{0.5}CoO_{3}$ and $%
Ca_{0.5}Sr_{0.5}RuO_{3}$. We stress that in these compounds there are no
signs of antiferromagnetic fluctuations (which are present in cuprates) and
the peculiar behavior is probably due to the EPI.

In that respect, recent experiments related to the \textit{optical sum-rule }%
with the transfer of the spectral weight from high to low frequency is an
additional example of controversies in this field coming from inadequate
interpretation of experimental results. We stress that the theory based on
EPI can explain the transfer of spectral weight by taking into account
strong the $\omega $ and $T$ dependence of $\Gamma _{tr}^{EPI}(\omega ,T)$
due to the EPI - for details see \cite{Maks-Karakaz}, \cite{BorisMPI} and
\cite{Kulic-AIP}.

The temperature behavior of \textit{in-plane resistivity} $\rho _{a-b}(T)$
is a direct consequence of quasi-$2D$ motion of quasiparticles and of
inelastic scattering which they experience. At present, there is no
consensus on the origin of linear temperature dependence of the in-plane
resistivity $\rho _{a-b}(T)$ in the normal state. Many researchers
(erroneously) believe that such a behavior can not be due to the EPI. The
inadequacy of this claim was already demonstrated by analyzing the dynamic
conductivity $\sigma (\omega )$. It is well-known that at temperatures $%
T>\Theta _{D}/5$ and for the Debay spectrum one has
\begin{equation}
\rho (T)\simeq 8\pi ^{2}\lambda _{tr}^{EP}\frac{k_{B}T}{\hbar \omega _{p}^{2}%
}=\rho ^{\prime }T.  \label{Eq17}
\end{equation}%
In cuprates the reach and broad spectrum of $\alpha _{tr}^{2}(\omega
)F(\omega )$ favors such a linear behavior in a broader $T$ region. The
measured transport coupling constant $\lambda _{tr}$ contains in principle
all scattering mechanisms, although usually some of them dominate. For
instance, the proponents of SFI mechanism assume that $\lambda _{tr}$ is
entirely due to scattering on spin fluctuations. However, by taking into
account specificities of cuprates the experimental results for the in-plane
resistivity $\rho _{a-b}(T)$ can be satisfactory explained by the EPI
mechanism. From tunnelling experiments \cite{Tunnelling-HTSC} one concludes
that $\lambda \approx 2-3$ and if one assumes that $\lambda _{tr}\approx
\lambda $ and $\omega _{pl}\geq (3-4)$ $eV$ (the value obtained from the
band-structure calculations) then Eq. (\ref{Eq17}) describes the
experimental situation rather well. The plasma frequency $\omega _{pl}$
which enters Eq.(\ref{Eq17}) can be extracted from optic measurements ($%
\omega _{pl,ex}$), i.e. from the width of the Drude peak at small
frequencies. However, since $\lambda _{tr}\approx 0.25\omega
_{pl}^{2}(eV)\rho ^{\prime }(\mu \Omega \mathrm{cm}/K)$\ there is an
experimental constraint on $\lambda _{tr}$. This gives \cite{Bozovic} that $%
\omega _{pl}\approx (2-2.5)$ $eV$ \ and $\rho ^{\prime }\approx 0.6$ in the
oriented YBCO films, while $\rho ^{\prime }\approx 0.3$ in single crystals
of BSCCO. These results impose a limit on $\lambda _{tr}\approx 0.4-0.8$.

The above analysis implies, that in order to explain $\rho (T)$ with small $%
\lambda _{tr}$ and high $T_{c}$ (which needs large $\lambda $) by EPI it is
necessary to have $\lambda _{tr}\leq (\lambda /3)$. This means that in
cuprates EPI is, due to some reasons, reduced in transport properties where $%
\lambda _{tr}\ll \lambda $. This reduction of $\omega _{p}^{2}$ and $\lambda
_{tr}$ means that the $\lambda $ and $\lambda _{tr}$ contain renormalization
(with respect to the $LDA$ results) due to various quasiparticle scattering
processes and interactions, which do not enter in the $LDA$ theory. In
subsequent chapters we shall argue that the strong suppression of $\lambda
_{tr}$ may have its origin in strong electronic correlations \cite{Kulic1},
\cite{Kulic2}, \cite{Kulic3}.

\textit{In conclusion}, optics and resistivity measurements in normal state
of cuprates are much more in favor of EPI than against it. However, some
intriguing questions still remains to be answered: (\textbf{i}) what are the
values of $\lambda _{tr}$ and $\omega _{pl}$; (\textbf{ii}) what is the
reason that $\lambda _{tr}\ll \lambda $ is realized in cuprates; (\textbf{iii%
}) what is the role of Coulomb scattering in $\sigma (\omega )$ and $\rho
(T) $. Later on we shall argue that ARPES measurements in cuprates give
evidence for a contribution of Coulomb scattering at higher frequencies,
where $\Gamma (\omega )\approx \Gamma _{0}+\lambda _{c}\omega $ for $\omega
>\omega _{\max }^{ph}$ with $\lambda _{c}\approx 0.4$. So, in spite of the
fact that EPI is suppressed in transport properties it is sufficiently
strong that it dominates in the self-energy in some frequency and
temperature range. It is possible that at higher temperatures Coulomb
scattering dominates $\rho (T)$, which certainly does not disqualify EPI as
the pairing mechanism in cuprates. For better understanding of $\rho (T)$ we
need a controllable theory for Coulomb scattering in strongly correlated
systems, which is at present lacking.

\subsection{Electronic and phonon Raman scattering in cuprates}

(a) \textit{Electronic Raman scattering} in various cuprates show a
remarkable correlation between the Raman cross-section $\tilde{S}_{\exp
}(\omega )$ and the optical conductivity $\sigma _{a-b}(\omega )$, i.e. $%
\tilde{S}_{\exp }(\omega )\sim \sigma _{a-b}(\omega )$ \cite{Kulic-Review}.
Previously it was demonstrated that $\sigma _{a-b}(\omega )$ depends on the
transport scattering rate $\Gamma _{tr}(\omega ,T)$. We have also
demonstrated that the EPI with the very broad spectral function $\alpha
^{2}F(\omega )$ (see Fig.2 below) explains in a natural way the $\omega ,T$
dependence of $\sigma _{a-b}(\omega )$ and $\Gamma _{tr}(\omega ,T)$.
Therefore, the Raman spectra in cuprates can be explained by the EPI in
conjunction with strong correlations. This conclusion is supported by
calculations of the Raman cross-section \cite{Rashkeev} which take into
account the EPI with $\alpha ^{2}F(\omega )$ extracted from the tunnelling
measurements on $YBa_{2}Cu_{3}O_{6+x}$ and $Bi_{2}Sr_{2}CaCu_{2}O_{8+x}$
\cite{Tunnelling-HTSC}, \cite{Kulic-Review}. Quite similar (to cuprates)
properties of the electronic Raman scattering (besides $\sigma (\omega )$, $%
R(\omega )$ and $\rho (T)$) were observed in experiments \cite{Bozovic} on
isotropic metallic oxides $La_{0.5}Sr_{0.5}CoO_{3}$ and $%
Ca_{0.5}Sr_{0.5}RuO_{3}$ where there are no signs of antiferromagnetic
fluctuations.

(b) Two important experimental results related to the \textit{phonon Raman
scattering }give evidence for strong EPI in cuprates: (i) there is a
pronounced asymmetric line-shape (of the Fano resonance) in the metallic
state. For instance in $YBa_{2}Cu_{3}O_{7}$ two Raman modes at $115$ $%
cm^{-1} $ (Ba dominated mode) and at $340$ $cm^{-1}$ (O dominated mode in
the CuO$_{2}$ planes)\ show pronounced asymmetry which is absent in $%
YBa_{2}Cu_{3}O_{6}$. This result points to the strong interaction of Raman
active phonons with continuum states (quasiparticles); (ii) The phonon
frequencies for some $A_{1g}$ and $B_{1g}$ are strongly renormalized in the
superconducting state, pointing again to the large EPI - see more in \cite%
{Kulic-Review}, \cite{Kulic-AIP}.

\subsection{Tunnelling spectroscopy and Eliashberg spectral function}

By measuring the I-V\ characteristic, i.e., the tunnelling conductance $%
G(V)=dI/dV$, one can determine the Eliashberg spectral function $\alpha
^{2}F(\omega )$ which determines the pairing coupling constant $\lambda
=2\int d\omega \alpha ^{2}F(\omega )/\omega $. A number of experiments \cite%
{Tunnelling-HTSC} on tunnelling- and break-junctions in cuprates gave $%
\alpha ^{2}F(\omega )$ which is spread over a broad range of frequencies ($%
0<\omega <80$ $meV$) and whose maxima coincide with the maxima in the phonon
density of states $F(\omega )$ - see Fig. 2. This is a rather strong proof
of importance of the EPI in pairing. Moreover, most of these experiments
give the range of the coupling constant $1<\lambda <2.5$.

\begin{figure}[tbp]
\resizebox{.5\textwidth}{!} {
\includegraphics*[width=8cm]{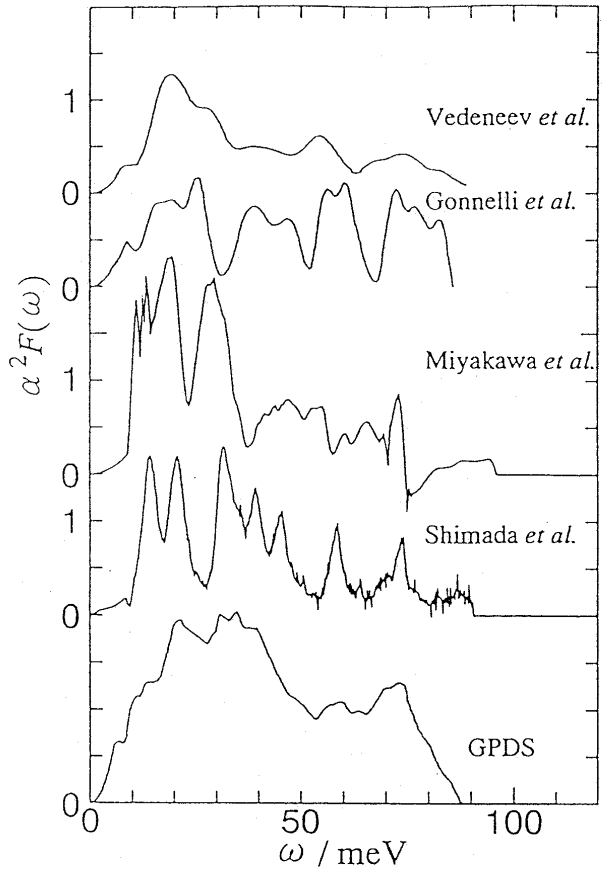}}
\caption{The spectral function $\protect\alpha ^{2}F(\protect\omega )$
obtained from measurements of $G(V)$ by various groups on various junctions.
The generalized density of states GPDS\ for $Bi2212$ is plotted at the
bottom. From Ref. \protect\cite{Tunnelling-HTSC}.}
\label{TunnelFig}
\end{figure}

Since practically all phonon modes contribute to $\lambda $ - see Fig. 2,
this means that on the average each particular phonon mode is moderately
coupled to electrons thus keeping the lattice stable. Additionally, they
have found that some low-frequency phonon modes corresponding to $Cu,Sr$ and
$Ca$ vibrations are rather strongly coupled to electrons. The high frequency
oxygen vibrations along the $c$-axis interact with quasiparticles
appreciably too. These results confirm the importance of axial phonon modes
in which the change of the Madelung energy is involved, thus supporting the
idea conveyed through this article of the importance of the ionic Madelung
energy in the EPI interaction of cuprates.

Recent STM experiments \cite{Davis} confirmed partly the results
of tunnelling measurements. They find near the anti-nodal point
strong EPI with 40 meV phonons and from Fourier-transformed
spectra the exclude the magnetic resonance as a possible origin.

In conclusion, the common results for tunnelling measurements in cuprates,
including $Ba_{1-x}K_{x}BiO_{3}$ too \cite{Huang}, is that no particular
mode can be singled out in the spectral function $\alpha ^{2}F(\omega )$ as
being the only one which dominates in pairing mechanism. This important
result means that the high T$_{c}$ is not attributable to a particular
phonon mode in the EPI mechanism, since all phonon modes contribute to $%
\lambda $. Having in mind that the phonon spectrum in cuprates is very broad
(up to $80$ $meV$ ), then the large EPI coupling constant ($\lambda \approx
2 $) in cuprates is not surprising at all. We stress, that compared to
neutron scattering experiments the tunnelling experiments are superior in
determining the EPI spectral function $\alpha ^{2}F(\omega )$ - see more in
\cite{Kulic-Review}.

\subsection{Isotope effect for various doping}

It is well known that in the pure EPI pairing mechanism the total isotope
coefficient $\alpha $ is given by%
\begin{equation}
\alpha =\sum_{i,p}\alpha _{i}^{(p)}=-\sum_{i,p}\frac{d\ln T_{c}}{d\ln
M_{i}^{(p)}},  \label{isotope}
\end{equation}
where $M_{i}^{(p)}$ is the mass of the i-th element in the p-th
crystallographic position. From this formula one can deduce that
the relative change of $T_{c}$, $\delta T_{c}/T_{c}$, for heavier
elements is rather small - for instance it is 0.02 for
$^{135}Ba\rightarrow ^{138}Ba$, 0.03 for $^{63}Cu\rightarrow
^{65}Cu$ and 0.07 for $^{138}La\rightarrow ^{139}La$. This means
that measurements of $\alpha _{i}$ for heavier elements are at/or
beyond the ability of experimental techniques. Therefore
most isotope effect measurements were done by substituting light atoms $%
^{16}O$ by $^{18}O$. In that respect one should have in mind the
tunnelling experiments discussed above, which tell us that
practically all phonons
contribute to the Eliashberg pairing function $\alpha ^{2}F(\mathbf{k}%
,\omega )$, and that oxygen modes give only moderate contribution to T$_{c}$%
. So, small oxygen isotope effect does not exclude the EPI
mechanism of pairing.

Measurements of the isotope coefficients $\alpha _{O}$ (and $\alpha _{Cu}$)
were performed on various hole-doped and electron-doped cuprates. A brief
summary of the main results is as follows \cite{Franck}: (\textbf{1}) The $O$
isotope coefficient $\alpha _{O}$ strongly depends on the hole concentration
in the hole-doped materials; in each family of cuprates ($%
YBa_{2}Cu_{3}O_{7-x}$, or $La_{2-x}Sr_{x}CuO_{4}$ etc.) a small oxygen
isotope effect is observed in the optimally doped (maximal $T_{c}$) samples.
For instance $\alpha _{O}\approx 0.02-0.05$ in $YBa_{2}Cu_{3}O_{7}$ with $%
T_{c,\max }\approx 91$ $K$, $\alpha _{O}\approx 0.1-0.2$ in $%
La_{1.85}Sr_{0.15}CuO_{4}$ with $T_{c,\max }\approx 35$ $K$; $\alpha
_{O}\approx 0.03-0.05$ in $Bi_{2}Sr_{2}CaCu_{2}O_{8}$ with $T_{c,\max
}\approx 76$ $K$; $\alpha _{O}\approx 0.03$ and even negative ($-0.013$) in $%
Bi_{2}Sr_{2}Ca_{2}Cu_{2}O_{10}$ with $T_{c,\max }\approx 110$ $K$; the
experiments on $Tl_{2}Ca_{n-1}BaCu_{n}O_{2n+4}$ ($n=2,3$) with $T_{c,\max
}\approx 121$ $K$ are still unreliable and $\alpha _{O}$ is unknown; $\alpha
_{O}<0.05$ in the electron-doped $(Nd_{1-x}Ce_{x})_{2}CuO_{4}$ with $%
T_{c,\max }\approx 24$ $K$. (\textbf{2}) For hole concentrations away from
the optimal one, T$_{c}$ decreases while $\alpha _{O}$ increases and in some
cases reaches large value $\alpha _{O}\approx 0.5$. This holds not only for
parent compounds but also for systems with substitutions, like $%
(Y_{1-x-y}\Pr_{x}Ca_{y})Ba_{2}Cu_{3}O_{7}$, $Y_{1-y}Ca_{y}Ba_{2}Cu_{4}O_{4}$
and $Bi_{2}Sr_{2}Ca_{1-x}Y_{x}Cu_{2}O_{8}$. Note that the decrease of T$_{c}$
is not a prerequisite for the increase in $\alpha _{O}$. This became clear
from the $Cu$ substituted experiments $YBa_{2}(Cu_{1-x}Zn_{x})_{3}O_{7}$
where the decrease of T$_{c}$ (by increasing the $Zn$ concentration) is
followed by only a small increase of $\alpha _{O}$. Only in the case of very
low $T_{c}<20$ $K$,$\ \alpha _{O}$ becomes large, i.e. $\alpha _{O}>0.1$. (%
\textbf{3)} The largest $\alpha _{O}$ is obtained even in the optimally
doped compounds like in systems with substitution, such as $%
La_{1.85}Sr_{0.15}Cu_{1-x}M_{x}O_{4}$, $M=Fe,Co$, where $\alpha _{O}\approx
1.3$ for $x\approx 0.4$ $\%$. (\textbf{4}) In $La_{2-x}M_{x}CuO_{4}$ there
is a $Cu$ isotope effect which is of the order of the oxygen one, i.e. $%
\alpha _{Cu}\approx \alpha _{O}$ giving $\alpha _{Cu}+\alpha _{O}\approx
0.25-0.35$ for optimally doped systems ($x=0.15$). In the case when $x=0.125$
with $T_{c}\ll T_{c,\max }$ one has$\ \alpha _{Cu}\approx 0.8-1$ with $%
\alpha _{Cu}+\alpha _{O}\approx 1.8$. The appreciate copper isotope effect
in $La_{2-x}M_{x}CuO_{4}$ tells us that vibrations of other than oxygen ions
could be important in giving high T$_{c}$. The latter property is more
obvious from tunnelling measurements which are discussed above. (\textbf{5})
There is a \textit{negative} Cu isotope effect in the oxygen-deficient
system $YBa_{2}Cu_{3}O_{7-x}$ where $\alpha _{Cu}$ is between $-0.14$ and $%
-0.34$ if T$_{c}$ lies in the $60$ $K$ plateau. (\textbf{6}) There are
reports on a \textit{small negative} $\alpha _{O}$ in some systems like $%
YSr_{2}Cu_{3}O_{7}$ with $\alpha _{O}\approx -0.02$ and in $BSCCO-2223$ ($%
T_{c}=110$ $K$) where $\alpha _{O}\approx -0.013$ etc. However, the systems
with negative $\alpha _{O}$ present considerable experimental difficulties,
as it is pointed out in \cite{Franck}.

The above enumerated results, despite experimental difficulties, tell us
that the EPI interaction is involved in the pairing mechanism of cuprates
and that the relation of $T_{c}$ to ionic masses is highly nontrivial and
non-BCS-like. By assuming that the experimental results on the isotope
effect reflect an intrinsic property of cuprates one can rise a question:
which theory can explain these results? Since at present there is no
consensus on the pairing mechanism in HTSC materials there is also no
definite theory for the isotope effect. Besides the calculation of the
coupling constant $\lambda $ any microscopic theory of pairing is confronted
also with the following questions: $\mathbf{(a)}$ why is the isotope effect
small in optimally doped systems and $\mathbf{(b)}$ why $\alpha $ increases
rapidly by further under(over)doping of the system? The answer to this
question is additionally complicated by the quasi-2D structure of cuprates
and the closeness of these materials to the Mott-Hubbard isolator. It is
known that the later gives rise to the smallness of the phase stiffness $%
K_{s}^{0}$ of the superconducting order parameter $\Delta =\mid \Delta \mid
e^{i\varphi }$. In 2D superconductors the energy increase due to phase
fluctuations in the conduction plane is given by \cite{Emery}%
\begin{equation}
F_{phase}(\varphi ,T)=\frac{1}{2}K_{s}^{0}\int d^{2}x(\nabla \varphi )^{2},
\label{phas-fluc}
\end{equation}%
where $K_{s}^{0}=\hbar ^{2}n_{s}(T)/4m^{\ast }$ is the 2D stiffness, $n_{s}$
is the 2D superconducting charge density and $m^{\ast }\approx 2m_{e}$ \ is
the effective mass. In the case of the Berezinskii-Kosterlitz-Thouless
transition (vortex-antivortex unbinding) one has $T_{c}=(\pi /2)K_{s}(T_{c})$%
, where $K_{s}$ is the renormalized stiffness. In that case the isotope
effect is completely different from the prediction of the BCS theory. In
that respect the very careful experiments on isotope effect of the \textit{%
penetration depth} in the a-b plane of YBCO show a significant increase upon
the substitution $O^{16}\rightarrow O^{18}$, i.e. $(\Delta \lambda _{ab}/$ $%
\lambda _{ab})=(^{18}\lambda _{ab}-^{16}\lambda _{ab})/^{16}\lambda
_{ab}=2.8 $ $\%$ at $4$ $K$ \cite{lambda-isotope}. Since $\lambda _{ab}\sim
m^{\ast }$ this could mean that there is a nonadiabatic increase of the
effective mass $m^{\ast }$.

\subsection{Angle-resolved photoemission spectroscopy in cuprates}

ARPES is nowadays a leading spectroscopy method in the solid state physics
\cite{Shen}. The method consists in shining light (photons) with energies
between $20-1000$ $eV$ on the sample and by detecting momentum ($\mathbf{k}$%
)- and energy($\omega $)-distribution of the outgoing electrons. The
resolution of ARPES has been significantly increased in the last decade with
the energy resolution of $\Delta E\approx 2$ $meV$ (for photon energies $%
\sim 20$ $eV$) and angular resolution of $\Delta \theta \approx 0.2{%
{}^{\circ }}$. The ARPES method is surface sensitive technique, since the
average escape depth ($l_{esc}$)\ of the outgoing electrons is of the order
of $l_{esc}\sim 10$ \AA , of course depending on the energy of incoming
photons. Therefore, very good surfaces are needed in order that the results
be representative of bulk samples. In that respect the most reliable studies
were done on the bilayer $Bi_{2}Sr_{2}CaCu_{2}O_{8}$ ($Bi2212$) and its
single layer counterpart $Bi_{2}Sr_{2}CuO_{6}$ ($Bi2201$), since these
materials contain weakly coupled $BiO$ planes with the longest inter-plane
separation in the cuprates. This results in a \textit{natural cleavage}
plane making these materials superior to others in ARPES experiments. After
a drastic improvement of sample quality in other families of HTSC materials,
the ARPES technique has became a central method in theoretical
considerations. Potentially, it gives valuable information on the
quasiparticle Green's function, i.e. on the quasiparticle spectrum and
life-time effects. The ARPES can indirectly give information on the momentum
and energy dependence of the pairing potential. Furthermore, the electronic
spectrum of the (above mentioned) cuprates is highly \textit{quasi-2D} which
allows an unambiguous determination of the initial state momentum from the
measured final state momentum, since the component parallel to the surface
is conserved in photoemission. In this case the ARPES probes (under some
favorable conditions) directly the single particle spectral function $A(%
\mathbf{k},\omega )$. In the following we discuss only those ARPES
experiments which give evidence for the importance of the EPI in cuprates -
see more in \cite{Shen}.

The \textit{photoemission} measures a nonlinear response function of the
electron system, and under some conditions it is analyzed in the so-called
\textit{three-step model}, where the total photoemission intensity $I_{tot}(%
\mathbf{k},\omega )\approx I\cdot I_{2}\cdot I_{3}$ is the product of three
independent terms: (\textbf{1}) $I$ - describes optical excitation of the
electron in the bulk; (\textbf{2}) $I_{2}$ - the scattering probability of
the travelling electrons; (\textbf{2}) $I_{3}$ - the transmission
probability through the surface potential barrier. The central quantity in
the three-step model is $I(\mathbf{k},\omega )$ and it turns out that it can
be written in the form for $\mathbf{k=k}_{\parallel }$ \cite{Shen}
\[
I(\mathbf{k},\omega )\simeq I_{0}(\mathbf{k},\upsilon )f(\omega )A(\mathbf{k}%
,\omega )
\]%
\[
I_{0}(\mathbf{k},\upsilon )\sim \mid \langle \psi _{f}\mid \mathbf{pA\mid }%
\psi _{i}\rangle \mid ^{2}
\]%
\begin{equation}
A(\mathbf{k},\omega )=-\frac{1}{\pi }\frac{\Sigma _{2}(\mathbf{k},\omega )}{%
[\omega -\xi (\mathbf{k})-\Sigma _{1}(\mathbf{k},\omega )]^{2}+\Sigma
_{2}^{2}(\mathbf{k},\omega )}  \label{intensity}
\end{equation}%
Here, $\langle \psi _{f}\mid \mathbf{pA\mid }\psi _{i}\rangle $ is the
dipole matrix element which depends on $\mathbf{k}$, polarization and energy
$\upsilon $ of the incoming photons. The knowledge of the matrix element is
of a great importance and its calculation from first principles was done
carefully in \cite{Bansil}. $f(\omega )=1/(1+\exp \{\omega /T\})$ is the
Fermi function, $A(\mathbf{k},\omega )=-$\textrm{Im}$G(\mathbf{k},\omega
)/\pi $, $G(\mathbf{k},\omega )$ and $\Sigma (\mathbf{k},\omega )=\Sigma
_{1}(\mathbf{k},\omega )+i\Sigma _{2}(\mathbf{k},\omega )$ are the spectral
function, the quasiparticle Green's function and the self-energy,
respectively.

In some period of the HTSC era there were a number of controversial ARPES
results and interpretations, due to bad samples and to the euphoria with
exotic theories. For instance, a number of (now well) established results
were \textit{questioned} in the first ARPES measurements, such as: the shape
of the Fermi surface, which is correctly predicted by the LDA band-structure
calculations; bilayer splitting in $Bi2212$, etc.

\begin{figure}[tbp]
\resizebox{.5\textwidth}{!} {
\includegraphics*[width=4cm]{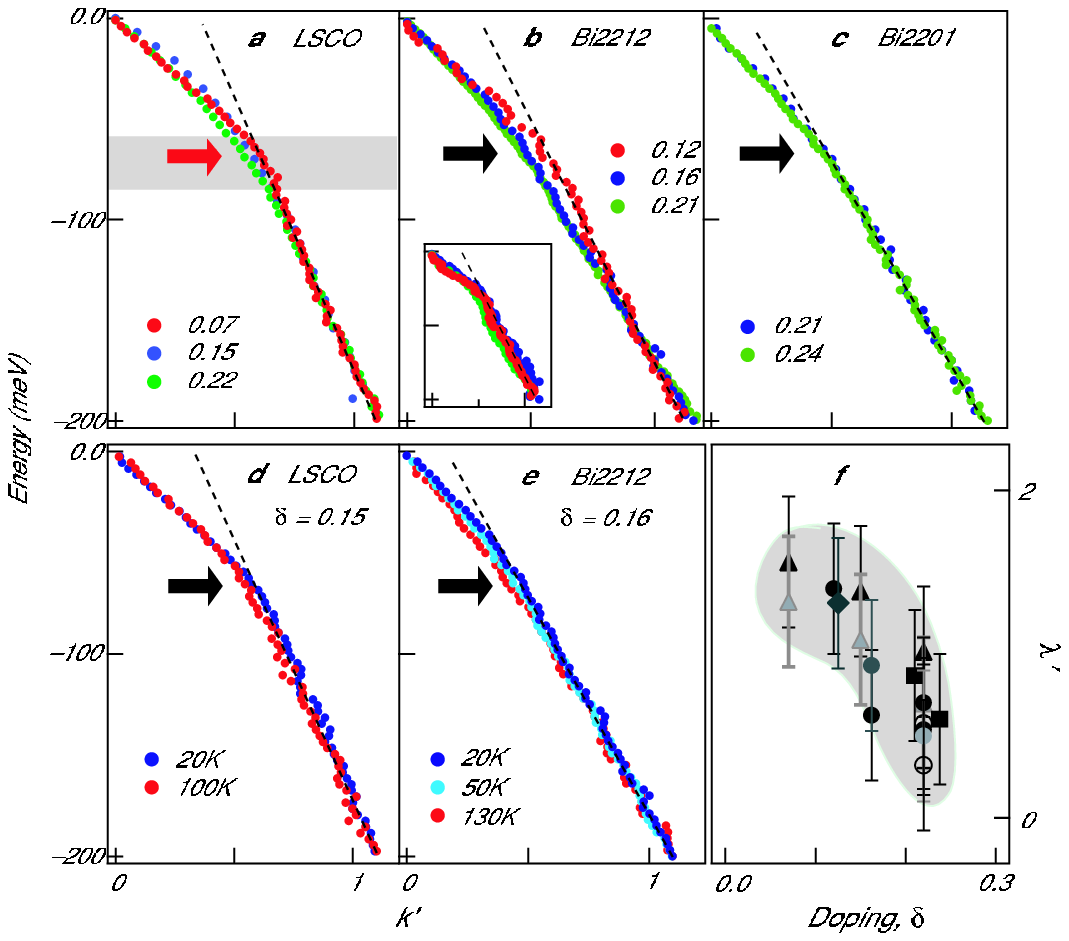}}
{\includegraphics*[width=9cm]{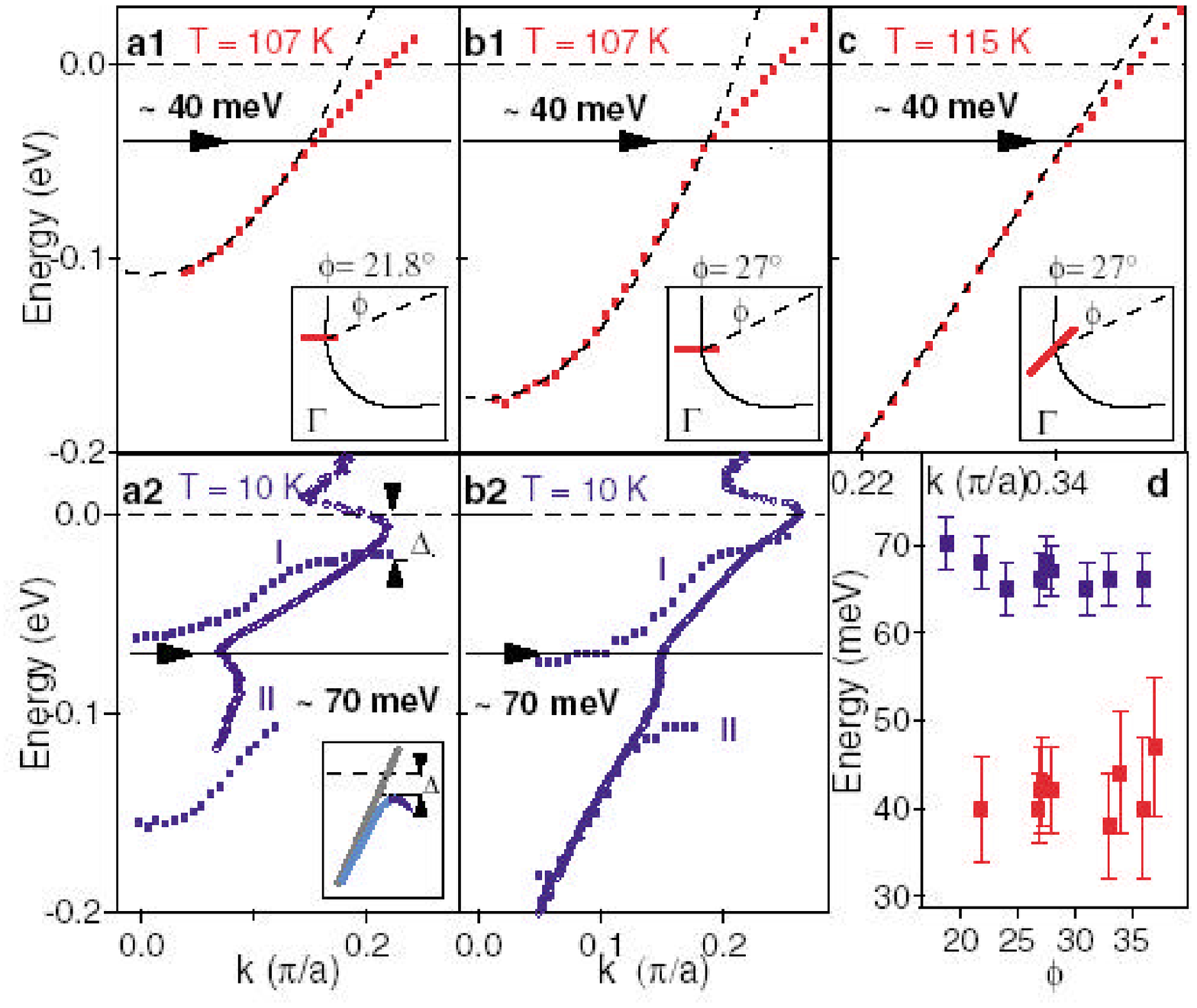}}
\caption{(\textbf{Top}) Quasiparticle dispersion of $Bi2212$, $Bi2201$ and $%
LSCO$ along the \textit{nodal} direction, plotted vs the momentum $k$ for $%
(a)-(c)$ different doings, and $(d)-(e)$ different $T$; black arrows
indicate the kink energy; the red arrow indicates the energy of the $q=(%
\protect\pi,0)$ oxygen stretching phonon mode; inset of $(e)$- T-dependent $%
\Sigma ^{\prime } $ for optimally doped $Bi2212$; $(f)$ - doping dependence
of $\protect\lambda^{\prime}$ along $(0,0)-(\protect\pi,\protect\pi)$ for
the different HTSC oxides. From Ref. \protect\cite{Lanzara}. (\textbf{Bottom}%
) Quasiparticle dispersion $E(k)$ in the normal state (a1, b1, c), at 107 K
and 115 K, along various directions $\protect\phi$ around the \textit{%
anti-nodal} point. The kink at $E=40meV$ is shown by the horizontal arrow.
(a2 and b2) is $E(k)$ in the superconducting state at 10 K with the shifted
kink to $70meV$. (d) kink positions as a function of $\protect\phi$ in the
anti-nodal region. From Ref. \protect\cite{Cuk}.}
\label{ARPESLanzFig}
\end{figure}

We summarize here some important ARPES results which were obtained recently,
first in the \textit{normal state} \cite{Shen}: (\textbf{1}$_{N}$) There is
a well defined Fermi surface in the metallic state - with the topology
predicted by the LDA but the bands are narrower than the LDA ones; (\textbf{2%
}$_{N}$) the spectral lines are broad with $\mid \Sigma _{2}(\mathbf{k}%
,\omega )\mid \sim \omega $ (or $\sim T$ for $T>\omega $); (\textbf{3}$_{N}$%
) there is a bilayer band splitting in $Bi2212$ (at least in the over-doped
state); (\textbf{4}$_{N}$) at temperatures $T_{c}<T<T^{\ast }$ and in the
under-doped cuprates there is a d-wave like pseudo-gap $\Delta _{pg}(\mathbf{%
k})\sim \Delta _{pg,0}(\cos k_{x}-\cos k_{y})$ in the quasiparticle
spectrum; (\textbf{5}$_{N}$) the pseudo-gap $\Delta _{pg,0}$ increases by
lowering doping; (\textbf{6}$_{N}$) the ARPES self-energy gives clear
evidence that the EPI interaction, with the \textit{characteristic phonon}
energy $\omega _{ph}$, is rather strong. At $T>T_{c}$ there are \textit{kinks%
} in the quasiparticle dispersion $\omega (\xi _{\mathbf{k}})$ in the
\textit{nodal} direction (along the $(0,0)-(\pi ,\pi )$ line) at $\omega
_{ph}^{(70)}\sim (60-70)$ $meV$ \cite{Lanzara}, see Fig.~\ref{ARPESLanzFig},
and near the \textit{anti-nodal point} $(\pi ,0)$ at 40 meV \cite{Cuk} - see
Fig.~\ref{ARPESLanzFig}. (\textbf{7}$_{N}$) The holes couple practically to
the whole spectrum of phonons. For instance, at least three group of phonons
(interacting with holes) were extracted recently from the ARPES self-energy
in $La_{2-x}Sr_{x}CuO_{4}$ \cite{Zhou-PRL}. (\textbf{8}$_{N}$) The EPI
coupling constant which is obtained by comparing the slope $d\omega /d\xi _{%
\mathbf{k}}$ of the quasiparticle energy $\omega (\xi _{\mathbf{k}})$ at
very small $\mid \xi _{\mathbf{k}}\mid \ll \omega _{ph}$ and $\mid \xi _{%
\mathbf{k}}\mid \gg \omega _{ph}$ gives $\lambda _{epi}\approx 1-2$, while
the Coulomb part is $\lambda _{c}\approx 0.4$. (\textbf{9}$_{N}$) Recent
ARPES spectra in the optimally doped \ $Bi2212$ near the nodal and
anti-nodal point \cite{Lanzara2} show a pronounced isotope effect in the
real part of the quasiparticle self-energy, thus pointing out the important
role of the EPI.

The corresponding ARPES results in the \textit{superconducting state} are
the following \cite{Shen}: (\textbf{1}$_{S}$) there is an anisotropic
superconducting gap in most HTSC compounds, which is predominately d-wave
like, i.e. $\Delta _{sc}(\mathbf{k})\sim \Delta _{0}(\cos k_{x}-\cos k_{y})$
with $2\Delta _{0}/T_{c}\approx 5-6$; (\textbf{2}$_{S}$) there are dramatic
changes in the spectral shapes near the anti-nodal point $(\pi ,0)$, i.e. a
\textit{sharp quasiparticle peak} develops at the lowest binding energy
followed by a dip and a broader hump, giving rise to the so called \textit{%
peak-dip-hump structure}; (\textbf{3}$_{S}$) the kink in the quasiparticle
energy around the nodal-point and at $(60-70)$ $meV$ is surprisingly \textit{%
not-shifted} in the superconducting state. To remind the reader, in the
standard Eliashberg theory the kink should be shifted to $\omega
_{ph}+\Delta _{0}$ at any point at the Fermi surface. (\textbf{4}$_{S}$) the
anti-nodal kink at $\omega _{ph}^{(40)}\sim 40$ $meV$ is shifted in the
superconducting state by $\Delta _{0}$, i.e. $\omega _{ph}^{(40)}\rightarrow
\omega _{ph}^{(40)}+\Delta _{0}=(65-70)meV$ since $\Delta _{0}=(25-30)meV$.
In the following sections we shall argue that all these properties can be
explained by EPI with FSP and the coupling constant $\lambda >1$.

\section{Theory of strong correlations without slave bosons}

The well established fact is that \textit{strong electronic correlations}
are pronounced in cuprates. However, the LDA theory fails to capture effects
of strong correlations by treating them as a local perturbation only. This
is an unrealistic approximation for cuprates, where strong correlations
introduce non-locality in charge interactions. One of the shortcomings of
the $LDA$ is that in the half-filled case (with $n=1$ and one particle per
lattice site) it predicts metallic state only, thus missing the existence of
the \textit{Mott insulating state.} In the latter, particles are localized
at lattice sites independently of the (non)existence of the $AF$. This
localization is due to the large Coulomb repulsion $U$ at given lattice
site, i.e. $U\gg W$ where $W$ is the band width. Some properties in the
metallic state can not be described by the simple canonical Landau-Fermi
liquid concept. For instance, recent ARPES photoemission measurements on the
hole-doped samples show a well defined Fermi surface in the one-particle
energy spectrum, which contains $1-\delta $ electrons in the Fermi volume ($%
\delta $ is the hole concentration), but the band width is $(2-3)$ times
smaller than the $LDA$ band structure calculations predict \cite{Shen}. The $%
"1-\delta "$ behavior of the Fermi volume is consistent with the Luttinger
theorem. However, experimental data on the dynamic conductivity (spectral
weight of the Drude peak), Hall measurements etc. indicate that in transport
properties a low density of hole-like charge carriers (which is proportional
to $\delta $) participates predominantly. These carriers experience strong
scattering and their inverse lifetime is proportional to the temperature (at
$T>T_{c}$) as we discussed earlier. It is worth mentioning here that the
local moments on the $Cu$ sites, which are localized in the parent $AF$
compound, are counted as a part of the Fermi surface area when the system is
doped by small concentration of holes in the metallic state. The latter fact
gives rise to a \textit{large Fermi surface} which scales with the number
(per site) of electrons $1-\delta $. At the same time the conductivity
sum-rule is proportional to the number of doped holes $\delta $, at least in
underdoped systems, i.e., $\int_{0}^{\omega _{c}}d\omega \sigma _{1}(\omega
)\sim \delta $) instead of $1-\delta $ as in the canonical Landau-Fermi
liquid. These two properties tell us that in cuprates we deal with a \textit{%
correlated state}, and the latter must be due to the specific electronic
structure of cuprates. The common ingredient of all cuprates is the presence
of $Cu$ atoms. In order to account for the absence of $Cu^{3+}$ ionic
configuration (the charge transfer $Cu^{2+}\rightarrow Cu^{3+}$ costs large
energy $U\sim 10$ $eV$, i.e. the occupation of the $Cu$ site with two holes
with opposite spins is unfavorable) P. W. Anderson proposed in his
remarkable paper \cite{Anderson} the Hubbard model as the basic model for
quasiparticle properties in these compounds. For some parameter values it
can be derived from the (minimal) microscopic \textit{three-band model }in
some parameter regime. Besides the hopping $t_{pd}$ between the $d$-orbital
of $Cu$ and $p$-orbital of $O$ ions (as well as $t_{pp}$ between O ions) -
the Emery model, it contains also the strong Coulomb interaction $U_{Cu}$ on
the Cu ions as well as interaction between p- and d-electrons. The main two
parameters are $U_{Cu}\sim (6-10)$ $eV$ and the charge transfer energy $%
\Delta _{pd}\equiv \epsilon _{d}^{0}-\epsilon _{p}^{0}\sim (2.5-4)$ $eV$,
where $\epsilon _{d}^{0},$ $\epsilon _{p}^{0}$ are bare energies of the d-
and p levels, respectively. In cuprates the case $U_{Cu}>>\Delta _{pd}$ is
realized, i.e. they belong to the class of \textit{charge transfer materials}%
. This allows us to project the complicated three-band Hamiltonian onto the
low-energy sector, and to obtain an effective single-band Hubbard
Hamiltonian with an effective hopping parameter $t$ and the effective
repulsion $U_{eff}\approx \Delta _{pd}$. It turns out that the case $%
U_{eff}>>t$ is realized, if $\Delta _{pd}\gg t=t_{pd}^{2}/\Delta _{pd}$. In
fact in HTSC cuprates the latter is not quite realized and probably the
slightly weaker condition $\Delta _{pd}>t$ is more appropriate, since $%
t_{pd}\approx 1.3$ $eV$ and $\Delta _{pd}\approx (2-3)$ $eV$. Nevertheless,
there is a widespread believe that the suitable effective and minimal
Hamiltonian, i.e. the toy model for \textit{low-energy physics of cuprates,}
is based on the Hubbard model which additionally comprises the long-range
Coulomb interaction $\hat{V}_{C}$ and EPI $\hat{V}_{EPI}$ - see \cite%
{Kulic-Review}
\begin{equation}
\hat{H}=-\sum_{i,j,\sigma }t_{ij}c_{i\sigma }^{\dagger }c_{j\sigma
}+U_{eff}\sum_{i}n_{i\uparrow }n_{i\downarrow }+\hat{V}_{C}+\hat{V}_{EPI}.
\label{Eq32}
\end{equation}%
Some experiments imply that the effective repulsion should be of the order
of $U_{eff}\approx 4$ $eV$, while the nearest neighbor and next-nearest
neighbor hopping $t$ and $t^{\prime }$ , respectively are estimated to be $%
t=0.3-0.5$ $eV$ and $t^{\prime }/t$ equal $-0.15$ in $LSCO$ and $-0.45$ in $%
YBCO$. Since $(U_{eff}/t)\gg 1$ the above Hamiltonian is again in the regime
of strong electronic correlations, where the double occupancy of a given
lattice site is strongly suppressed, i.e. $\langle n_{i\uparrow
}n_{i\downarrow }\rangle \ll 1$. The latter restricts charge fluctuations of
electrons (holes) on a given lattice site, since $n_{i}=0,1$ is allowed
only, while processes with $n_{i}=2$ are (practically) forbidden. Note, that
in (standard) weakly correlated metals all charge fluctuation processes ($%
n_{i}=0,1,2$) are allowed, since $U\ll W$ in these systems. Due to
suppression of double occupancy one expects that the screening
properties in cuprates are strongly affected - this will be
demonstrated below. From the Hamiltonian in Eq. (\ref{Eq32}),
which is the 2D model for the low-energy physics in the $CuO_{2}$
plane, comes out that in the \textit{undoped} system there is one
particle per lattice - the so called half-filled case (in the band
language) with $\langle n_{i}\rangle =1$. It is an insulator
because of large $U$ and even antiferromagnetic insulator at $T=0$
$K$. The effective exchange interaction (with $J=4t^{2}/U$)
between spins is
Heisenberg-like. By doping the system with holes (the hole concentration is $%
\delta (<1)$) means that particles are taken out from the system in which
case there is on the average $\langle n_{i}\rangle =1-\delta $ particles per
lattice site. Above some (small) critical doping $\delta _{c}$ (which is of
the order $0.01$ in doped cuprates) the $AF$ order vanishes and the system
becomes a strongly correlated metal. For some \textit{optimal} doping $%
\delta _{op}(\sim 0.1)$ the system is metallic with a large Fermi surface
and can exhibit even high-T$_{c}$ superconductivity in the presence of the
EPI, as it will demonstrated below. The latter interaction and its interplay
with strong correlations is the central subject in the following sections.

\subsection{Strong correlations in the t-J model in terms of Hubbard
operators $X^{\protect\alpha \protect\beta }$}

Since $U_{eff}>>t$ one can let with a good accuracy $U_{eff}\rightarrow
\infty $, i.e. the system is in the \textit{strongly correlated regime }%
where the doubly occupancy $n_{i}=2$ is excluded. One of the ways to cope
with such strong correlations is to introduce the Hubbard operators $%
X^{\alpha \beta }$ which automatically take into account the exclusion of
double occupancy. The creation and annihilation Hubbard operators $\hat{X}%
_{i}^{\sigma 0}$ and $\hat{X}_{i}^{0\sigma }=(\hat{X}_{i}^{\sigma
0})^{\dagger }$, with $\hat{X}_{i}^{\sigma 0}=c_{i\sigma }^{\dagger
}(1-n_{i,-\sigma })$, are fermion-like and they fulfill the condition that $%
n_{i,\sigma }+n_{i,-\sigma }\leq 1$ on each lattice site. The latter means
that there is no more than one electron (hole) per lattice site, i.e. the
double occupancy is forbidden. The bosonic-like Hubbard operators $\hat{X}%
_{i}^{\sigma _{1}\sigma _{2}}=\hat{X}_{i}^{\sigma _{1}0}\hat{X}_{i}^{0\sigma
_{2}}$ (with \ $\sigma _{1}\neq \sigma _{2}$) describe spin fluctuations at
the i-th site. Here, the spin projection parameter $\sigma =\uparrow
,\downarrow $ ( $-\sigma =\downarrow ,\uparrow $) and the operator $\hat{X}%
_{i}^{\sigma \sigma }$ has the meaning of the electron (hole) number on the
i-th site. In the following we shall use the convention that when $\hat{X}%
_{i}^{\sigma \sigma }\mid 1\rangle =1\mid 1\rangle $ there is a fermionic
particle (electron) on the $i$-th site, while for $\hat{X}_{i}^{\sigma
\sigma }\mid 0\rangle =0\mid 0\rangle $ the site is empty, i.e. there is a
hole on it. It is useful to introduce the hole number operator $\hat{X}%
_{i}^{00}=\hat{X}_{i}^{0\sigma }\hat{X}_{i}^{\sigma 0}$, i.e. if $\hat{X}%
_{i}^{00}\mid 0\rangle =1\mid 0\rangle $ the $i$-th site is empty, i.e.
there is one hole on it, while for $\hat{X}_{i}^{00}\mid 1\rangle =0\mid
1\rangle $ it is occupied by an electron and there is no hole. These
operators fulfill the \textit{non-canonical} commutation relations

\begin{equation}
\left[ \hat{X}_{i}^{\alpha \beta },\hat{X}_{j}^{\gamma \lambda }\right]
_{\pm }=\delta _{ij}\left[ \delta _{\gamma \beta }\hat{X}_{i}^{\alpha
\lambda }\pm \delta _{\alpha \lambda }\hat{X}_{i}^{\gamma \beta }\;\right] .
\label{Eq34}
\end{equation}%
Here, $\alpha ,\beta ,\gamma ,\lambda =0,\sigma $ and $\delta _{ij}$ is the
Kronecker symbol. The (anti)commutation relations in Eq. (\ref{Eq34}) are
rather different from the canonical Fermi and Bose (anti)commutation
relations. Since $U_{eff}=\infty $ the double occupancy is excluded, i.e. $%
\hat{n}_{i\uparrow }\hat{n}_{i\downarrow }\mid \psi \rangle (=\hat{X}%
_{i}^{22}\mid \uparrow \downarrow \rangle )=0$, and by construction the $%
\hat{X}^{\alpha \beta }$ operators satisfy the \textit{local constraint}
(the completeness relation)
\begin{equation}
\hat{X}_{i}^{00}+\sum_{\sigma =1}^{N}\hat{X}_{i}^{\sigma \sigma }=1.
\label{Eq35}
\end{equation}%
This condition tells us that at a given lattice site there is either one
hole ($\hat{X}_{i}^{00}\mid hole\rangle =1\mid hole\rangle $) or one
electron ($\hat{X}_{i}^{\sigma \sigma }\mid elec\rangle =1\mid elec\rangle $%
). Note, if Eq. (\ref{Eq35}) is obeyed, then both commutation and
anticommutation relations hold also in Eq. (\ref{Eq34}) at the same lattice
site, which is due to the projection properties of Hubbard operators $\hat{X}%
^{\alpha \beta }\hat{X}^{\gamma \mu }=\delta _{\beta \gamma }\hat{X}^{\alpha
\mu }$.

For further purposes, i.e. for studying low-energy excitations in a
controllable way, the number of spin projections is \textit{generalized }to
be $N$ , i.e. $\sigma =1,2,...N$. By projecting out the double occupied
(high energy) states from the Hamiltonian in Eq. (\ref{Eq32}) one obtains
the generalized t-J model \cite{Kulic-AIP}%
\[
\hat{H}_{t-J}=\hat{H}_{t}+\hat{H}_{J}=-\sum_{i,j,\sigma }t_{ij}\hat{X}%
_{i}^{\sigma 0}\hat{X}_{j}^{0\sigma }
\]

\begin{equation}
+\sum_{i,j,}J_{ij}(\mathbf{\hat{S}}_{i}\cdot \mathbf{\hat{S}}_{j}-\frac{1}{4}%
\hat{n}_{i}\hat{n}_{j}\;)+\hat{H}_{3}.  \label{Eq-36}
\end{equation}

The first term describes electron hopping by ensuring that the double
occupancy is excluded. The second term describes the Heisenberg-like
exchange energy of almost-localized electrons. $\hat{H}_{3}$ contains
three-site hopping which is believed not to be important and is usually
omitted. For effects related to charge fluctuation processes it is plausible
to omit it, while for spin-fluctuation processes this approximation may be
questionable. The spin and number operators are given by $\mathbf{\hat{S}}=%
\hat{X}_{i}^{\bar{\sigma}_{1}0}(\vec{\sigma})_{\bar{\sigma}_{1}\bar{\sigma}%
_{2}}\hat{X}_{i}^{0\bar{\sigma}_{2}};\ \hat{n}_{i}=\hat{X}_{i}^{\bar{\sigma}%
\bar{\sigma}}$ where summation over bar indices is assumed \cite{Kulic-AIP}%
.\

The rather "awkward" commutation relations of $\hat{X}^{\alpha \beta }$ were
one of the reason that most researchers in the field prefer to use some
representations of the Hubbard operators $\hat{X}^{\alpha \beta }(\hat{f},%
\hat{b})$ in terms of fermion and boson operators $\hat{f},\hat{b}$,
respectively. For instance, in the very popular \textit{slave-boson
representation} one has $\hat{X}_{i}^{0\sigma }=\hat{f}_{i,\sigma }\hat{b}%
_{i}^{\dagger }$ with the constraint $\hat{b}_{i}^{\dagger }\hat{b}%
_{i}+\sum_{\sigma =1}^{N}\hat{f}_{i,\sigma }^{\dag }\hat{f}_{i,\sigma }=1$
where $\hat{f}_{\sigma }$ destroys a spinon with spin $\sigma $ and $\hat{b}%
^{\dagger }$ creates a holon. Since $\hat{f}_{\sigma },\hat{b}$ obey the
canonical Fermi and Bose commutation rules it is, at first glance, very
convenient to use this approach since for these operators well-elaborated
many-body techniques (Feynman diagrams) exist. However, the constraint makes
the spinon and the slave boson strongly coupled, and a naive decoupling of
the equations of motion is risky, especially in dimensions $d\geqslant 2$,
sometimes giving fictitious results. Therefore, the present author and his
collaborators decided to work directly with the Hubbard operators, thus
always keeping the composite object (bound spinon and holon in the language
of SB). However, the central question in this approach is - \textit{how to
study the quasiparticle dynamic in absence of the canonical Fermi and Bose
commutation rules}? The way out was found in the very elegant Baym-Kadanoff
technique in obtaining Dyson equations, the method which is independent on
the operator algebra - see more in \cite{Kulic-Review} and references
therein. Applied to strong correlations we call it the \textit{X-method}. It
turns out that the method is very efficient and one can formulate very
elegant and controllable 1/N expansion which also allows us to study the
EPI\ interaction in a consistent way. We stress that until now there has
been no transparent and efficient way to treat the EPI within the
slave-boson method.

The basic idea behind the \textit{X-method} \cite{Ruckenstein} is that the
Dyson's equation for the electron Green's function can be effectively
obtained by introducing the \textit{external potential } $u^{\sigma
_{1}\sigma _{2}}(1)$ (source) with the Hamiltonian $\hat{H}_{s}$%
\[
\int \hat{H}_{s}\,d\tau =\int \sum_{\sigma _{1},\sigma _{2}}u^{\sigma
_{1}\sigma _{2}}(1)\hat{X}^{\sigma _{1}\sigma _{2}}(1)\,d1
\]%
\begin{equation}
\equiv u^{\bar{\sigma}_{1}\bar{\sigma}_{2}}(\bar{1})\hat{X}^{\bar{\sigma}_{1}%
\bar{\sigma}_{2}}(\bar{1})  \label{H-source}
\end{equation}%
Here, $1\equiv (\mathbf{l},\tau )$ and $\int (..)\,d1\equiv \int (..)\,d\tau
\sum_{\mathbf{l}}$ and $\tau $ is the Matsubara time. In the following,
integration over the bared variables ($\bar{1},\bar{2}..$) and a summation
over bared spin variable ($\bar{\sigma}..$) is understood. The sources $%
u^{\sigma _{1}\sigma _{2}}(1)$ are useful in generating higher correlation
functions entering the self-energy. The electronic Green's function is
defined by \cite{Kulic-Review}, \cite{Ruckenstein}
\begin{equation}
G^{\sigma _{1}\sigma _{2}}(1,2)=\frac{-\langle \hat{T}\left( \hat{S}\hat{X}%
^{0\sigma _{1}}(1)\hat{X}^{\sigma _{2}0}(2)\right) \rangle }{\langle \hat{T}%
\hat{S}\rangle },  \label{Eq38}
\end{equation}%
where $\hat{T}$ is the time-ordering operator and $\hat{S}=\hat{T}\exp
\{-\int \hat{H}_{s}(1)\,d1\}$. We define the \textit{quasiparticle Green's} $%
g^{\sigma _{1}\sigma _{2}}$ and the \textit{Hubbard spectral weight} $%
Q^{\sigma _{1}\sigma _{2}}$

\[
g^{\sigma _{1}\sigma _{2}}(1,2)=G^{\sigma _{1}\overline{\sigma _{2}}%
}(1,2)Q^{-1,\overline{\sigma }_{2}\sigma _{2}}(2)
\]

\begin{equation}
Q^{\sigma _{1}\sigma _{2}}(1)=\delta ^{\sigma _{1}\sigma _{2}}\langle \hat{X}%
^{00}(1)\rangle +\langle \hat{X}^{\sigma _{1}\sigma _{2}}(1)\rangle .
\label{g-Q}
\end{equation}%
It is useful to define the vertex functions%
\begin{equation}
\gamma _{\sigma _{3}\sigma _{4}}^{\sigma _{1}\sigma _{2}}(1,2;3)=-\frac{%
\delta g^{-1,\sigma _{1}\sigma _{2}}(1,2)}{\delta u^{\sigma _{3}\sigma
_{4}}(3)}  \label{gamma}
\end{equation}%
which in what follows play very important role in the quasiparticle
dynamics. It turns out that the self-energy $\Sigma _{g}$ (note $%
g^{-1}=g_{0}^{-1}-\Sigma _{g}$ - \cite{Kulic-Review}) can be expressed via
the \textit{charge vertex }$\gamma _{c}(1,2;3)\equiv \gamma _{\bar{\sigma}%
\bar{\sigma}}^{\sigma \sigma }(1,2;3)$ and the \textit{spin vertex }$\gamma
_{s}(1,2;3)\equiv \gamma _{\bar{\sigma}\sigma }^{\bar{\sigma}\sigma }(1,2;3)$%
. In the paramagnetic state one has $g^{\sigma _{1}\sigma _{2}}=\delta
^{\sigma _{1}\sigma _{2}}g$, $Q^{\sigma _{1}\sigma _{2}}=\delta ^{\sigma
_{1}\sigma _{2}}Q$ and $\Sigma ^{\sigma _{1}\sigma _{2}}=\delta ^{\sigma
_{1}\sigma _{2}}\Sigma _{g}$

\[
\Sigma _{g}(1-2)=-\frac{t_{0}(1-2)}{N}Q(1)
\]%
\[
+\delta (1-2)\frac{J_{0}(1-\overline{1})}{N}<\hat{X}^{\sigma \sigma }(%
\overline{1})>
\]%
\[
-\frac{t_{0}(1-\overline{1})}{N}g(\overline{1}-\bar{2})\gamma _{c}(\overline{%
2},2;1)
\]%
\begin{equation}
\frac{+t_{2}(1,\overline{1},\overline{3})}{N}g(\bar{1}-\overline{2})\gamma
_{s}(\overline{2},2;\overline{3})+\Sigma _{Q}(1-2),  \label{Eq44}
\end{equation}%
where $t_{2}(1,2,3)=\delta (1-2)t_{0}(1-3)-\delta (1-3)J_{0}(1-2)$. The
notation $t_{0}(1-2)$ (and $J_{0}(1-2)$) means $t_{0}(1-2)=t_{0,i_{1}j_{2}}%
\delta (\tau _{1}-\tau _{2})$. The first two terms in Eq. (\ref{Eq44})
represent the effective (reduced) kinetic energy of quasiparticles in the
lower Hubbard band. As we shall see below they give rise to the
quasiparticle \textit{band narrowing} and to the \textit{shift }of the band
center, respectively. The third and fourth terms describe the kinematic and
dynamic interaction of quasiparticles with charge and spin fluctuations,
respectively, while the term $\Sigma _{Q}(1,2)$ takes into account the
counterflow of surrounding quasiparticles which takes place because of the
local constraint (absence of double occupancy). $\Sigma $ depends on the
vertex functions $\gamma _{c}(1,2;3)$ and $\gamma _{s}(1,2;3)$ and it does
not contain a small expansion parameter, like the interaction energy in
weakly interacting systems, because the hopping parameter $t$ describes at
the same time the kinetic energy and \textit{kinematic interaction} of
quasiparticles. This means that there is no controllable perturbation
technique for $\Sigma _{g}$, due to the lack of a small parameter. In the
past various decoupling procedures and mean-field like techniques were
utilized, such as the steepest descent method in the path integral
technique, etc. However, to the author's best knowledge, all decoupling
techniques are not only non-systematic and uncontrollable, but moreover they
suffer from inability to extract the coherent part of the quasiparticle
spectrum. As we shall argue in what follows the coherent quasiparticle
band-width (in the case $J=0$) is proportional to $\delta $, i.e. $%
W_{qp}\sim \delta \cdot t$. All decoupling technique obtain the band-width $%
W_{d}\sim (1-\alpha \cdot \delta )\cdot t$ \cite{Avella}, with $\alpha \cdot
\delta $ finite in the limit $\delta \rightarrow 0$. So obtained $W_{d} $ is
in fact the total band-width for coherent and incoherent quasiparticle
motions, i.e. $W_{d}=W_{qp}+W_{inc}$. Thus the decoupling techniques are
inappropriate in studying the coherent quasiparticle motion.

What is then the advantage of the \textit{X-method }and the self-energy in
Eq. (\ref{Eq44})? This method allows formulation of a controllable $1/N$
expansion for $\Sigma _{g}$ by including also the EPI, as shown below. For
that purpose it is necessary to generalize the local constraint condition
\cite{Ruckenstein}
\begin{equation}
\hat{X}_{i}^{00}+\sum_{\sigma =1}^{N}\hat{X}_{i}^{\sigma \sigma }=\frac{N}{2}%
,  \label{Eq45}
\end{equation}%
where $N/2$ replaces the unity in Eq. (\ref{Eq35}). Only in this case the
1/N expansion is systematic. It is apparent from Eq. (\ref{Eq45}) that for $%
N=2$ it coincides with Eq. (\ref{Eq35}) and has the meaning that at most
half of all spin states at a given lattice site can be occupied. The
particle spectral function $A(\mathbf{k},\omega )=-ImG(\mathbf{k},\omega
)/\pi $ must obey the generalized Hubbard sum rule which respects the new
local constraint in Eq. (\ref{Eq45})%
\[
\int d\omega A(\mathbf{k},\omega )=\frac{1+(N-1)\delta }{2}.
\]%
The $N>2$ generalization of the local constraint allows us to make a \textit{%
controllable }$1/N$ \textit{expansion} of the self-energy with respect to
the small quantity $1/N$ (when $N\gg 1$). Physically this procedure means
that we are selecting a class of diagrams in the self-energy and response
functions which might be important in some parameter regime. In \cite{Kulic1}%
, \cite{Kulic2}, \cite{Kulic3} it was shown that there is a
systematic $1/N$ expansion for vertices $\gamma _{c}$ and $\gamma
_{s}$ as well as for other functions
\begin{eqnarray}
g &=&g_{0}+g_{1}/N+...,  \label{1/N} \\
Q &=&Nq_{0}+Q_{1}+Q_{2}/N...,  \nonumber \\
\Sigma &=&\Sigma _{0}+\Sigma _{1}/N+...  \nonumber
\end{eqnarray}%
etc.

\subsubsection{The main results of the X-method}

The first nontrivial terms in \textit{leading }$O(1)$\textit{-order }are $%
\Sigma _{0}$ and $g_{0}$ - for details see in \cite{Kulic-Review}, \cite%
{Kulic1}, \cite{Kulic2}, \cite{Kulic3}, which in fact describe the \textit{%
coherent part} of the quasiparticle self-energy and the Green's function
\begin{eqnarray}
g_{0}(\mathbf{k},\omega ) &\equiv &G_{0}(\mathbf{k},\omega )/Q_{0}=\frac{1}{%
\omega -(\epsilon _{0}(\mathbf{k})-\mu },  \nonumber \\
\epsilon _{0}(\mathbf{k}) &=&\epsilon _{c}-q_{0}t_{0}(\mathbf{k})-\frac{1}{%
N_{L}}\sum_{\mathbf{p}}J_{0}(\mathbf{k}+\mathbf{p})n_{F}(\mathbf{p}),
\nonumber \\
\epsilon _{c} &=&\frac{1}{N_{L}}\sum_{\mathbf{p}}t_{0}(\mathbf{p})n_{F}(%
\mathbf{p}),  \nonumber \\
Q_{0} &=&<\hat{X}_{i}^{00}>=Nq_{0}=N\frac{\delta }{2}.
\end{eqnarray}%
where $N_{L}$ is the number of lattice site, $\epsilon _{0}(\mathbf{k})$ is
the \textit{quasiparticle energy} for the coherent motion and $\epsilon _{c}$
is the \textit{level shift.} Here, $t_{0}(\mathbf{k})$ and $J_{0}(\mathbf{k}%
) $ are Fourier transforms of $t_{0,ij}$ and $J_{0,ij}$, respectively. For $%
t-t^{\prime }$ one has $t_{0}(\mathbf{k})=2t_{0}(\cos k_{x}+\cos
k_{y})+2t^{\prime }\cos k_{x}\cos k_{y}$ and $J_{0}(\mathbf{k})=2J_{0}(\cos
k_{x}+\cos k_{y})$, since the lattice constant is let $a=1$. Since $%
Q_{0}(\sim N\delta )$ this means that the \textit{quasiparticle
residuum} vanishes at vanishing of doping ($\delta \rightarrow
0$). This is physically plausible since for $\delta \rightarrow 0$
the coherent motion in the t-J model is blocked. The chemical
potential $\mu $ is obtained from the condition $1-\delta
=2\sum_{\mathbf{p}}n_{F}(\mathbf{p})$ which \ gives
large Fermi surface, see discussion below. Note that $\Sigma _{Q}$ in Eq. (%
\ref{Eq44}) is of the $O(1/N)$ order.

We summarize the main results obtained by the \textit{X-method} to the
leading $O(1)$ order and compare them with the corresponding ones in the
slave boson ($SB$)-method \cite{Kulic1}, \cite{Kulic2}, \cite{Kulic3}: ($%
\mathbf{1)}$ up to the $O(1)$ order, $g_{0}(\mathbf{k},\omega )$ describes
the \textit{coherent motion} of quasiparticles in the metallic state, whose
contribution to the total spectral weight of the particle Green's function $%
G_{0}(\mathbf{k},\omega )$ is $Q_{0}=N\delta /2$, i.e. $G_{0}(\mathbf{k}%
,\omega )=Q_{0}g_{0}(\mathbf{k},\omega )$ to the leading order, so we
dealing with a Landau-Fermi liquid. This means that the quasiparticle
residuum (and the metallic state) vanishes for zero doping $\delta =0$ and
the system is in the Mott insulating state. The quasiparticle energy is
dominated by the exchange parameter if $J_{0}>\delta t_{0}$, i.e. for very
low doping, since in cuprates $(J_{0}/t_{0})<1/3$. For $J_{0}=0$ there is
\textit{band narrowing} since $\delta <1$ and the quasiparticle band-width
is proportional to $\delta $, i.e. $W=z\cdot \delta \cdot t_{0}$. These
results are identical to the corresponding ones in the SB method. This does
not mean that the next leading terms coincide in both methods. This is
actually not the case for a number of quantities (response functions). This
is understandable since the X-method keeps the composite object (correlated
motion of spinon and holon), while in the SB method they are decoupled. Only
after invoking the existence of gauge fields (also fictitious particles) one
can keep spinon and holon together in the SB model, if at all. ($\mathbf{2)}$
The X-method respects the local constraint at each lattice site in every
step of calculations, which guarantees the correct study of response
functions. ($\mathbf{3)}$ In the (important) paper \cite{Greco} - which is
based on the theory elaborated in \cite{Kulic1}, \cite{Kulic2}, \cite{Kulic3}%
, it was shown that there is large discrepancy in approximate calculations
within the SB and X methods. For instance, in the superconducting state the
anomalous self-energy (which is of $O(1/N)$-order in the $1/N$ expansion) of
the $X$- and $SB$ methods \textit{differ substantially}. As a result, the $%
SB $ method erroneously predicts \textit{superconductivity} due to the
kinematic interaction in the $t$-$J$ model (for $J=0$) with large $T_{c}$,
while the X-method gives extremely small $T_{c}$ $(\approx 0)$ \cite{Greco}.
The reason\ for this discrepancy is that the calculations done within the $%
SB $ method miss a class of compensating diagrams, which are in contrast
taken into account automatically within the X method. So, although the two
approaches yield some similar results in leading $O(1)$ order their
implementation in the next leading $O(1/N)$ order show that they are indeed
different - see discussion below. Note that the $1/N$ expansion in the
X-method is well-defined and transparent. ($\mathbf{4)}$ The renormalization
of the EPI coupling constant by strong correlations is different in the two
approaches even in the large $N$-limit, as shown below; ($\mathbf{5}$) The
optical conductivity $\sigma _{1}(\omega ,\mathbf{q}=0)\equiv \sigma
_{1}(\omega )$ and the optics sum-rule exhibit very interesting behavior as
a function of doping concentration
\begin{equation}
\int_{0}^{\omega _{c}}d\omega \sigma (\omega )=\frac{\pi }{4}\delta \cdot
Ne^{2}a^{2}\epsilon _{c}.  \label{sum-rule}
\end{equation}%
It is seen that the optics sum-rule is proportional to the number of holes $%
\delta $, instead of $n=1-\delta $ as it would be in the weakly interacting
case \cite{Kulic-Review}. Here $a$ is the lattice constant. ($6$) Note, that
the volume below the Fermi surface in the case of strong correlations scales
with $n=1-\delta $, i.e. the \textit{Fermi surface is large} like in the
conventional Fermi liquid. The above analysis clearly shows significant
difference (($5$) vs ($6$)) in response functions of strongly correlated
systems and the canonical Landau-Fermi liquid.

\section{The forward scattering peak in the EPI of cuprates}

The EPI coupling constant in $LTSC$ materials is calculated by the
local-density functional ($LDA$) method. The latter is suitable for ground
state properties of crystals and it is based on an effective electronic
crystal potential $V_{g}$. Since in principle $V_{g}$ differs significantly
from $\Sigma _{0}(\mathbf{k},\omega =0)$ (the many body effective potential
- see \cite{Kulic-Review}) then the $LDA$ coupling constant $g_{EP}^{(LDA)}$
can be also very different from the real coupling constant $g_{EP}$.
Strictly speaking the EPI does not have meaning in the $LDA$ method - see
discussion in \cite{Kulic-Review}, because the latter treats ground state
properties of materials, while the EPI is due to excited states and
inelastic processes in the system. We shall not deal with this problem here
- see \cite{Kulic-Review}.

\subsection{LDA method for the EPI in cuprates}

The LDA method considers electrons in the ground state (there is a
generalization to finite $T$), whose energy can be calculated by knowing the
spectrum \{$\epsilon _{k}$\} of the Kohn-Sham (Schr\"{o}dinger like)
equation
\begin{equation}
\lbrack \frac{\mathbf{\hat{p}}^{2}}{2m}+V_{g}(\mathbf{r})]\psi _{k}(\mathbf{r%
})=\epsilon _{k}\psi _{k}(\mathbf{r}),  \label{Eq30}
\end{equation}%
which depends on the \textit{effective one-particle potential}
\begin{equation}
V_{g}(\mathbf{r})=V_{ei}(\mathbf{r})+V_{H}(\mathbf{r})+V_{XC}(\mathbf{r}).
\label{Eff-pot}
\end{equation}%
Here, $V_{ei}$ is the electron-lattice potential, $V_{H}$ is the Hartree
term and $V_{XC}$ describes exchange-correlation effects - see \cite%
{Kulic-Review}. Because the EPI depends on the excited states (above the
ground state) of the system, this means that in principle the LDA method can
not describe it - see \cite{Kulic-Review}. However, by using an analogy with
the microscopic Migdal-Eliashberg theory one can define the EPI coupling
constant $g^{(Mig)}=g\Gamma _{c}/\varepsilon $ also in the LDA theory, see
\cite{Kulic-Review}. It reads%
\[
g_{\alpha ,ll^{\prime }}^{(LDA)}(\mathbf{k},\mathbf{k}^{\prime
})=\sum_{n}g_{\alpha ,nll^{\prime }}^{(LDA)}(\mathbf{k},\mathbf{k}^{\prime
})
\]%
\begin{equation}
=\langle \psi _{\mathbf{k}l}\mid \sum_{n}\frac{\delta V_{g}(\mathbf{r})}{%
\delta R_{n\alpha }}\mid \psi _{\mathbf{k}^{\prime }l^{\prime }}\rangle ,
\label{Eq-31}
\end{equation}%
where $n$ means summation over the lattice sites, $\alpha =x,y,z$ and the
wave function $\psi _{\mathbf{k}l}$ is the solution of the Kohn-Sham
equation. Formally one has $\delta V_{g}/\delta \mathbf{R}_{n}=\Gamma
_{LDA}\varepsilon _{e}^{-1}\nabla V_{ei}$. Even in such a simplified
approach it is difficult to calculate $g_{\alpha ,ll^{\prime
}}^{(LDA)}=g_{\alpha ,n}^{RMTA}+g_{\alpha ,n}^{nonloc}$ because it contains
the \textit{short-range} \textit{(local) coupling}
\begin{equation}
g_{\alpha ,n}^{RMTA}\sim g_{\alpha ,n}^{RMTA}(\mathbf{k},\mathbf{k}^{\prime
})\sim \langle Y_{lm}\mid \hat{r}_{\alpha }\mid Y_{l^{\prime }m^{\prime
}}\rangle  \label{RMT}
\end{equation}%
with $\Delta l=1$, and the \textit{long-range coupling}
\begin{equation}
g_{\alpha ,n}^{nonloc}(\mathbf{k},\mathbf{k}^{\prime })\sim \langle
Y_{lm}\mid (\mathbf{R}_{n}^{0}-\mathbf{R}_{m}^{0})_{\alpha }\mid
Y_{l^{\prime }m^{\prime }}  \label{Nonloc}
\end{equation}%
with $\Delta l=0$. In most calculations the local term $g_{\alpha ,n}^{RMTA}$
is calculated only, which is justified in simple metals only but not in the
HTSC oxides. In cuprates the local term gives a very small EPI coupling $%
\lambda ^{RMTA}\sim 0.1$, which is apparently much smaller than the
experimental value $\lambda >1$, and which gives rise to a pessimistically
small T$_{c}$ \cite{Mazin}. The small $\lambda ^{RMTA}$ was also one of the
reasons for discarding the EPI as pairing mechanism in HTSC oxides. At the
beginning of the HTSC era the electron-phonon spectral function $\alpha
^{2}F(\omega )$\ was calculated for the case $La_{2-x}Sr_{x}CuO_{4}$ in \cite%
{Weber} by using the first-principles band structure calculations and the
nonorthogonal tight-binding theory of lattice dynamics. The value $\lambda
=2.6$ was obtained and with the assumed $\mu ^{\ast }=0.13$ this gave $%
T_{c}=36$ $K$. However, these calculations also predicted a lattice
instability for the oxygen breathing mode in $La_{1.85}Sr_{0.15}CuO_{4}$
that has never been observed. Moreover, the same method was applied\ to $%
YBa_{2}Cu_{3}O_{7}$\ in \cite{Weber} where it was found $\lambda =0.5$,
which at best provides $T_{c}=(19-30)$ $K$.\ In fact the calculations in
\cite{Weber} did not take into account the Madelung coupling (i.e. they
neglected the matrix elements with $\Delta l=0$).

However, because of the weak screening of the ionic (long-range) Madelung
coupling in HTSC oxides - especially for vibrations along the c-axis - it is
necessary to include the nonlocal term $g_{\alpha ,n}^{nonloc}$. This task
was achieved within the LDA approach by the Pickett's group \cite{Krakauer},
where the EPI coupling for $La_{2-x}M_{x}CuO_{4}\ $is calculated within the
\textit{frozen phonon} method for $\mathbf{q}=0$ modes. By extrapolating the
result for $\mathbf{q}=0$ to finite $q$ they obtained $\lambda =1.37$ and $%
\omega _{\log }\approx 400$ $K$, and for $\mu ^{\ast }=0.1$ they predicted
that $T_{c}=49$ $K$ ($T_{c}\approx \omega _{\log }\exp \{-1/[(\lambda
/(1+\lambda ))-\mu ^{\ast }]\}$). The justification of this procedure was
questioned in \cite{Bohnen} were smaller $\lambda $ was obtained. For more
details see Ref. \cite{Kulic-Review} and references therein. Next, some
calculations, based on the tight-binding parametrization of the band
structure in $YBa_{2}Cu_{3}O_{7}$, gave rather large EPI coupling $\lambda
\approx 2$ and $T_{c}=90$ $K$\ \cite{Zhao-tc}.

We point out, that model calculations which take into account the long-range
ionic Madelung potential appropriately \cite{Jarlborg}, \cite{Barisic1} also
gave rather large coupling constant $\lambda \sim 2$, which additionally
hints to the importance of long-range forces in the EPI - see \cite%
{Kulic-Review} and references therein..

Since in HTSC oxides the plasma frequency along the c-axis, $\omega
_{pl}^{c} $, is of the order (or even less) of some characteristic c-axis
vibration modes, it is necessary to include the \textit{nonadiabatic effects}
in the EPI coupling constant, i.e. its frequency dependence $g_{\alpha
,n}\sim g^{0}/\varepsilon _{cc}(\omega )$. This non-adiabaticity was partly
accounted for by the Falter group \cite{Falter} by calculating the
electronic dielectric function along the c-axis $\varepsilon _{cc}(\mathbf{k}%
,\omega )$ in the RPA approximation. The result was that $g_{\alpha ,n}$
increased appreciably beyond its (well screened) metallic part, which gave a
large increase of the EPI coupling not only in the phonon modes but also in
the plasmon one. This question deserves much more attention than what it
received in the past.

The electron band structure and the Fermi surface can be satisfactory
described by the tight binding model, which is characterized with various
hopping parameters $t_{ij}$ and the local atomic (ionic) levels $\epsilon
_{i}$. Accordingly, there are there two kinds of the EPI coupling. The
\textit{covalent part} of the EPI is due to strong \textit{covalency} of the
$Cu$ and $O$ orbital in the $CuO_{2}$ planes. In this case, the EPI coupling
constant is characterized by the parameter (\textquotedblright
field\textquotedblright ) $E^{cov}\sim \partial t_{p-d}/\partial R\sim
q_{cov}t_{p-d}$, where $t_{p-d}$ is the hopping integral between $%
Cu(d_{x^{2}-y^{2}})$ and $O(p_{x,y})$ orbital and the length $q_{cov}^{-1}$
characterizes the spacial exponential fall-off of the hopping integral $%
t_{p-d}$. As the phonon Raman scattering shows, the covalent EPI is unable
to explain the strong phonon renormalization (the self-energy features) in
the $B_{1g}$ mode in $YBa_{2}Cu_{3}O_{7}$ by superconductivity, since in
this mode the O-ions vibrate along the $c-axis$ in opposite directions and
for this mode $\partial t_{p-d}/\partial R$ is zero in the first order in
the phonon displacement. Therefore, the EPI in this mode is certainly due to
the \textit{ionic contribution} which comes from the change in Madelung
energy, as it was first proposed in \cite{Jarlborg}, \cite{Barisic1}.
Namely, the Madelung interaction creates an electric field perpendicular to
the $CuO_{2}$ planes, which is due to the surrounding ions that form an
asymmetric environment. In that case the site energies $\epsilon _{i}^{0}$
contain the matrix element $\epsilon ^{ion}=\langle \psi _{i}\mid V(\mathbf{r%
})\mid \psi _{i}\rangle $, where $\mid \psi _{i}\rangle $ is the atomic wave
function at the $i$-th site, while the potential $V(\mathbf{r})$ steams from
the surrounding ions. In the simple and transition metals the surrounding
ions are well screened and therefore the change of $\epsilon ^{ion}$ in the
presence of phonons is negligible, in contrast to cuprates which are almost
\textit{ionic compounds, }in particular along the $c$-axis where the change
of $\epsilon ^{ion}$ is appreciable and characterized by the field strength $%
E^{ion}=V_{M}/d_{n}$. Here, $V_{M}$ is the characteristic Madelung potential
due to the surrounding ions and $d_{n}$ is the distance between the
neighboring ions. Immediately after the discovery of cuprates it was assumed
in many papers \cite{Weber} that the covalent part dominates the EPI in
these materials. However, the calculation that considered only the covalent
effects \cite{Weber} gave a rather small T$_{c}$ ($\sim $10-20 K in $YBCO$,
and 20-30 K in $La_{1.85}Sr_{0.15}CuO_{4}$). It turns out that in cuprates
the opposite inequality $E^{ion}\gg E^{cov}$ is realized for most c-axis
phonon modes in spite of the fact that $q_{cov}>1/d_{n}$ - see more in \cite%
{Kulic-Review} and references therein. This is supported by detailed
theoretical studies of $YBCO$ \cite{Barisic1}, where\ the change in the
\textit{ionic Madelung energy} due to the out of plane oxygen vibration in
the $B_{1g}$ mode is calculated . Similar to $YBCO$, the large
superconductivity-induced phonon self-energy effects in $%
HgBa_{2}Ca_{3}Cu_{4}O_{10+x}$ and in $(Cu,C)Ba_{2}Ca_{3}Cu_{4}O_{10+x}$ for
the $A_{1g}$ modes are also due to the ionic (Madelung) coupling. In these
modes oxygen ions move also along the c-axis and the ionicity of the
structure is involved in the EPI. This type of \textit{long-range} \textit{%
EPI} is absent in usual isotropic metals ($LTSC$ superconductors), where the
large Coulomb screening makes EPI local. Similar ideas were recently
incorporated into the Eliashberg equations in \cite{Abrikosov4}. The weak
screening along the $c$-axis, which is due to the very small hopping
integral for carrier motion, is reflected in the very small plasma frequency
$\omega _{p}^{(c)}$ along this axis. Since for some optical phonon modes one
has $\omega _{ph}>\omega _{p}^{(c)}$, nonadiabatic effects in the screening
are important. The latter can give rise to much larger EPI coupling constant
for this modes \cite{Falter}. These ideas are also supported by the recent
ARPES\ measurements on the single-layer $Bi_{2}Sr_{2}CuO_{6}$ (Bi2201) where
the coupling of electrons with phonons in the region $30-60$ $meV$ in the
overdoped compound is significantly reduced compared to the optimally doped
case. This can be explained by the larger electronic screening along the
c-axis in overdoped compound.

To summarize, ARPES, electron and phonon Raman scattering measurements in
the normal and superconducting state of cuprates gave the following
important results: $(a)$ phonons interact strongly with the electronic
continuum, i.e. EPI is substantial; $(b)$ the ionic contribution (the
Madelung energy) to EPI interaction for c-axis phonon modes gives
substantial contribution to the (large) EPI coupling constant ($\lambda >1$).

\subsection{Renormalization of the electron-phonon interaction by strong
correlations}

Based on the above discussion the minimal (toy) model Hamiltonian for HTSC
cuprates contains besides the t-J terms also the electron-phonon interaction%
\[
\hat{H}=\hat{H}_{tJ}+\sum_{i,\sigma }\epsilon _{a,i}^{0}\hat{X}_{i}^{\sigma
\sigma }+\hat{H}_{ph}
\]%
\[
+\hat{H}_{EP}^{ion}+\hat{H}_{EP}^{cov}+\hat{V}_{LC},
\]%
\[
\hat{H}_{EP}^{ion}=\sum_{i,\sigma }\hat{\Phi}_{i}(\hat{X}_{i}^{\sigma \sigma
}-\langle \hat{X}_{i}^{\sigma \sigma }\rangle ),
\]%
\begin{equation}
\hat{H}_{EP}^{cov}=-\frac{1}{N}\sum_{i,j,\sigma }\frac{\partial t_{0,ij}}{%
\partial (\mathbf{R}_{i}^{0}-\mathbf{R}_{j}^{0})}(\mathbf{\hat{u}}_{i}-%
\mathbf{\hat{u}}_{j})\hat{X}_{i}^{\sigma 0}\hat{X}_{j}^{0\sigma }\;.
\label{H-tot}
\end{equation}%
where $\hat{H}_{EP}^{ion},$ $\hat{H}_{EP}^{cov}$ are the ionic and covalent
contribution to the EPI, respectively. We consider first the ionic term
where $\hat{\Phi}_{i}(\mathbf{\hat{u}}_{L,\kappa })$ describes the change of
the atomic energy $\epsilon _{a,i}^{0}$ due to the change of long-range
Madelung energy in the presence of phonon displacements $\mathbf{\hat{u}}%
_{L,\kappa }$ of other atoms. $L$ and $\kappa $ enumerate unit lattice
vectors and atoms in the unit cell, respectively. Note that the theory is
formulated for the general nonlinear form of $\hat{\Phi}_{i}(\mathbf{\hat{u}}%
_{L,\kappa })$ dependence, and the following analysis holds in principle
also for an anharmonic EPI. The term proportional to $\langle \hat{X}%
_{i}^{\sigma \sigma }\rangle $ in Eq. (\ref{H-tot}) is introduced in order
to have $\langle \hat{\Phi}_{i}\rangle =0$ in the equilibrium state. We
stress that the X-method allows controllable calculation of the covalent
contribution \cite{Kulic-AIP}. In contrast, the treatment of EPI by the $SB$
method is complicated and not well defined, giving sometimes wrong results -
see more in \cite{Kulic-Review}.

After technically lengthy calculations, which are performed in \cite{Kulic1}%
, \cite{Kulic2}, the expression for the ionic contribution to the EPI part
(frequency-dependent) of the self-energy reads%
\[
\Sigma _{EP}^{(dyn)}(1,2)=-V_{EP}(\bar{1},\bar{2})\gamma _{c}(1,\bar{3};\bar{%
1})g_{0}(\bar{3},\bar{4})\gamma _{c}(\bar{4},2;\bar{2})
\]%
\[
V_{EP}(1,2)=\varepsilon _{e}^{-1}(1-\bar{1})V_{EP}^{0}(\bar{1}-\bar{2}%
)\varepsilon _{e}^{-1}(\bar{2}-2)
\]

\begin{equation}
V_{EP}^{0}(1-2)=-\langle T\hat{\Phi}(1)\hat{\Phi}(2)  \label{Eq57}
\end{equation}%
The propagator $V_{EP}^{0}(1-2)$ (which includes also the coupling constant)
of the bare EPI comprises in principle also the anharmonic contribution.
From Eq. (\ref{Eq57}) it is seen that in strongly correlated systems the
ionic part of the EPI is proportional to the square of the \textit{%
three-point charge vertex} $\gamma _{c}(1,2;3)$ (due to correlations). The
self-energy is given by%
\[
\Sigma _{EP}^{(dyn)}(\mathbf{k},\omega )=\int_{0}^{\infty }d\Omega \langle
\alpha ^{2}F(\mathbf{k,k}^{\prime },\Omega )\rangle _{\mathbf{k}^{\prime
}}R(\omega ,\Omega )
\]%
\[
R(\omega ,\Omega )=-2\pi i(n_{B}(\Omega )+\frac{1}{2})
\]%
\begin{equation}
+\psi (\frac{1}{2}+i\frac{\Omega -\omega }{2\pi T})-\psi (\frac{1}{2}-i\frac{%
\Omega +\omega }{2\pi T})  \label{sig-dynam}
\end{equation}%
where the (momentum-dependent) Eliashberg spectral function $\alpha ^{2}F(%
\mathbf{k,k}^{\prime },\omega )$ is defined below, $n_{B}(\Omega )$ is the
Bose function and $\psi (x$) is the di-gamma function - \cite{Kulic-Review},
\cite{Kulic1}, \cite{Kulic2}.

\subsection{Forward scattering peak in the charge vertex $\protect\gamma %
_{c} $}

\begin{figure}[tbp]
\resizebox{.5\textwidth}{!} {
\includegraphics*[width=10cm]{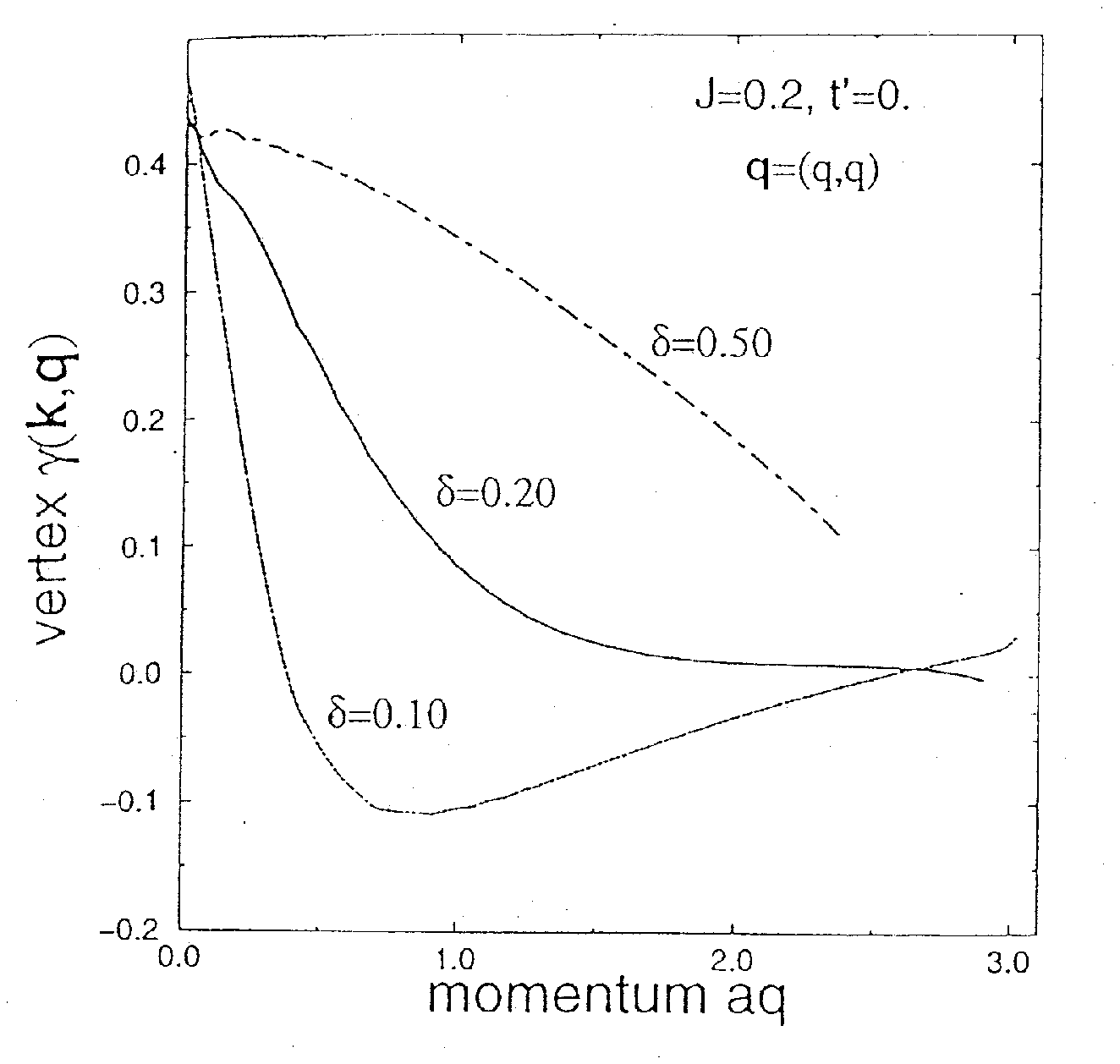}}
{\includegraphics*[width=8cm]{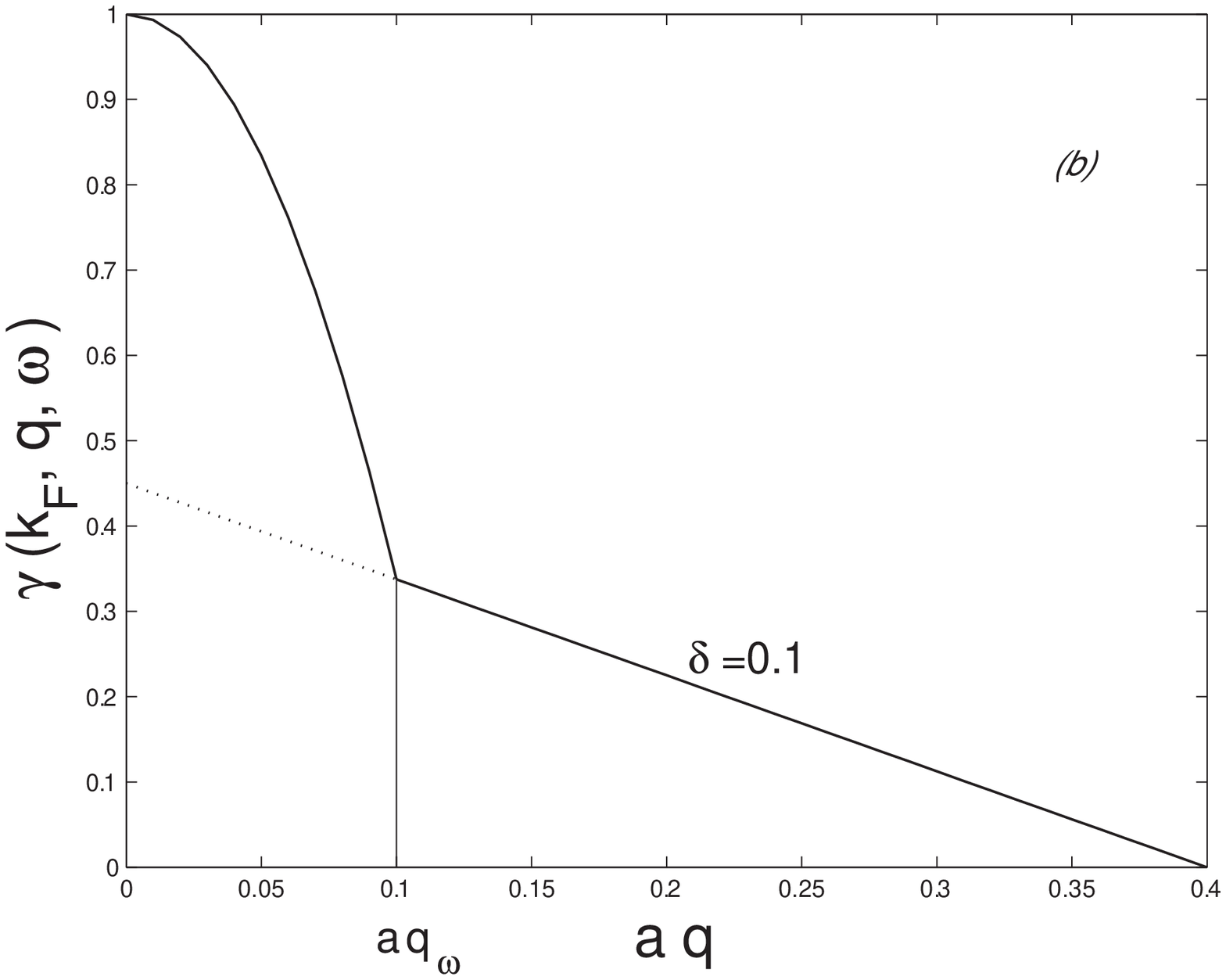}}
\caption{(a) Adiabatic ($\protect\omega =0$) vertex function $\protect\gamma %
(\mathbf{k}_{F},\mathbf{q})$ of the t-J model as a function of the momentum $%
aq$ with $\mathbf{q}=(q,q)$ for three different doping levels $\protect%
\delta $. From Ref. \protect\cite{Kulic3}. (b) Non-adiabatic ($\protect%
\omega \neq 0$) vertex function $\protect\gamma (\mathbf{k}_{F},\mathbf{q},%
\protect\omega )$ schematically with $q_{\protect\omega }=\protect\omega %
/v_{F}$.}
\label{VertexFig}
\end{figure}

The three-point \textit{charge vertex}\textbf{\ }$\gamma _{c}(1,2;3)$ plays
an important role in renormalization of all charge processes, such as EPI in
Eq. (\ref{Eq57}), Coulomb scattering and the scattering on non-magnetic
impurities. $\gamma _{c}(1,2;3)$ (and its Fourier transform $\gamma _{c}(%
\mathbf{k},q)$, $q=(\mathbf{q},iq_{n})$, $q_{n}=2\pi nT$,) was calculated in
the $t-J$ model in leading $O(1)$ order \cite{Kulic1}, \cite{Kulic2}, \cite%
{Kulic3}, \cite{Kulic-Review}. It is the solution of the following linear
integral equation%
\[
\gamma _{c}(1,2;3)=\delta (1-2)\delta (1-3)
\]%
\[
+t_{0}(1-2)g_{0}(1-\bar{1})g_{0}(\bar{2}-1^{+})\gamma _{c}(\bar{1},\bar{2}%
;3)
\]%
\[
+\delta (1-2)t_{0}(1-\bar{1})g_{0}(\bar{1}-\bar{2})g_{0}(\bar{3}-1)\gamma
_{c}(\bar{2},\bar{3};3)
\]%
\begin{equation}
-J_{0}(1-2)g_{0}(1-\bar{1}))g_{0}(\bar{2}-2)\gamma _{c}(\bar{1},\bar{2};3)
\label{3-point}
\end{equation}%
The (charge) vertex function $\gamma _{c}(\mathbf{k},q)$ describes a
specific screening of the charge potential due to strong correlations. In
the presence of an external perturbation ($u$) there is a \textit{change of
the band-width}, as well as the \textit{change of the local chemical
potential}, which comes from the suppression of double occupancy. These
processes are contained in the second and the third term of Eq. (\ref%
{3-point}). The central result is that for momenta $\mathbf{k}$\ laying at
(and near) the Fermi surface, $\gamma _{c0}(\mathbf{k},\mathbf{q},\omega =0)$
has a strong \textbf{\ }\textit{forward scattering peak} at $\mathbf{q}=0$,
which is very pronounced for lower doping $\delta $ $(\ll 1)$. On the other
hand, the backward (at large $\mathbf{q}$) scattering is substantially
suppressed, see Fig.~\ref{VertexFig}a. Such a behavior of the vertex
function means that a quasiparticle moving in the strongly correlated medium
digs up a \textit{giant correlation hole} with the radius $\xi _{ch}\approx
a/\delta $, where $a$ is the lattice constant. The effect of the $J$-term is
small, since $\gamma _{c}$ is determined by charge fluctuations which are
spread over the whole bare bandwidth $W_{B}\sim zt_{0}\gg J_{0}$.

We stress that $\gamma _{c0}$ in Fig.~\ref{VertexFig}a is calculated in the
\textit{adiabatic approximation, }i.e. for $\omega =0$ in $\gamma _{c0}(%
\mathbf{k},\mathbf{q},\omega )$. However, in the \textit{nonadiabatic regime}
$\omega >\mathbf{q\cdot v}_{F}(\mathbf{q})$ the vertex function reaches its
maximal value $\gamma _{c}(\mathbf{k}_{F},\mathbf{q=0,}\omega \mathbf{)}=1$,
see Fig.~\ref{VertexFig}b. This nonadiabaticity occurs whenever the phase
velocity ($\omega /q$) of charge excitations is larger than the Fermi
velocity. This effect is realized in case of high energy phonons. This means
that due to nonadiabatic effects strong correlations do not suppress the EPI
coupling with optical phonons at $q=0$, in contrast to a number of claims
\cite{Scalapino-Hanke} that they do.

Since the real physics is characterized by $N=2$ one can ask - how \textit{%
reliable }are the calculations of $\gamma _{c}(\mathbf{k}_{F},\mathbf{q})$
by the $1/N$ expansion method and in the limit $N\rightarrow \infty $. Here
let us mention that VL (Ginzburg) used to paraphrase a Landau's remark "that
the many-body theory is an experimental science". This means that the
correctness of the theory is tested by experiments. In this respect, the
exact diagonalization of charge correlation function $N(\mathbf{q},\omega )$
in the t-J model \cite{Tohyama} shows clearly that low-energy charge
scattering processes at large momenta $\mid \mathbf{q}\mid \approx 2k_{F}$\
are strongly suppressed compared to scattering at small transferred momenta (%
$\mid \mathbf{q}\mid \ll 2k_{F}$). This unambiguously\textbf{\ }confirms the
results obtained by the X-method in \cite{Kulic1}, \cite{Kulic2}, \cite%
{Kulic3} on the suppression of backward scattering in the vertex. Second,
very recent Monte Carlo (numerical) calculations of the vertex function $%
\gamma _{c}(\mathbf{k}_{F},\mathbf{q})$ in the Hubbard model with finite $U$
\cite{Scalapino-Hanke} show clear development of the forward scattering peak
in $\gamma _{c}(\mathbf{k}_{F},\mathbf{q})$, that is more pronounced for
larger U. This result definitely confirms the predictions of appearance of
the forward scattering peak in $\gamma _{c}(\mathbf{k}_{F},\mathbf{q})$ in
Refs. \cite{Kulic1}, \cite{Kulic2}, \cite{Kulic3}, \cite{Kulic-Review}. For
analytical theories numerical calculations in some models play the role of
an experiment, then the above theory of strong correlations and EPI is in
that sense confirmed experimentally in accord with the Landau remark.

Finally, it is possible to calculate the contribution of \textit{covalent EPI%
} coupling $\hat{H}_{EP}^{cov}$ to the self-energy. In this case, the vertex
renormalization by strong correlations is determined by the four-point
\textit{vertex}\textbf{\ }\textit{function}\textbf{\ }$\gamma _{t}(1,2;3,4)$%
. The latter is an important ingredient of EPI not only in HTSC cuprates but
also in heavy fermion systems. The calculations in \cite{Kulic-AIP} show
that $\gamma _{t}$ differs from that obtained by the mean-field
approximation SB model, what implies that the previous studies of EPI in
heavy fermion materials, in which the "covalent" coupling $\sim \gamma _{t}$
dominates, should be reconsidered by using the correct form of $\gamma _{t}$.

\subsection{Pairing and transport EPI coupling constants}

Depending on the symmetry of superconducting order parameter $\Delta (%
\mathbf{k},\omega )$ ($s-$, d-wave pairing) various averages (over the Fermi
surface) of $\alpha ^{2}F(\mathbf{k,k}^{\prime },\omega )$ enter the
Eliashberg equations. Assuming that the superconducting order parameter
transforms according to the representation $\Gamma _{i}$ ($i=1,3,5$) of the
point group $C_{4v}$ of a square lattice (in the $CuO_{2}$ planes), the
appropriate symmetry-projected spectral function is given by

\[
\alpha ^{2}F_{i}(\mathbf{\tilde{k},\tilde{k}}^{\prime },\omega )=\frac{%
N_{sc}(0)}{8}\sum_{\nu ,j}\mid g_{0,scr}(\mathbf{\tilde{k},\tilde{k}-}T_{j}%
\mathbf{\tilde{k}}^{\prime },\nu )\mid ^{2}\times
\]%
\begin{equation}
\times \delta (\omega -\omega _{\nu }(\mathbf{\tilde{k}-}T_{j}\mathbf{\tilde{%
k}}^{\prime }))\mid \gamma _{c}(\mathbf{\tilde{k},\tilde{k}-}T_{j}\mathbf{%
\tilde{k}}^{\prime })\mid ^{2}D_{i}(j),  \label{Eq60}
\end{equation}%
where$\ \mathbf{\tilde{k}}$ and $\mathbf{\tilde{k}}^{\prime }$ are the
momenta on the Fermi line in the irreducible Brillouin zone which is equal
to $1/8$ of the total Brillouin zone. Here, $g_{0,scr}(\mathbf{k,p},\nu )$
is the EPI coupling constant for the $\nu $-the mode, where the screening by
long-range Coulomb interaction is included, i.e. $g_{0,scr}(\mathbf{k,p},\nu
)=g_{0}(\mathbf{k,p},\nu )/\varepsilon _{e}(\mathbf{p})$. The density of
states $N_{sc}(0)$ is renormalized by strong correlations, where $%
N_{sc}(0)=N_{0}(0)/q_{0}$ and $q_{0}=\delta /2$ in the t-t$^{\prime }$ model
($J=0$). In the $t-J$ model $N_{sc}(0)$ has another form which does not
diverge for $\delta \rightarrow 0$, but one has $N_{sc}(0)(\sim
1/J_{0})>N_{0}(0)$, where the bare density of states $N_{0}(0)$ is
calculated, for instance by the $LDA$ scheme. $T_{j}$ , where $j=1,..8$,
denotes the eight point-group transformations that form the symmetry group
of the square lattice. This group has five irreducible representations which
we distinguish by the label $i=1,2,...5$. In the following the
representations $i=1$ \ and $i=3$ , which correspond to the $s-$ and d-wave
symmetry of the full rotation group, respectively, will be of importance. $%
D_{i}(j)$ is the representation matrix of the $j$-$th$ transformation for
the representation $i$. For each symmetry one obtains the corresponding
spectral function $\alpha ^{2}F_{i}(\omega )=\langle \langle \alpha
^{2}F_{i}(\mathbf{\tilde{k},\tilde{k}}^{\prime },\omega )\rangle _{\mathbf{%
\tilde{k}}}\rangle _{\mathbf{\tilde{k}}^{\prime }}$, which (in the first
approximation) determines the transition temperature for the order parameter
with the symmetry $\Gamma _{i}$. In the case $i=3$ the electron-phonon
spectral function $\alpha ^{2}F_{3}(\omega )$ in the $d$\textit{-channel} is
responsible for d-wave superconductivity represented by the irreducible
representation $\Gamma _{3}$ (or sometimes labelled as $B_{1g}$).

\begin{figure}[tbp]
\resizebox{.5\textwidth}{!} {
\includegraphics*[width=10cm]{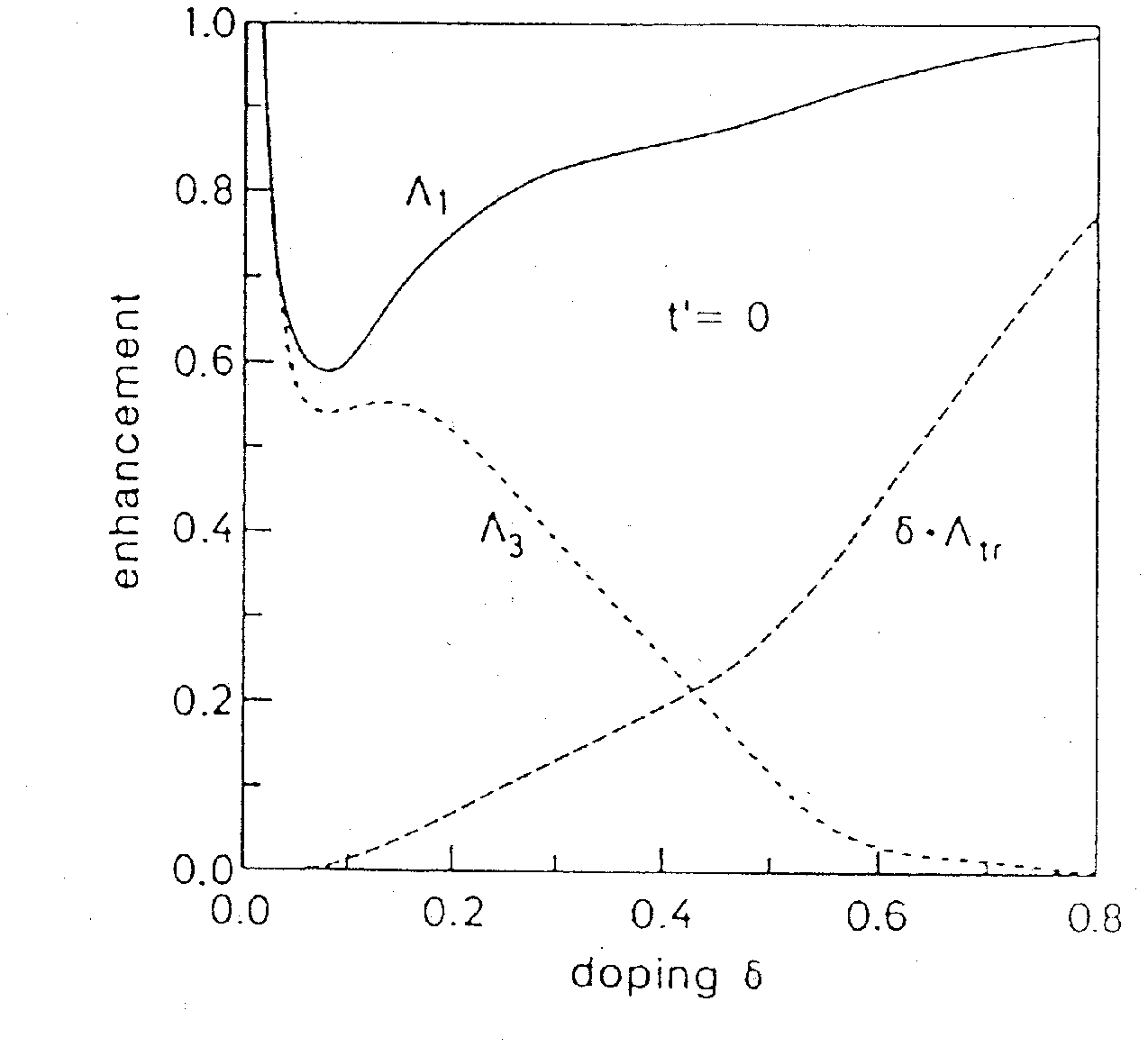}}
{\includegraphics*[width=8cm]{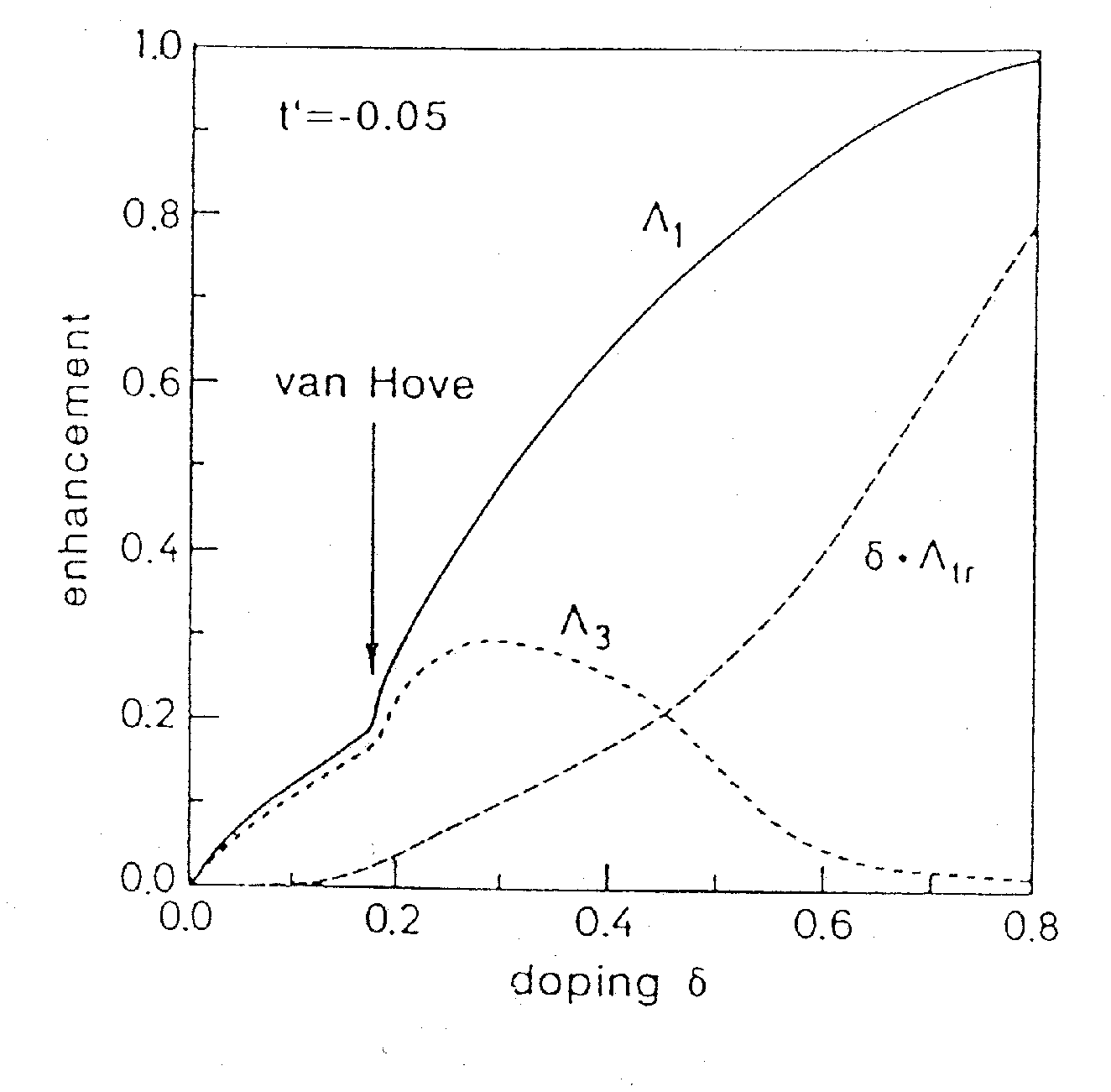}}
\caption{\textbf{(a)} Normalized s-wave $\Lambda _{1}$, d-wave $\Lambda _{3}$
and transport$\protect\delta \cdot \Lambda _{tr}$ coupling constants as a
function of doping $\protect\delta $ for $\ t^{\prime }=0$ and $J=0$.
\textbf{(b)} $\Lambda _{1}$ and $\Lambda _{3}$ and $\protect\delta \cdot
\Lambda _{tr}$ as a function of doping $\protect\delta $ for $t^{\prime
}=-0.05$ and $J=0$. After Ref. \protect\cite{Kulic2}.}
\label{CouplingFig}
\end{figure}

Performing similar calculations (as above) for the phonon-limited
resistivity one finds that the latter is related to the \textit{transport
spectral function} $\alpha ^{2}F_{tr}(\omega )=\langle \langle \alpha ^{2}F(%
\mathbf{k,k}^{\prime },\omega )[\mathbf{v}(\mathbf{k})-\mathbf{v}(\mathbf{k}%
^{\prime })]^{2}\rangle _{\mathbf{k}}\rangle _{\mathbf{k}^{\prime
}}/2\langle \langle \mathbf{v}^{2}(\mathbf{k})\rangle _{\mathbf{k}}\rangle _{%
\mathbf{k}^{\prime }}$, where $\mathbf{v}(\mathbf{k})$ is the Fermi
velocity. The effect of strong correlations on EPI was discussed in \cite%
{Kulic1} and more extensively in \cite{Kulic2}, \cite{Kulic3} within the
model where the phonon frequencies $\omega (\mathbf{\tilde{k}-\tilde{k}}%
^{\prime })$ and $g_{0,scr}(\mathbf{k,p},\lambda )$ are weakly momentum
dependent. In order to illustrate the effect of strong correlations on $%
\alpha ^{2}F_{i}(\omega )$ we consider the latter functions at zero
frequency ($\omega =0$) which are then reduced to the (so called)
\textquotedblright enhancement\textquotedblright\ functions%
\begin{equation}
\Lambda _{i}=\frac{N_{sc}(0)}{8N_{0}(0)}\sum_{j=1}^{8}\langle \langle \mid
\gamma _{c}(\mathbf{\tilde{k},\tilde{k}-}T_{j}\mathbf{\tilde{k}}^{\prime
})\mid ^{2}\rangle _{\mathbf{\tilde{k}}}\rangle _{\mathbf{\tilde{k}}^{\prime
}}D_{i}(j).  \label{Lamda}
\end{equation}%
Note that in the case $J=0$ one has $N_{sc}(0)/N_{0}(0)=q_{0}^{-1}$, where $%
q_{0}$ is related to the doping concentration, i.e. $q_{0}=\delta /2$.
Similarly, the correlation effects on the resistivity $\rho (T)(\sim $ $%
\Lambda _{tr})$ renormalize the transport coupling constant $\Lambda _{tr}$
(similarly to $\Lambda _{i}$ it is defined via $\alpha ^{2}F_{tr}(\omega )$%
). Note that for quasiparticles with an isotropic band the absence of
correlations implies that $\Lambda _{1}=\Lambda _{tr}=1$, $\Lambda _{i}=0$
for $i>1$. The averages in $\Lambda _{1},\Lambda _{3}$ and $\Lambda _{tr}$,
shown in Fig.~\ref{CouplingFig}, were performed numerically in \cite{Kulic2}
by using a realistic anisotropic band dispersion\ in the t-t$^{\prime }$-J
model and the corresponding charge vertex. The three curves are multiplied
by a common factor so that $\Lambda _{1}$ approaches $1$ in the empty-band
limit $\delta \rightarrow 1$, when strong correlations are absent. Note that
T$_{c}$ in the weak-coupling limit and in the $i$-$th$ channel scales like $%
T_{c}^{(i)}\approx \langle \omega \rangle \exp (-1/(\Lambda _{i}-\mu
_{i}^{\ast })$, where $\mu _{i}^{\ast }$ is the Coulomb pseudopotential in
the i-th channel and $\langle \omega \rangle $ is the averaged phonon
frequency.

Several interesting results, which are shown in Fig.~\ref{CouplingFig},
should be stressed. \textit{First}, in the empty-band limit $\delta
\rightarrow 1$ the d-wave coupling constant $\Lambda _{3}$ is much smaller
than the s-wave coupling constant $\Lambda _{1}$, i.e. $\Lambda _{3}\ll
\Lambda _{1}$. Furthermore, the totally symmetric function $\Lambda _{1}$
decreases with decreased doping. \textit{Second}, in both models $\Lambda
_{1}$ and $\Lambda _{3}$ \textit{approach one another} at some small doping $%
\delta \approx 0.1-0.2$, where $\Lambda _{1}\approx $ $\Lambda _{3}$ but
still $\Lambda _{1}>$ $\Lambda _{3}$. By taking into account residual
Coulomb repulsion of quasiparticles which usually have $\mu _{d}^{\ast }\ll
\mu _{s}^{\ast }$ one finds that the s-wave superconductivity (which is
governed by the coupling constant $\Lambda _{1}$) is strongly suppressed by $%
\mu _{s}^{\ast }$, while the d-wave superconductivity (governed by $\Lambda
_{3}$) is only weakly affected by $\mu _{d}^{\ast }$. In that case the
d-wave superconductivity due to EPI becomes more stable than the s-wave
superconductivity at sufficiently small doping $\delta $, i.e. $%
T_{c}^{(d)}>T_{c}^{(s)}$. Interference experiments \cite{Tsuei-recent} show
that this occurs in underdoped, optimally doped and overdoped cuprates. This
means, that EPI is responsible for the strength of pairing in cuprates,
while the d-wave superconductivity is triggered by the residual Coulomb
interaction (including also spin-fluctuation scattering). \textit{Third}, in
the nonadiabatic regime when the phonon frequency fulfills the condition $%
\omega _{ph}>\mathbf{q\cdot v}_{F}(\mathbf{q})$ the enhancement function $%
\gamma _{c}^{2}(\mathbf{k}_{F},\mathbf{p,}\omega _{ph}\mathbf{)}/q_{0}$ is
substantially larger compared to the adiabatic one, which is shown in Fig.~%
\ref{CouplingFig}. This means that the strength of EPI coupling is
differently affected by strong correlations for different phonons. For a
given frequency the coupling to phonons with momenta $p<$ $p_{c}=\omega
/v_{F}$ is \textit{enhanced,} while the coupling to those with $%
p>p_{c}=\omega /v_{F}$ is substantially reduced due to suppression of
backward scattering by strong correlations. \textit{Fourth}, the \textit{%
transport coupling constant} $\Lambda _{tr}$ (not properly normalized in
Fig.~\ref{CouplingFig} - see the correction in \cite{Kulic3}) is reduced in
the presence of strong correlations, especially for lower doping where one
has approximately $\Lambda _{tr}<\Lambda /3$ \cite{Kulic3}. This is an
important result since it resolves the experimental puzzle - that $\lambda
_{tr}$ (which enters resistivity $\rho (T)\sim \lambda _{tr}T$) is \textit{%
much smaller} than the coupling constant $\lambda $ (which enters the
self-energy $\Sigma $ and $T_{c}$), i.e. $\lambda _{tr}<<\lambda $. To
remind the reader, $\lambda _{tr}\approx \lambda $ is realized in almost all
LTSC. One of the important differences between LTSC\ and HTSC in cuprates
lies in strong correlations present in cuprates which causes the forward
scattering peak in charge scattering processes. \textit{Fifth}, the forward
scattering peak in EPI of strongly correlated systems is a general
phenomenon \textit{affecting electronic coupling to all phonons}. This means
that the bare EPI coupling constant for each phono-mode ($\nu $) must be
multiplied by the vertex function, i.e. $g_{\nu }^{0}(\mathbf{k},\mathbf{q}%
)\rightarrow \gamma (\mathbf{k},\mathbf{q})g_{\nu }^{0}(\mathbf{k},\mathbf{q}%
)$.

As we already said, Monte Carlo (numerical) calculations of the Hubbard
model at finite $U$ - performed by Scalapino's group \cite{Scalapino-Hanke},
show that the forward scattering peak in the EPI coupling constant (and the
charge vertex) develops as U is increased. These numerical results prove the
essential correctness of EPI theory based on the X method. The latter effect
is more pronounced at lower doping. Results similar to Monte Carlo ones were
obtained recently within the framework of R\"{u}ckenstein-Kotliar (four
slave-boson) model \cite{Pietro-SB}.

We stress that, contrary to the X-method where the systematic $1/N$
calculations of the EPI self-energy is uniquely done, this is still a
problem for the $SB$ method where an $1/N$ expansion of the partition
function $Z(T,\mu )$ is usually performed \cite{Kim}. The expression for the
vertex function in the SB method is different from that in the X method \cite%
{Kulic1}, \cite{Kulic2}. It seems that the existing SB treatment of EPI
omits a class of diagrams which causes an incorrect dependence of the
coupling constant $\lambda $ on doping. As a result, the vertex function in
the $SB$ approach is peaked not at $q=0$ but at some finite $q_{\max }$,
where $q_{\max }=0$ only for doping $\delta =0$.

\section{Novel effects due to the forward scattering peak in EPI}

There are a number of effects which are predicted by the theory with FSP in
EPI\ and in nonmagnetic impurity scattering - the \textit{FSP theory}. We
have already explained the effects of the forward scattering peak on the
EPI. We discuss briefly some other predictions of the FSP theory including
aspects not covered in \cite{Kulic-Review}.

\subsection{Theory of ARPES self-energy in cuprates}

Recent ARPES experiments of the Shen group \cite{Lanzara}, \cite{Cuk}, \cite%
{Zhou-PRL}, \cite{ARPES-Bi2201} gave additional evidence that phonons are
involved in pairing mechanism of cuprates. Furthermore, the clear isotope
effect was found in the real part of the ARPES self-energy $\func{Re}\Sigma (%
\mathbf{k},\omega )$ \cite{Lanzara2}, which confirms the importance of EPI\
interaction in cuprates.

\subsubsection{ARPES kink and non-shift puzzle}

Recently, it was reported that in all metallic cuprates there is a \textit{%
kink} in the quasiparticle dispersion $\omega (\xi _{\mathbf{k}})$ in the
nodal direction (along the $(0,0)-(\pi ,\pi )$ line) at around $\omega
_{ph}^{(70)}\sim (60-70)$ $meV$ \cite{Lanzara}, see the property (\textbf{6}$%
_{N}$) in Sec.II.F and Fig.~\ref{ARPESLanzFig}. In the superconducting state
this kink is not shifted (the property (\textbf{3}$_{S}$) in Sec. II.F)\
contrary to the standard Eliashberg theory - we call this the \textit{ARPES
non-shift puzzle}. In the case of the anti-nodal point $(\pi ,0)$ there is a
kink at 40 meV in the normal state which is shifted by the maximal gap in
the superconducting state (the property (\textbf{4}$_{S}$) \cite{Cuk}, see
Fig.~\ref{ARPESLanzFig}. This means that any theory which aspires to explain
the pairing in cuprates must solve the \textit{non-shift puzzle}. It is
worth mentioning that in spite of the apparent existence of the non-shift
puzzle since 2001, practically there have been no publications that treat
this problem except the paper \cite{Kulic-Dolgov} where a plausible theory
was given. The approach in \cite{Kulic-Dolgov} is based on the \textit{FSP
theory }for EPI and other charge scattering processes, which is discussed
above. In order to explain the non-shift puzzle by the FSP theory the
following simplifications are made: \textit{(i)} electron-phonon interaction
is dominant in HTSC, and its spectral function $\alpha ^{2}F(\mathbf{k},%
\mathbf{k}^{\prime },\Omega )\approx \alpha ^{2}F(\varphi ,\varphi ^{\prime
},\Omega )$ ($\varphi $ is the angle on the Fermi surface) has a pronounced
forward scattering peak due to strong correlations. Its width is very narrow
$\mid \mathbf{k}-\mathbf{k}^{\prime }\mid _{c}\ll k_{F}$ even in overdoped
systems \cite{Kulic1}, \cite{Kulic2}, \cite{Kulic3}. To the leading order,
one can put $\alpha ^{2}F(\varphi ,\varphi ^{\prime },\Omega )\sim \delta
(\varphi -\varphi ^{\prime })$; \textit{(ii) }for simplicity the slightly
broadened Einstein spectrum is assumed; \textit{(iii)} the dynamic part
(beyond the Hartree-Fock) of the Coulomb interaction is characterized by the
spectral function $S_{C}(\mathbf{k},\mathbf{k}^{\prime },\Omega )$. The
ARPES non-shift puzzle implies that $S_{C}$ is either peaked at small
transfer momenta $\mid \mathbf{k}-\mathbf{k}^{\prime }\mid $, or it is so
small that the shift is weak and beyond the experimental resolution of
ARPES. We assume that the former case is realized, although this is not
crucial because ARPES indicates that electron-phonon coupling is much larger
than the Coulomb coupling, i.e. $\lambda _{ph}>>\lambda _{C}$; \textit{(iv)}
The scattering potential from non-magnetic impurities has pronounced forward
scattering peak, which is also due to strong correlations \cite{Kulic1},
\cite{Kulic2}, \cite{Kulic3}. The latter is characterized by two rates $%
\gamma _{1(2)}$. The case $\gamma _{1}=\gamma _{2}$ mimics the extreme
forward impurity scattering. In this case, d-wave pairing is unaffected by
impurities \cite{Kulic-Review} - the \textit{Anderson theorem for
unconventional pairing}. On the other hand the case $\gamma _{2}=0$
describes the isotropic exchange scattering - see discussion in \cite%
{Kulic-Dolgov}.

\begin{figure}[tbp]
\resizebox{.5 \textwidth}{!} {
\includegraphics*[width=8cm]{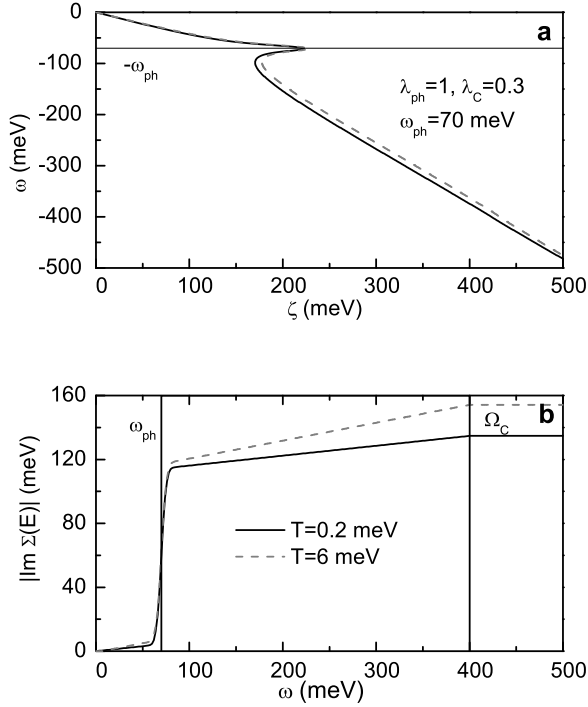}}
\caption{Nodal direction: (a) The quasiparticle-spectrum $\protect\omega (%
\protect\xi _{\mathbf{k}})$ and (b) the imaginary self-energy $Im\Sigma (%
\protect\xi =0,\protect\omega )$ in the nodal direction ($\protect\varphi =%
\protect\pi /4$) in the superconducting ($T=0.2$ $meV$) and normal ($T=6$ $%
meV$) state. $\Omega _{C}=400$ $meV$ is the cutoff in $S_{C}$. After Ref.
\protect\cite{Kulic-Dolgov}.}
\label{pssNodalFig}
\end{figure}

The Green's function is given by
\begin{equation}
G_{k}=-\frac{i\tilde{\omega}_{k}+\xi _{\mathbf{k}}}{\tilde{\omega}%
_{k}^{2}+\xi _{\mathbf{k}}^{2}+\tilde{\Delta}_{k}^{2}}  \label{green-func}
\end{equation}%
where $k=(\mathbf{k},\omega )$. In the FSP theory the equations for $\tilde{%
\omega}_{k}$ and $\tilde{\Delta}_{k}$ are local \ on the Fermi surface, i.e.
in $\mathbf{k}$ space and \cite{Kulic-Dolgov}

\begin{equation}
\tilde{\omega}_{n,\varphi }=\omega _{n}+\pi T\sum_{m}\frac{\lambda
_{1,\varphi }(n-m)\tilde{\omega}_{m,\varphi }}{\sqrt{\tilde{\omega}%
_{m,\varphi }^{2}+\tilde{\Delta}_{m,\varphi }^{2}}}+\Sigma _{n,\varphi }^{C},
\label{Eq23}
\end{equation}

\begin{equation}
\tilde{\Delta}_{n,\varphi }=\pi T\sum_{m}\frac{\lambda _{2,\varphi }(n-m)%
\tilde{\Delta}_{m,\varphi }}{\sqrt{\tilde{\omega}_{m,\varphi }^{2}+\tilde{%
\Delta}_{m,\varphi }^{2}}}+\tilde{\Delta}_{n,\varphi }^{C},  \label{Eq24}
\end{equation}%
where $\lambda _{1(2),\varphi }$ is given by
\begin{eqnarray}
\lambda _{1(2),\varphi }(n-m) &=&\lambda _{ph,\varphi }(n-m)+\delta
_{mn}\gamma _{1(2),\varphi }  \label{lambda-total} \\
\lambda _{ph,\varphi }(n) &=&2\int_{0}^{\infty }d\Omega \frac{\alpha
_{ph,\varphi }^{2}F_{\varphi }(\Omega )\Omega }{(\Omega ^{2}+\omega _{n}^{2})%
}  \nonumber
\end{eqnarray}%
with the electron-phonon coupling function $\lambda _{ph,\varphi }(n)$.
Since EPI and $\Sigma _{n,\varphi }^{C}$ in Eq. (\ref{Eq23}-\ref{Eq24}) has
a \textit{local form} as a function of the angle $\varphi $ due to FSP, then
the equation for $\tilde{\omega}_{n,\varphi }$ has also a local form. This
means that different points on the Fermi surface are decoupled. In that case
$\tilde{\omega}_{n,\varphi }$ depends on the local value (on the Fermi
surface) of the gap $\tilde{\Delta}_{n,\varphi }\approx \Delta _{0}\cos
2\varphi $. \textit{This property alone is important in solving the ARPES
non-shift puzzle}. So, at the \textit{nodal point} ($\varphi =\pi /4$) one
has $\tilde{\Delta}_{n,\varphi }=0$ and the quasiparticle spectrum given by%
\begin{equation}
E-\xi _{\mathbf{k}}-\Sigma _{k}(E,\tilde{\Delta}_{n,\varphi }=0)=0
\label{spect-self}
\end{equation}%
is \textit{unaffected} by superconductivity, i.e. the kink is not-shifted in
the superconducting state, as shown in Fig.~\ref{pssNodalFig}a. This is
exactly what is seen in the experiment of the Shen group \cite{Lanzara},
\cite{Zhou} - see Fig.~\ref{ARPESLanzFig}. The FSP theory also predicts: (1)
that $\Sigma _{2}(\equiv \func{Im}\Sigma )$ has a knee-like shape for $%
\omega <\omega _{ph}^{(70)}$, see Fig.~\ref{pssNodalFig}b, exactly as seen
recently by ARPES \cite{Lanzara}, \cite{Zhou}; (2) that for $\omega >\omega
_{ph}^{(70)}$ the EPI contribution to $\Sigma _{2}$ is constant while its
slope in this region is determined by the Coulomb scattering giving a small
coupling constant $\lambda _{C}<0.4$. Such a behavior is just what has been
observed in ARPES spectra \cite{Lanzara}, \cite{Zhou}, thus confirming the
correctness of the FSP theory. Let us point out that the experimental ARPES
curves for $\Sigma _{1}$ and $\Sigma _{2}$ are smoother than (our)
theoretical curves, since we study the problem in the very simplified model
with the slightly broadened Einstein spectrum. Since the tunnelling and
ARPES experiments indicate that the broad spectrum of phonons contribute to
EPI, then the quantitative analysis must rely on the realistic Eliashberg
spectral function which is at present unknown.

\begin{figure}[tbp]
\resizebox{.5 \textwidth}{!}
{\includegraphics*[width=8cm]{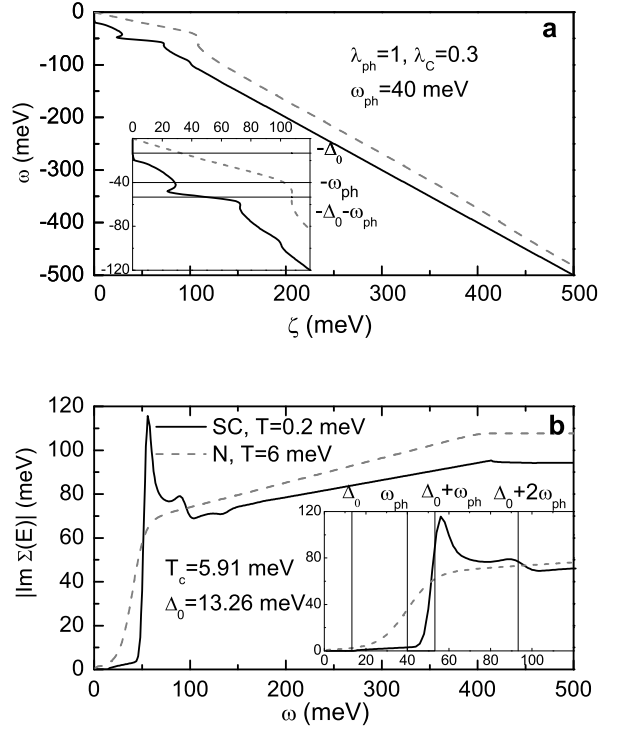}}
\caption{Anti-nodal point: \textbf{(a)} The quasiparticle-spectrum $\protect%
\omega (\protect\xi_{\mathbf{k}})$ and \textbf{(b)} the imaginary
self-energy $Im \Sigma (\protect\xi =0,\protect\omega )$ in the anti-nodal
direction ($\protect\varphi =0;\protect\pi /2$) in the superconducting ($%
T=0.2$ $meV$) and normal ($T=6$ $meV$) state. After Ref. \protect\cite%
{Kulic-Dolgov}.}
\label{pssAntiNodalFig}
\end{figure}

In the case of \textit{antinodal point} ($\varphi \approx \pi /2$) the
calculations show, that there is a singularity at $40$ $meV$ (E$_{\sin g}$)
in the quasiparticle spectrum in the normal state - see Fig.~\ref%
{pssAntiNodalFig}. The analytic and numeric calculations of Eq. (\ref{Eq23})
show that this singularity is \textit{shifted} by $\Delta _{0}$ in the
superconducting state, i.e. $E_{\sin g}\rightarrow E_{\sin g}+\Delta _{0}$.
This is exactly what is seen in the recent experiment on BSCCO \cite{Cuk} -
see Fig.~\ref{ARPESLanzFig}, where the singularity of the normal state
spectrum at $40$ $meV$ is shifted to $(65-70)$ $meV$ in the superconducting
state, since $\Delta _{0}\approx (25-30)$ $meV$. We stress that the coupling
constants $\lambda _{ph}$ $(>1)$ and $\lambda _{C}$ $(\sim 0.4)$ were
estimated from $\func{Re}\Sigma $ by knowing that $\func{Re}\Sigma
=-(\lambda _{ph}+\lambda _{C})\omega $ for $\omega \ll \omega _{ph}^{\max }$
and $\func{Re}\Sigma =-\lambda _{C}\omega $ for $\omega _{ph}^{\max }\ll
\omega <W$ (the band width). The FSP theory explains in the natural way also
the \textit{peak-dip-hump structure }in $A(\mathbf{k},\omega )$ if $\lambda
_{ph}>1$, as observed experimentally- see more in \cite{Kulic-Dolgov}.

In conclusion, the different behavior of the nodal and the anti-nodal kinks
in the superconducting state can be consistently described by the existence
of the forward scattering peak in EPI and other charge scattering processes.
We stress that the FSP\ theory is up to now the only one that succeeded in
resolving the non-shift.

\subsubsection{ARPES isotope effect}

Recent ARPES spectra in the optimally doped Bi-2212 taken near the nodal and
the anti-nodal points \cite{Lanzara2} show after oxygen isotope substitution
$^{16}O\rightarrow ^{18}O$ a pronounced isotope effect in the real part of
the quasiparticle self-energy. This result, if confirmed, will be a crucial
evidence for the importance of EPI in cuprates.

\textbf{A. Isotope effect along the nodal direction. }The ARPES results for
the isotope effect in optimally doped $Bi2212$ samples are as follows \cite%
{Lanzara2}: (\textit{1}) there is a kink in the quasiparticle spectrum at $%
\omega _{k,70}\simeq 70$ $meV$ which is unshifted in the SC state; (\textit{2%
}) there is a red shift $\delta \omega _{k,70}\sim -(10-15)$ $meV$ of the
kink for the $^{16}O\rightarrow ^{18}O$ substitution; (\textit{3}) the
isotope shift of the self-energy is more pronounced at large energies $%
\omega =100-300$ $meV$.

\begin{figure}[tbp]
\resizebox{.5\textwidth}{!} {
\includegraphics*[width=7cm]{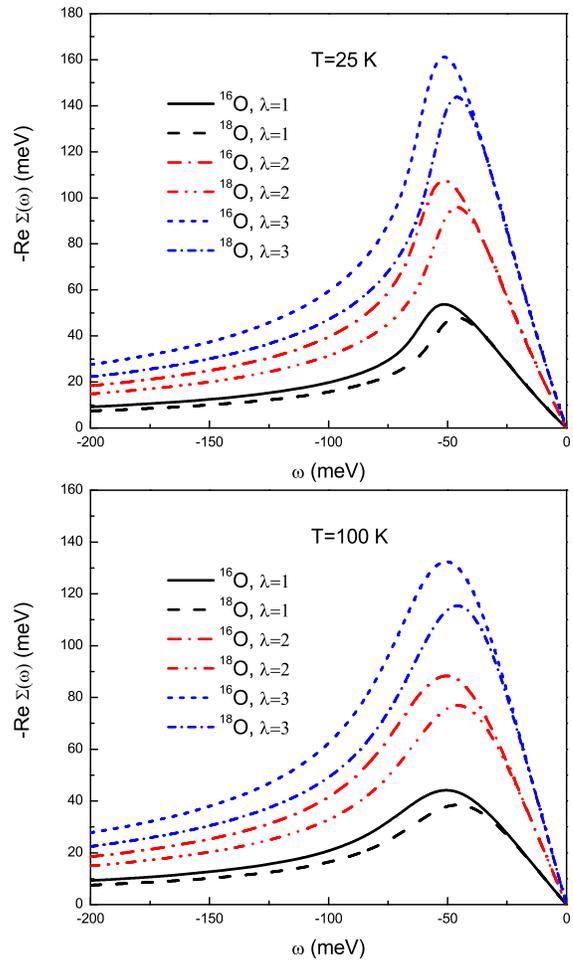}}
\caption{The shift of $Re\Sigma (\mathbf{k},\protect\omega )$ in the nodal
direction upon the isotope substitution $^{16}O\rightarrow ^{18}O$
calculated using the Debye model with $\protect\omega_{O^{16}}=60$ $meV$ for
various EPI coupling constants $\protect\lambda$ at $T=25$ $K$ (top) and $%
T=100$ $K$ (bottom). From Ref. \protect\cite{MDK}.}
\label{MDKFig}
\end{figure}

The results (\textit{1})-(\textit{2}) can be explained by the EPI theory
with FSP in a natural way. As it is already discussed, the theory predicts
that the kink in the superconducting state is non-shifted with respect to
that in the normal state, what is due to the FSP in the EPI. Related to the
isotope effect, the theory predicts \cite{MDK} that in the \textit{nodal
direction} after the substitution $^{16}O\rightarrow ^{18}O$ \ ( $\omega
_{O^{16}}=\omega _{kink}\sim 70$ $meV$) there is a red-shift of $Re\Sigma (%
\mathbf{k},\omega )$ in the normal and superconducting state, due to the
isotope dependence of the maximum of $Re\Sigma (\mathbf{k},\omega )$ at $%
\omega \approx \omega _{kink}$. The isotope effect is more pronounced at
higher energies. This is shown in Fig.~\ref{MDKFig} for the Debye model with
the spectral function $\alpha ^{2}F(\omega )=\lambda (\omega /\omega
_{O})^{2}$ with $\omega _{O}=(k/M_{O})^{1/2}$ ($M_{O}$ is the oxygen mass) -
see the left panels in Fig.~\ref{MDKFig}.

\begin{figure}[tbp]
\resizebox{.5\textwidth}{!}
{\includegraphics*[width=7cm]{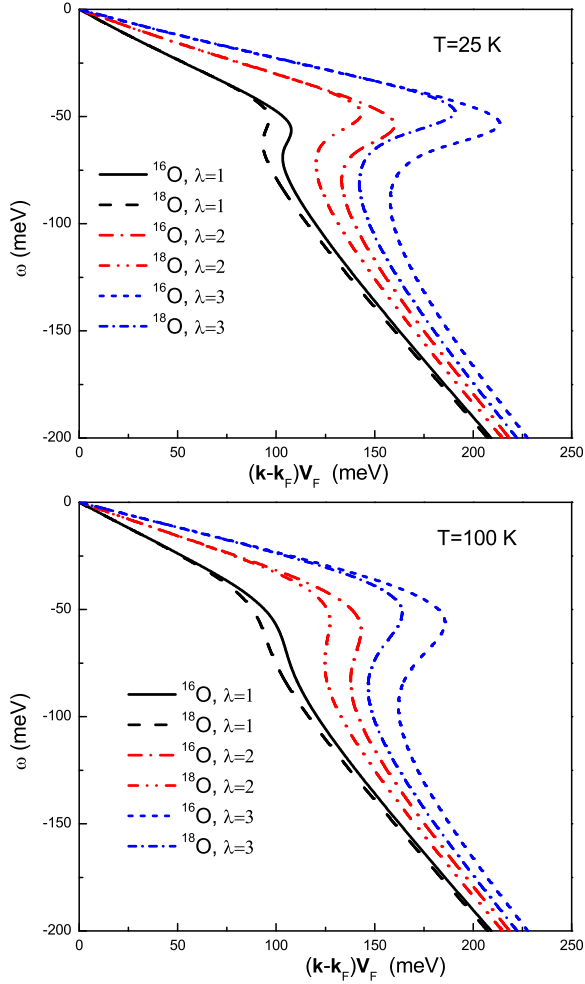}}
\caption{The
quasiparticle energy $\protect\omega (k-k_{F} )$ in the nodal
direction upon the isotope substitution $^{16}O\rightarrow ^{18}O$
calculated using the Debye model with $\protect\omega_{O^{16}}=60
$ $meV$ for various EPI coupling constants $\protect\lambda$ at
$T=25$ $K$ (top) and $T=100$ $K$ (bottom). From Ref.
\protect\cite{MDK}.} \label{MDKFigKink}
\end{figure}

The pronounced isotope effect is also present in the quasiparticle
dispersion $\omega (k)$ with the read shift of the kink energy
upon the substitution $^{16}O\rightarrow ^{18}O$. This is shown in
the right panels of Fig. ~\ref{MDKFigKink}. The property
(\textit{3}) is difficult to explain by the Eliashberg-like theory
we have used above. Some speculations that the pronounced isotope
effect at high energies is of polaronic origin wait for further
theoretical elaborations.

B. \textbf{Isotope effect near the anti-nodal point. }The main experimental
results for the ARPES isotope effect near the \textit{anti-nodal} point $%
\mathbf{k}_{AN}\approx (0,\pi )$ are \cite{Lanzara2}: (\textit{1}) there is
a kink in the quasiparticle spectrum around $\omega _{k,40}\approx 40$ $meV$
which is shifted in the SC state by $\Delta (\mathbf{k}_{AN},\omega )\approx
30$ $meV$; (\textit{2}) there is a red shift of the kink in the SC state
with the energy change $\delta \omega _{\mathbf{k}}\sim -5$ $meV$ at $T=25$ $%
K$, which is smaller than the corresponding one in the nodal direction; (%
\textit{3}) there is an inverse IE at higher energies $\omega >\omega
_{k,40} $, i.e. $\omega (\xi _{k};^{18}O)>\omega _{O}(\xi _{k};^{16}O)$; (%
\textit{4}) in the normal state at $T=100$ $K$ there is practically no red
shift for the $^{16}O\rightarrow ^{18}O$ substitution.

\begin{figure}[tbp]
\resizebox{.5 \textwidth}{!} {
\includegraphics*[width=6cm]{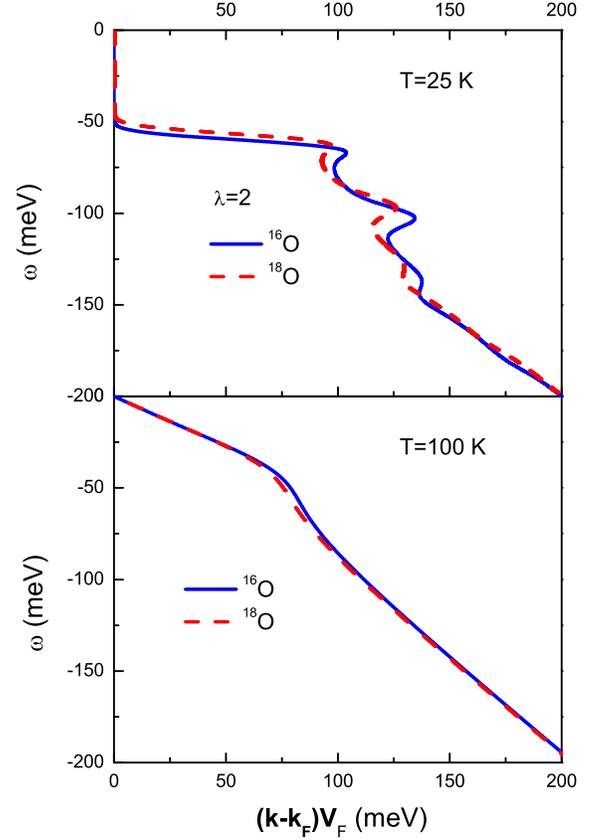}}
\caption{ The quasiparticle energy $\protect\omega (k-k_{F} )$ in the
\textit{anti-nodal region} under the isotope substitution $^{16}O\rightarrow
^{18}O$ in the Debye model with $\protect\omega_{O^{16}}=40$ $meV$ for the
EPI coupling constants $\protect\lambda _{ep,\mathbf{k}_{AN}}=2$ at T=25 K
(top) and T= 100 K (bottom). $k_{F}$ and $v_{F}$ are the Fermi momentum and
velocity, respectively.}
\label{MDKFig-3}
\end{figure}

Theoretical calculations of the self-energy in the \textit{anti-nodal}
region have been done for the Debye spectrum \cite{MDK}, only. Since the
density of states near the anti-nodal point is higher than at the nodal
point, we assume a larger (than in the nodal case) EPI coupling constant $%
\lambda _{ep,\mathbf{k}_{AN}}=2$. In Fig. \ref{MDKFig-3} (top) it is seen
that in the \textit{SC} state (at $T=25$ $K$) the kink in the quasiparticle
energy is shifted by the gap $\Delta (\mathbf{k}_{AN})$ and that there is a
moderate red-shift ( $\sim -(5-8)$ $meV$) of the kink for the $%
^{16}O\rightarrow ^{18}O$ substitution. Note that this value is only
accidentally in agreement with the experimental one ( $\sim -5$ $meV$).
However, the theory predicts smaller (compared to experiments) inverse
isotope effect ($\omega (\xi _{k};^{18}O)>\omega _{O}(\xi _{k};^{16}O)$)
which occurs at higher energies. It seems that the origin of this behavior
lies in the smaller gap value $\Delta (\mathbf{k}_{AN})$ of $^{18}O$ (dashed
line in Fig. \ref{MDKFig-3}) than of $^{16}O$ (bold line). In spite of the
fact that these results resemble qualitatively the experimental results (%
\textit{1})-(\textit{3}) there is quantitative difference between the theory
and experiment at high energies, the origin of which is unknown at present.

The FSP theory predicts that the isotope effect at the anti-nodal point is
much less pronounced in the normal state (at $100$ $K$) as shown in Fig. \ref%
{MDKFig-3}. These results resemble qualitatively the experimental results in
\cite{Lanzara2}. The origin of the small shift of the kink in the normal
state lies in the lower antinodal kink energy and in the smearing effects at
higher temperatures.

\subsubsection{Collapse of the anti-nodal elastic scattering in the
superconducting state}

A number of ARPES experiments on optimally doped Bi-2212 give evidence for
the significant elastic scattering which varies along the Fermi surface,
being the smallest along the nodal direction and the largest near the
anti-nodal point, see \cite{Shen} and references therein. The experiments
indicate that the quasiparticle spectral function $A(\mathbf{k},\omega )$ at
the anti-nodal point is broad in the normal state and strongly sharpens in
the superconducting state. On the other side, at the nodal point $A(\mathbf{k%
},\omega )$ has a standard Lorentzian form which is almost unchanged in the
superconducting state.

This dramatic sharpening of the spectral function near the anti-nodal point $%
(\pi ,0)$ at $T<T_{c}$ can be explained by assuming that there is a forward
scattering peak in the elastic impurity scattering as it was done in \cite%
{Kulic-Dolgov}. It is easy to see from Eq. (\ref{Eq23}-\ref{Eq24}), which
describe the extreme forward scattering, that in the case of the FSP in
impurity potential one has $\gamma _{1}=\gamma _{2}$. As a result, one finds
that the impurity scattering rate $\Gamma _{imp}(\mathbf{k},\omega )=\Gamma
_{n}(\mathbf{k},\omega )+\Gamma _{a}(\mathbf{k},\omega )=0$ for $\mid \omega
\mid <\Delta _{0}$ for any kind of pairing (s- p- d-wave etc.) since the
normal ($\Gamma _{n}$)\ and the anomalous ($\Gamma _{a}$) scattering rates
compensate each other in the superconducting state - the \textit{collapse of
the elastic scattering rate}. This result is a consequence of the Anderson
theorem \cite{KuOudo}-\cite{Oleg}. In such a case d-wave pairing is
unaffected by impurities - there is a negligible reduction in $T_{c}$, as
was first elaborated in \cite{KuOudo} and \cite{Oleg}, and further studied
in \cite{Kee-Tc}. The physics behind this result is rather simple. Forward
scattering means that electrons scatter into a very small region in the
k-space, so that at the most part of the Fermi surface there is no mixing of
states with different signs of the order parameter $\Delta (\mathbf{k})$,
and the detrimental effect of impurities is reduced. Only for states near
the nodal points there is mixing but since $\Delta (\mathbf{k})$ is a small
there is only small reduction in $T_{c}$ \cite{Licht}.The collapse of the
elastic scattering rate in the ARPES spectra was recently considered in \cite%
{Zhu}, where a number of interesting results are reported.

The idea of the forward scattering peak in the non-magnetic impurity
scattering was applied recently in \cite{Varma-Hall} in the study of the
anomalous temperature dependence of the Hall angle in optimally doped HTSC.
By taking into account (besides the inelastic scattering) also the
small-angle elastic scattering on non-magnetic impurities in \cite%
{Varma-Hall} it was explained the experimental finding in $%
YBa_{2}Cu_{3-x}Zn_{x}O_{7-\delta }$ that $cot\Theta _{H}(T)=\sigma
_{xx}/\sigma _{xy}\sim T^{2}$ ($\Theta _{H}$ is the Hall angle, H is the
applied magnetic field) for $100K<T<300K$. This is contrary to the expected
linear behavior (for $cot\Theta _{H}(T)$), since in that temperature
interval one has $\rho (T)\sim T$.

In conclusion, in order to explain ARPES results in cuprates it is necessary
to take into account: (1) EPI interaction, since it dominates in the
quasiparticle scattering in the frequency region responsible for pairing in
cuprates; (2) effects of elastic nonmagnetic impurities with FSP; (3) the
Coulomb interaction which dominates at higher energies $\omega >\omega _{ph}$%
. In this respect, the presence of ARPES kink and the knee-like shape of the
spectral width are the \textit{smoking-gun experiments} that strongly
constraint possible theories.

\subsection{Nonmagnetic impurities and robustness of d-wave pairing}

In the presence of strong correlations\ the impurity potential is also
renormalized, as it is mentioned above, and the effective potential in the
Born approximation is given by $u_{i}^{2}(\mathbf{q})=\gamma _{c}^{2}(k_{F},%
\mathbf{q})u_{i,0}^{2}(\mathbf{q})$, where $u_{i,0}(\mathbf{q})$ is the
single impurity scattering potential in the absence of strong correlations.
This was proved first in \cite{Kulic1} and elaborated quantitatively in \cite%
{KuOudo} and \cite{Oleg}. Since the charge vertex $\gamma _{c}(\mathbf{p}%
_{F},\mathbf{q})$ is peaked at $\mathbf{q=0}$ the potential $u_{i}(\mathbf{q}%
)$ is also peaked at $\mathbf{q=0}$. This means that the scattering
amplitude contains not only the s-channel (as usually assumed in studying
impurity effects in cuprates), but also the d-channel, etc. Based on this
property the FSP theory succeeded in explaining some experimental facts,
such as: $(\mathbf{i})$ suppression of the residual resistivity $\rho _{i}$
\cite{Kulic1}, \cite{Kulic2}. This effect was observed in optimally doped $%
YBCO$, where the resistivity $\rho (T)$ at $T=0$ $K$ has a rather small
value $<10$ $\mu \Omega cm$.; $(\mathbf{ii})$ robustness of d-wave pairing
\cite{KuOudo}, \cite{Oleg} in the presence of impurities and other defects
in $CuO_{2}$ planes, such as Zn, Ga, O defects, etc., see \cite{Tolpygo}.

Early theories of the effect of nonmagnetic impurities on $T_{c}$ in
cuprates \cite{Ferenbacher} always assumed that $u_{i}(\mathbf{q})=const$,
i.e. that only the s-wave scattering channel is present. Such a theory
predicts that $T_{c}(\rho _{i,c})=0$ at a much smaller residual resistivity $%
\rho _{i,c}^{(s)}\sim 50$ $\mu \Omega $cm, while the experimental range is $%
200$ $\mu \Omega $cm$<\rho _{i,c}^{\exp }<1500$ $\mu \Omega $cm \cite%
{Tolpygo}. The latter experimental fact means that d-wave pairing in HTSC is
\textit{much more robust} than what the standard theory predicts, and it is
one of the smoking gun experiments in testing the concept of FSP in the
charge scattering potential. It is worth mentioning that in a number of
papers the pair-breaking effect of non-magnetic impurities in HTSC was
analyzed in terms of the impurity concentration $n_{i}$, i.e. the dependence
$T_{c}(n_{i})$. However, $n_{i}$ is not the parameter which governs this
pair-breaking effect. The more appropriate parameter for discussing the
robustness of d-wave pairing is the impurity scattering amplitude $\Gamma
(\theta ,\theta ^{\prime })$, which can be related to the measured residual
resistivity $\rho _{i}$ and to $T_{c}(\rho _{i})$. The robustness of d-wave
pairing in HTSC can be revealed only by studying the experimental curve $%
T_{c}(\rho _{i})$, as was first recognized experimentally in \cite{Tolpygo}
and theoretically in \cite{Kulic1}, \cite{KuOudo}.

\begin{figure}[tbp]
\resizebox{.5\textwidth}{!} {
\includegraphics*[width=10cm]{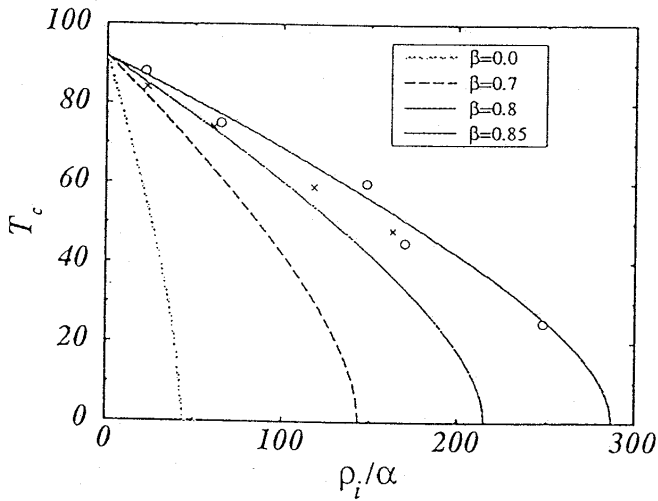}}
\caption{The critical temperature $T_{c}$ $[K]$ of a d-wave superconductor
as a function of the experimental parameter $\protect\rho _{i}/\protect%
\alpha _{c}[K]$, where $\protect\rho _{i}$ is the residual resistivity and $%
\alpha =8\pi^2 \lambda_{tr}/\omega_{pl}^2$. The case
$\protect\beta =0$ corresponds to the prediction of the standard
d-wave theory with isotropic
scattering \protect\cite{Ferenbacher}. The experimental data \protect\cite%
{Tolpygo} are indicated by crosses for $YBa_{2}(Cu_{1-x}Zn_{x})_{3}O_{7-%
\protect\delta }$, and by circles for $Y_{1-y}\Pr_{y}Ba_{2}Cu_{3}O_{7-%
\protect\delta }$. From Ref. \protect\cite{KuOudo}.}
\label{ImpurityFig}
\end{figure}

The robustness of d-wave pairing in HTSC was explained first in \cite{KuOudo}%
, where the FSP theory \cite{Kulic1}, \cite{Kulic2} was applied to this
problem. We shall not go into details - see \cite{KuOudo}, \cite%
{Kulic-Review} - but just give here the general formula for the $T_{c}(\rho
_{i})$ dependence in anisotropic (including unconventional) superconductors.
We assume that in an unconventional (anisotropic) superconductor the order
parameter has the form $\Delta (\theta )=\Delta _{0}Y(\theta )$ and
generally one has $\langle Y(\theta )\rangle \neq 0$ ($\langle Y^{\ast
}(\theta )Y(\theta )\rangle =1$). In the case of FSP, the momentum dependent
impurity scattering amplitude is given by $\Gamma (\theta ,\theta ^{\prime
})=\Gamma _{s}(\theta ,\theta ^{\prime })+\Gamma _{d}Y_{d}(\theta
)Y_{d}(\theta ^{\prime })+....$. and the critical temperature $T_{c}$ by%
\[
\ln \frac{T_{c}}{T_{c0}}=\Psi (\frac{1}{2})-\Psi (\frac{1}{2}+(1-\beta )x)
\]%
\begin{equation}
-\langle Y(\theta )\rangle ^{2}[\Psi (\frac{1}{2})-\Psi (x+\frac{1}{2})].
\label{Eq65}
\end{equation}%
Here, $T_{c0}$ is the bare critical temperature in the absence of
impurities, $x=\Gamma _{s}/4\pi T_{c}$, $\beta =\Gamma _{d}/\Gamma
_{s}$ and $\langle Y(\theta )\rangle $ means averaging over the
Fermi surface. If one takes into account that the inelastic
scattering, which is characterized by the coupling constant
$\lambda _{Z}$, screens the impurity scattering, then the
parameter $x$ in Eq. (\ref{Eq65}) should be replaced by
$x/(1+\lambda _{Z})$. Note that Eq. (\ref{Eq65}) holds
independently on the scattering strength $\Gamma _{s}$, $\Gamma
_{d}$, i.e. it holds in the Born limit as well as in the unitary
limit. The residual resistivity $\rho _{i}$ can be related to the
transport scattering rate by $\rho _{i}=4\pi \Gamma _{tr}/\omega
_{pl}^{2}$, while the s-wave amplitude is related to $\Gamma
_{tr}$ by $\Gamma _{s}=p\Gamma _{tr}$. The parameter $p>1$ can be
obtained from the microscopic model (in the t-J model $p\approx
2-3$) or can be treated as a fitting parameter, see more in
\cite{Kulic-Review}. In the case of an unconventional pairing in
the tetragonal lattice one has $\langle
Y(\theta )\rangle =0$ and the last term drops. For the s-scattering only ($%
\Gamma _{d}=0$) one has $\beta =0$ and $T_{c}(\rho _{i})$ should be
suppressed very strongly contrary to the experimental results \cite{Tolpygo}%
, see Fig.~\ref{ImpurityFig}.

The FSP theory of impurity scattering in the t-J model\ \cite{KuOudo} gives
that the s-channel and d-channel almost equally contribute to the impurity
scattering amplitude, since $\beta \approx 0.75-0.85$ for doping $\delta
\approx 0.1-0.2$. The dependence of $\beta (\delta )$ has been calculated
for the t-J model, see \cite{KuOudo}, \cite{Kulic-Review}. Since the
d-channel in scattering is not detrimental for d-wave pairing the FSP theory
predicts that $T_{c}(\rho _{i})$ vanishes at much larger $\rho _{i,c}$, i.e.
$\rho _{i,c}^{(FSP)}>>\rho _{i,c}^{(s)}$, in good agreement with
experiments, as shown in Fig.~\ref{ImpurityFig}. In the extreme forward
scattering limit, $T_{c0}$ would be unrenormalized, i.e. the Anderson
theorem holds also for unconventional superconductors. In this case impurity
scattering does not mix superconducting states with different sites of $%
\Delta (\mathbf{k})$.

Here, we have studied the effects of impurities and other defects placed in
the $CuO_{2}$\ planes. The corresponding experiments can be explained by FSP
in the impurity potential, which is due to strong correlations. In the case
when impurities are placed out of these planes their effect is weaker and
also there is a forward scattering peak due to weak screening along the
c-axis. The proposed theory in \cite{KuOudo}, \cite{Oleg} and Eq.((\ref{Eq65}%
) holds for both cases equally.

\subsection{Nonadiabatic corrections of $T_{c}$}

Cuprates are characterized not only by strong correlations but also by the
relatively small Fermi energy $E_{F}$, which is not much larger than the
characteristic (maximal) phonon frequency $\omega _{ph}^{\max }$, i.e. $%
E_{F}\simeq 0.1-0.3$ $eV$, $\omega _{ph}^{\max }\simeq 80$ $meV$. The
situation is even more pronounced in \textit{fullerene compounds} $%
A_{3}C_{60}$, with $T_{c}=20-35$ $K$, where $E_{F}\simeq 0.2$ $eV$ and $%
\omega _{ph}^{\max }\simeq 0.16$ $eV$. \ This fact implies a possible
breakdown of the Migdal's theorem \cite{AllenMit}, which asserts that the
relevant vertex corrections due to the $E-P$ interaction are small if $%
(\omega _{D}/E_{F})\ll 1$. In that respect a comparison of the intercalated
graphite $KC_{8}$ and the fullerene $A_{3}C_{60}$ compounds, given in \cite%
{Grimaldi}, is very instructive, because both compounds have a number of
similar properties. However, the main difference in these systems lies in
the ratio $\omega _{D}/E_{F}$, since $(\omega _{D}/E_{F})\ll 1$ in $KC_{8}$,
while it is rather large $(\omega _{D}/E_{F})\sim 1$ in $A_{3}C_{60}$. Since
the ratio $\omega _{D}/E_{F}$ is not negligible in the fullerene compounds
and in cuprates it is necessary to correct the Migdal-Eliashberg theory by
\textit{vertex corrections due to EPI}. It is known that these vertex
corrections lower T$_{c}$ in systems with isotropic EPI. However, in systems
with FSP and with the cut-off $q_{c}<<k_{F}$ in the pairing potential these
vertex corrections cause an appreciable increase in T$_{c}$. The
calculations by the Pietronero group \cite{Grimaldi} provided two important
results: (\textbf{1}) there is a drastic increase of $T_{c}$ by lowering $%
Q_{c}=q_{c}/2k_{F}$, for instance $T_{c}(Q_{c}=0.1)\approx 4T_{c}(Q_{c}=1)$;
(\textbf{2}) Even small values of $\lambda <1$ can give large T$_{c}$. The
latter results open a new possibility in reaching high $T_{c}$ in systems
with appreciable ratio $\omega _{D}/E_{F}$ and with FSP. The difference
between the Migdal-Eliashberg and non-Migdal theories can be explained
qualitatively in the framework of an approximative McMillan formula for $%
T_{c}$ (for not too large $\lambda $) which reads $T_{c}\approx \langle
\omega \rangle e^{-1/[\tilde{\lambda}-\mu ^{\ast }]}$. The Migdal-Eliashberg
theory predicts $\tilde{\lambda}_{ME}\approx \lambda /(1+\lambda )$ while
the non-Migdal theory \cite{Grimaldi} gives $\tilde{\lambda}_{nM}\approx
\lambda (1+\lambda )$. For instance $T_{c}\sim 100$ $K$ in cuprates can be
explained by the Migdal-Eliashberg theory if $\lambda \sim 2$, while in the
non-Migdal theory much smaller coupling constant is needed, i.e. $\lambda
\sim 0.5$.

\section{Pseudogap and the electron-phonon interaction}

The pseudogap (PG) problem is a very intriguing one and it is not surprising
that a variety of theoretical models have been proposed to explain it. We
are not going to discuss all these in detail but only mention some that may
have experimental support. The \textit{first} one is based on the assumption
that the PG phase represents the state with \textit{pre-formed pairs} \cite%
{Randeria}, where the true critical temperature $T_{c}$ is smaller than the
mean-field one $T_{c}^{MF}$. In the region $T_{c}<T<T_{c}^{MF}$ pre-formed
pairs give rise to a dip in the density of states $N(\omega )$. This
approach is physically plausible having in mind that cuprates are
characterized by a short coherence length, quasi-two dimensionality and
proximity to the Mott-Hubbard insulating state. As we already discussed the
latter gives rise to small phase stiffness $K_{s}^{0}$ of the
superconducting order parameter $\Delta =\mid \Delta \mid e^{i\varphi }$. In
2D systems the Berezinskii-Kosterlitz-Thouless transition temperature, $%
T_{c}=(\pi /2)K_{s}^{0}(T_{c})$ ($\sim \delta $ for small doping), can be
much smaller than $T_{c}^{MF}$. From the experimental side there is some
support for this approach, at least in not very broad temperature region
close to $T_{c}$. For instance, the specific heat measurements \cite{Junod}
(with smaller jumps at $T_{c}$)\ point to a \textit{non-mean field}
character of the superconducting phase transition, particularly for
underdoped compounds. As we have already mentioned in the Introduction,
ARPES measurements show that for $T_{c}<T<T^{\ast }$ PG is d-wave like $%
\Delta _{pg}(\mathbf{k})\approx \Delta _{pg,0}(\cos k_{x}-\cos k_{y})$, like
the superconducting gap, and that $\Delta _{pg,0}$ increases at lower doping.

The \textit{second} approach assumes that PG is due to existence of an
additional competing order, but usually without the true long-range order.
For instance, the \textquotedblright spin-density wave\textquotedblright\
alias for strong antiferromagnetic fluctuations can also produce dip in $%
N(\omega )$ \cite{Sadovskii}. Related approaches are based on RVB, orbital
currents, d-wave order, etc., which are not discussed here.

We would like to pay attention to two possible effects, \ which are due to
FSP, that can also give rise to PG in the density of states. The first is
related to EPI interaction with FSP that can produce a dip in $N(\omega )$
for $\omega <\Omega $, where $\Omega $ \ is the characteristic phonon
frequency. The second is related to a novel type of fluctuations - internal
Cooper pair fluctuations, due to the long-range pairing forces (alias for
FSP) - due to EPI with FSP.

\subsubsection{Pseudogap due to phonons}

\begin{figure}[tbp]
\resizebox{.5\textwidth}{!} {
\includegraphics*[width=10cm]{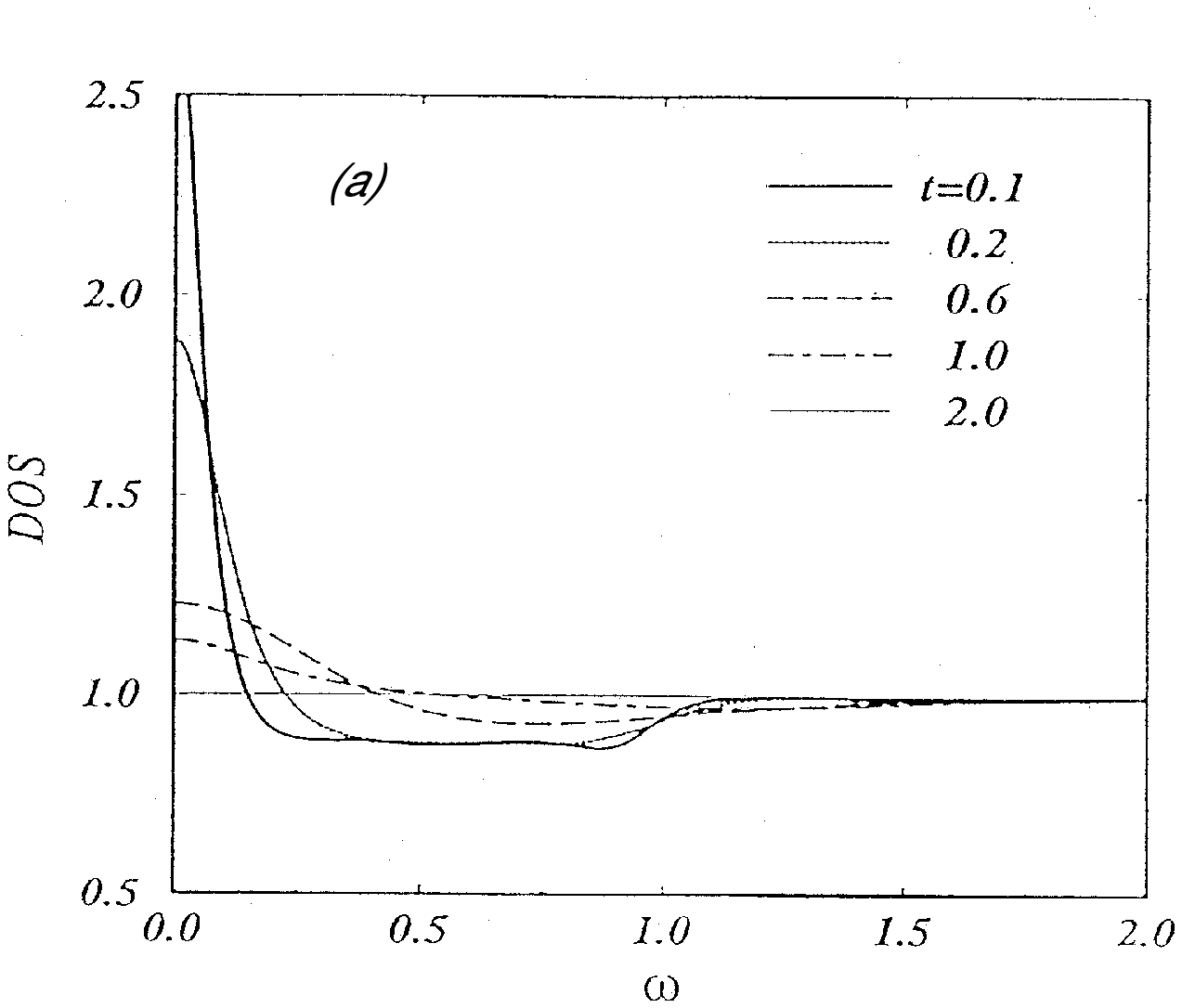}}
{\includegraphics*[width=8cm]{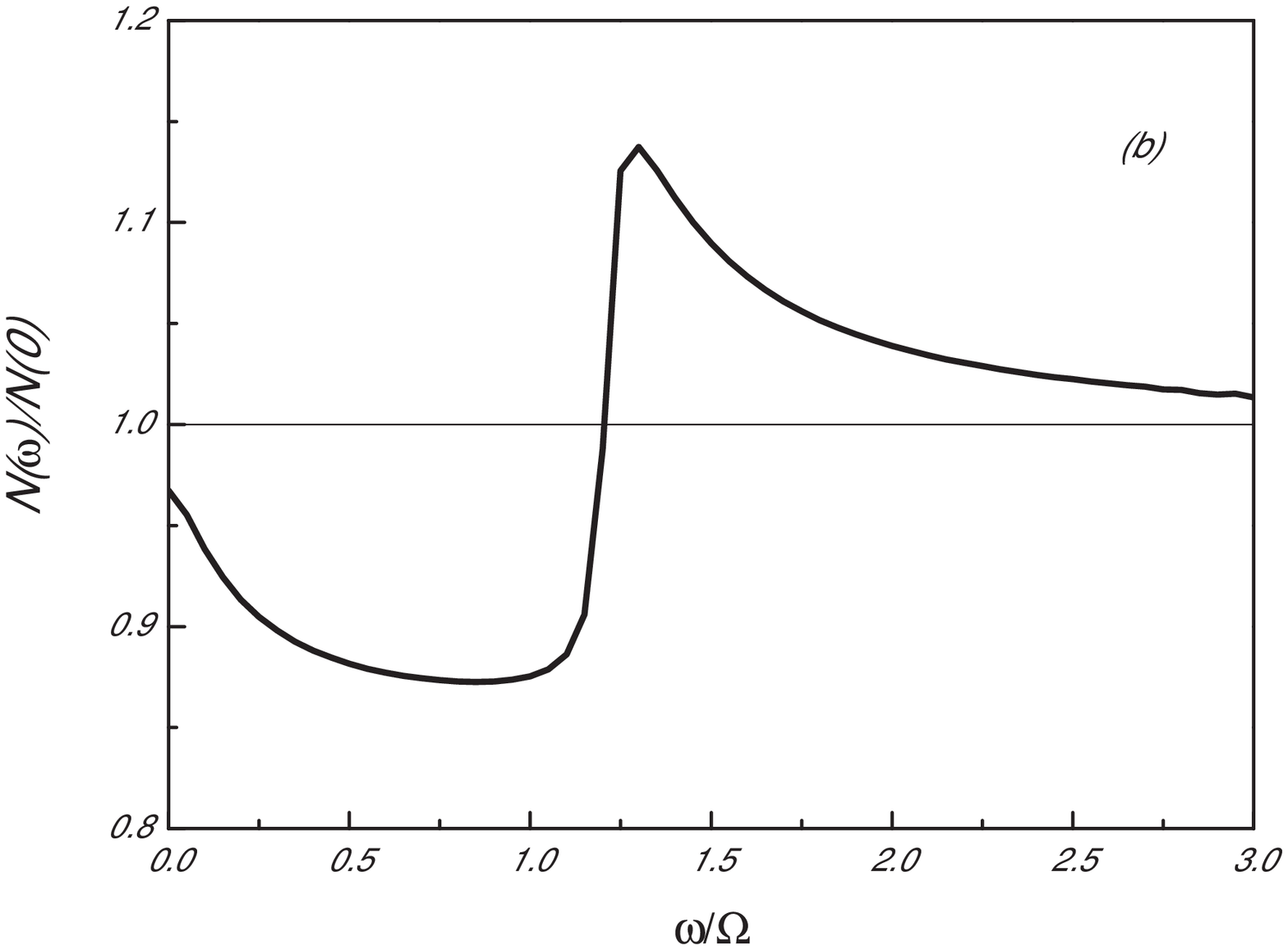}}
\caption{(a) The density
of states $N(\protect\omega )$ in the FSP model for EPI with the
dimensionless coupling $l(=V_{EP}/\protect\pi \Omega )=0.1$ for
various $t(=\protect\pi T/\Omega )$ - case $\protect\xi _{\mathbf{k},\protect%
\omega }=\protect\chi (\mathbf{k,}\protect\omega )\protect\xi _{\mathbf{k}}$%
. (b) $N(\protect\omega )$ for the same parameter l but for $\protect\xi _{%
\mathbf{k},\protect\omega }=\protect\xi _{\mathbf{k}}$. From Ref.
\protect\cite{DDKuOudo2}.}
\label{NomegaFSPFig}
\end{figure}

The EPI, in the Einstein model with the phonon frequency $\Omega $, was also
studied in the extreme limit of FSP \cite{DDKuOudo2}, where in leading order
the spectral function is \textit{singular}, i.e. $\alpha ^{2}F(\mathbf{k},%
\mathbf{k}^{\prime },\omega )\sim \delta (\mathbf{k}-\mathbf{k}^{\prime
})\delta (\omega -\Omega )$. In such a singular case besides renormalization
of frequency ($\omega \rightarrow Z(\mathbf{k,}\omega )\omega $) there is
also significant renormalization of the quasiparticle energy ($\xi _{\mathbf{%
k}}\rightarrow \chi (\mathbf{k,}\omega )\xi _{\mathbf{k}}$) in the
Eliashberg equations, giving a very interesting behavior of the \textit{%
density of states} $N(\omega )$ \cite{DDKuOudo2}. \textit{First}, there is
an increase of $N(\omega =0)(>N_{bare}(\omega =0))$ at $\omega =0$, where $%
N_{bare}(\omega =0)$ is the bare density of states in the absence of EPI,
see Fig.~\ref{NomegaFSPFig}a. This fact is mainly due to renormalization of
the quasiparticle energy ($\xi _{\mathbf{k}}\rightarrow \chi (\mathbf{k,}%
\omega )\xi _{\mathbf{k}}$).

In such a case there is a pseudogap feature in the region $(\Omega
/5)<\omega \leq \Omega $, where $N(\omega )<N_{bare}(\omega )$. This
pseudogap feature disappears at temperature comparable to the phonon energy $%
\Omega $. Note that the usual isotropic and short-ranged EPI does not
renormalize the density of states in the normal state, i.e. $N(\omega
)=N_{bare}(\omega )$.

It is interesting that without renormalization of the quasiparticle energy ($%
\xi _{\mathbf{k}}\rightarrow \xi _{\mathbf{k}}$, i.e. $\chi (\mathbf{k,}%
\omega )=1$) there is also pseudogap behavior. The case $\chi =1$ is
probably realized in systems with finite width of FSP ($\delta k\ll k_{F}$)
and with \textit{particle-hole symmetry}. In such a case one has $N(\omega
=0)<N_{bare}(\omega =0)$ and the pseudogap exist for $\omega <\Omega $, as
it is seen in Fig.~\ref{NomegaFSPFig}b.

Let us mention that for $\chi (\mathbf{k,}\omega )\neq 1$ the transport
properties are very peculiar due to the pseudogap behavior of $N(\omega )$
and the existence of peak at $N(\omega =0)$. For instance, the resistivity $%
\rho (T)$ is linear in $T$ starting at very low temperatures, i.e. $\rho
(T)\sim T$ for $(\Omega /30)\leq T$ and this linearity extends up to several
$\Omega $. The dynamic conductivity $\sigma _{1}(\omega )$ shows the
(extended) Drude-like behavior with the Drude width $\Gamma _{tr}\sim T$,
for $\omega <T$. The above numbered properties are in a qualitative
agreement with experimental results in cuprates - see more in \cite%
{DDKuOudo2}, \cite{Kulic-Review}.

In this extreme FSP limit one can also calculate the mean-field critical
temperature $T_{c0}^{MF}$. In leading order of $\Omega /T_{c0}^{MF}$ ($\gg 1$%
) one has $T_{c0}^{MF}\approx N(0)V_{EP}=\lambda N(0)/4$, where $\lambda
=N(0)V_{EP}$. In this case the maximal superconducting gap is given by $%
\Delta _{0}=2T_{c0}^{MF}$ which is reached on the Fermi surface, while away
from it the gap decreases, i.e. $\Delta _{k}=\Delta _{0}\sqrt{1-(\xi
_{k}/\Delta _{0})^{2}}$. The expression for $T_{c0}^{MF}$ tells us that it
can be large even for $\lambda <0.1$, since in cuprates the bare density of
states is $N_{bare}(0)\sim 1$ $states/eV$. It is apparent that in this order
\textit{there is no isotope effect }in $T_{c0}^{MF}$ since $\alpha =0$.
However, such an extreme limit is never realized in nature, but for a
qualitative understanding of the self-energy it is a good starting point,
since the effects of the finite width ($k_{c}$) of $\alpha ^{2}F(\mathbf{k},%
\mathbf{k}^{\prime },\omega )$, whenever $k_{c}\ll k_{F}$, change mostly the
quantitative picture - see \cite{DDKuOudo2}. In case of finite width of FSP
when $k_{c}v_{F}\ll \Omega $ the reduction of $T_{c}$ is given by $%
T_{c}^{MF}=T_{c0}^{MF}(1-7\zeta (3)k_{c}v_{F}/4\pi ^{2}T_{c0}^{MF})$. There
is also reduction of $T_{c}^{MF}$ due to retardation effects giving rise to
finite isotope effect $\alpha \neq 0$. \cite{DDKuOudo2}. Such a singular
model has repercussions on other properties. For, instance interesting
calculations within more realistic FSP theory with the finite width $k_{c}$,
but $k_{c}\ll k_{F}$, were done in \cite{VarelogFEP}, where the FSP theory
for EPI and the SFI theory (based on spin-fluctuation mechanism of pairing)
were compared. It was shown in this interesting paper that the FSP theory
can explain the appreciable increase of anisotropy gap-ratio $R\equiv \Delta
(\pi ,0)/\Delta (\pi /2,\pi /2)$ when $T\rightarrow T_{c}$, while the SFI
theory fails. Furthermore, the FSP theory of EPI can explain the pronounced
effect of orthorhombicity ($a\neq b$) in $YBCO$ on the gap ratio $\Delta
_{a}/\Delta _{b}$, the anisotropy of penetration depth $\lambda
_{a}^{2}/\lambda _{b}^{2}$ and the supercurrent ratio in the c-axis Pb-YBCO
junction. On the other hand, the SFI theory is ineffective here, since it
predicts at least one order of magnitude smaller effects \cite{VarelogFEP},
\cite{Kulic-Review}.

\subsubsection{Pseudogap due to internal Cooper pair fluctuations}

As it is stated above, the FSP theory also predicts existence of a
long-range pairing force due to the renormalization of EPI by strong
correlations. This opens an additional possibility for PG. Due to FSP in
EPI,\ the mean-field critical temperature has a non-BCS dependence, i.e. $%
T_{c}^{MF}=V_{EP}/4$ and it is inevitably reduced by \textit{phase and
internal Cooper pair fluctuations}. The latter are always present in systems
with long-range attractive forces, i.e. with FSP, as it was pointed out
first in \cite{Yang}. It was shown there that such a long-ranged
superconductor exhibits an additional class of fluctuations in which the
internal structure of Cooper pair is soft. This leads to PG behavior in
which the actual transition temperature $T_{c}$ is greatly depressed from
its mean-field value $T_{c}^{MF}$. We stress that these fluctuations are not
the standard phase fluctuations in superconductors. Because of the
importance, as well as due to internal beauty, of the approach given in \cite%
{Yang} we summarize briefly its main results.

In the following the weak coupling limit for the pairing Hamiltonian is
assumed

\[
\hat{H}=\sum_{\sigma }\int d\mathbf{x}\hat{\psi}_{\sigma }^{\dagger }(%
\mathbf{x})\xi (\hat{\mathbf{p}})\hat{\psi}_{\sigma }(\mathbf{x})
\]%
\begin{equation}
-\int d\mathbf{x}d\mathbf{\mathbf{x}}^{\prime }V(\mathbf{x-\mathbf{x}}%
^{\prime })\hat{\psi}_{\uparrow }^{\dagger }(\mathbf{x})\hat{\psi}%
_{\downarrow }^{\dagger }(\mathbf{x}^{\prime })\hat{\psi}_{\downarrow }(%
\mathbf{x}^{\prime })\hat{\psi}_{\uparrow }(\mathbf{x}).  \label{H-long}
\end{equation}

In the mean-field approximation (MFA) the order parameter $\Delta (\mathbf{x}%
,\mathbf{\mathbf{x}}^{\prime })$ is given by $\Delta (\mathbf{x},\mathbf{%
\mathbf{x}}^{\prime })=V(\mathbf{x-\mathbf{x}}^{\prime })\langle \mathbf{%
\hat{\psi}}_{\downarrow }(\mathbf{x}^{\prime })\mathbf{\hat{\psi}}_{\uparrow
}(\mathbf{x})\rangle $, which depends in fact on the internal coordinate $%
\mathbf{r}=\mathbf{x}-\mathbf{\mathbf{x}}^{\prime }$ and the center of mass $%
\mathbf{R}=(\mathbf{x}+\mathbf{\mathbf{x}}^{\prime }\mathbf{\mathbf{)/2}}$,
i.e. $\Delta (\mathbf{x},\mathbf{\mathbf{x}}^{\prime })=\Delta (\mathbf{r},%
\mathbf{\mathbf{R}})$. In standard superconductors (LTSC) with a \textit{%
short-range pairing potential} $V_{sr}(\mathbf{x-\mathbf{x}}^{\prime
})\approx V_{0}\delta (\mathbf{x-\mathbf{x}}^{\prime })$ one has $\Delta (%
\mathbf{r},\mathbf{\mathbf{R}})=\Delta (\mathbf{\mathbf{R}})$ and in that
case only the spatial ($\mathbf{\mathbf{R}}$) fluctuations of the order
parameter might be pronounced. In the case of long-range pairing potential
there are additional fluctuations of the internal degrees of freedom ($%
\mathbf{r}$). In the following we sketch the analysis given in \cite{Yang}.
When the range of pairing potential is large, i.e. $r_{c}>\xi $ (the
superconducting coherence length), fluctuations of the internal Cooper pair
wave-function are important since they give rise to a large reduction of the
mean-field quantities. In order to understand the physics of internal
wave-function fluctuations we study a much simpler Hamiltonian - the so
called reduced BCS Hamiltonian,
\begin{equation}
\hat{H}_{BCS}=\sum_{\mathbf{k}\sigma }\xi _{\mathbf{k}}\hat{c}_{\mathbf{k}%
\sigma }^{\dagger }\hat{c}_{\mathbf{k}\sigma }-\sum_{\mathbf{k},\mathbf{k}%
^{\prime }}V_{\mathbf{k}-\mathbf{k}^{\prime }}\hat{c}_{\mathbf{k}\uparrow
}^{\dagger }\hat{c}_{-\mathbf{k}\downarrow }^{\dagger }\hat{c}_{-\mathbf{k}%
^{\prime }\downarrow }\hat{c}_{\mathbf{k}^{\prime }\downarrow }.
\label{Hred}
\end{equation}%
Since we shall study excitations around the ground state we assume that
there are no unpaired electrons which allows us to study the problem in the
\textit{pseudo-spin Hamiltonian }(see \cite{Yang} and references therein)%
\textit{\ }
\[
\hat{H}_{ps}=\sum_{\mathbf{k}\sigma }2\xi _{\mathbf{k}}\hat{S}_{\mathbf{k}%
\sigma }^{z}-\frac{1}{2}\sum_{\mathbf{k},\mathbf{k}^{\prime }}V_{\mathbf{k}-%
\mathbf{k}^{\prime }}(\hat{S}_{\mathbf{k}}^{+}\hat{S}_{\mathbf{k}^{\prime
}}^{-}+\hat{S}_{\mathbf{k}^{\prime }}^{+}\hat{S}_{\mathbf{k}}^{-})
\]%
\begin{equation}
=\sum_{\mathbf{k}\sigma }2\xi _{\mathbf{k}}\hat{S}_{\mathbf{k}\sigma
}^{z}-\sum_{\mathbf{k},\mathbf{k}^{\prime }}V_{\mathbf{k}-\mathbf{k}^{\prime
}}(\hat{S}_{\mathbf{k}}^{x}\hat{S}_{\mathbf{k}^{\prime }}^{x}+\hat{S}_{%
\mathbf{k}^{\prime }}^{y}\hat{S}_{\mathbf{k}}^{y}),  \label{Hpseudo}
\end{equation}%
where the pseudo-spin 1/2 operators $\hat{S}_{\mathbf{k}\sigma }^{z}$, $\hat{%
S}_{\mathbf{k}\sigma }^{+}=(\hat{S}_{\mathbf{k}\sigma }^{-})^{\dagger }$ are
given by
\[
\hat{S}_{\mathbf{k}\sigma }^{z}=\frac{1}{2}(\hat{c}_{\mathbf{k}\uparrow
}^{\dagger }\hat{c}_{\mathbf{k}\uparrow }-\hat{c}_{-\mathbf{k}\downarrow
}^{\dagger }\hat{c}_{-\mathbf{k}\downarrow }-1),
\]%
\begin{equation}
\hat{S}_{\mathbf{k}\sigma }^{+}=\hat{c}_{\mathbf{k}\uparrow }^{\dagger }\hat{%
c}_{-\mathbf{k}\downarrow }^{\dagger }.  \label{ps-spin}
\end{equation}%
We see that Eq. (\ref{Hpseudo}) belongs to the class of \textit{Heisenberg
ferromagnetic Hamiltonians} (since $V_{\mathbf{k}-\mathbf{k}^{\prime }}>0$)
with the lattice in the Brillouin zone. It is well-known for the Heisenberg
model that in the case of long-range forces $V_{\mathbf{k}-\mathbf{k}%
^{\prime }}=const$ in a large part of the k-space, then the phase
fluctuations are negligible and thermodynamic properties are well described
in MFA
\begin{equation}
\hat{H}_{MFA}=-\sum_{k}\mathbf{h}_{\mathbf{k}}\mathbf{\hat{S}}_{\mathbf{k}}
\label{Hmfa}
\end{equation}%
with the mean-field $\mathbf{h}_{\mathbf{k}}$ given by
\begin{equation}
\mathbf{h}_{\mathbf{k}}=-2\xi _{\mathbf{k}}\mathbf{z+}\sum_{\mathbf{k}%
^{\prime }}V_{\mathbf{k}-\mathbf{k}^{\prime }}\langle \hat{S}_{\mathbf{k}%
^{\prime }}^{x}\mathbf{x+}\hat{S}_{\mathbf{k}^{\prime }}^{y}\mathbf{y\rangle
,}  \label{mf}
\end{equation}%
where $\mathbf{x,y}$ and $\mathbf{z}$ are unit vectors. Since $x$- and $y$%
-axis are equivalent one can look for $\mathbf{h}_{\mathbf{k}}$ in the form $%
\mathbf{h}_{\mathbf{k}}=-2\xi _{\mathbf{k}}\mathbf{z+}2\Delta _{\mathbf{k}}%
\mathbf{x}$, where the order parameter $\Delta _{\mathbf{k}}$ is the
solution of equation
\begin{equation}
\Delta _{\mathbf{k}}=\sum_{\mathbf{k}^{\prime }}V_{\mathbf{k}-\mathbf{k}%
^{\prime }}\langle \hat{S}_{\mathbf{k}^{\prime }}^{x}\rangle =\sum_{\mathbf{k%
}^{\prime }}\frac{V_{\mathbf{k}-\mathbf{k}^{\prime }}\Delta _{\mathbf{k}%
^{\prime }}}{2E_{\mathbf{k}}}\tanh \frac{\beta E_{\mathbf{k}}}{2},
\label{gap-eq}
\end{equation}%
with $E_{\mathbf{k}}=\sqrt{\xi _{\mathbf{k}}^{2}+\Delta _{\mathbf{k}}^{2}}$.
We stress that the long-range force ($V_{\mathbf{k}-\mathbf{k}^{\prime
}}=const$) in the k-space produces in the real space \textit{short-range
BCS-like force }$V_{BCS}(\mathbf{x-\mathbf{x}}^{\prime })\approx V_{0}\delta
(\mathbf{x-\mathbf{x}}^{\prime })$. This situation is realized in LTSC
superconductors where the metallic screening makes the pairing potential
short-ranged.

For the \textit{long-range attractive forces} the function $V_{\mathbf{k}-%
\mathbf{k}^{\prime }}$ is sharply peaked at $\mid \mathbf{k}-\mathbf{k}%
^{\prime }\mid =0$, for instance in the extreme FSP case (see Section 6.2)
one has $V_{\mathbf{k}-\mathbf{k}^{\prime }}=V_{0}\delta (\mathbf{k-\mathbf{k%
}}^{\prime })$. In such a case we deal with the short-range Heisenberg model
in the k-space. It is well known that in such a system low-laying
excitations, such as the Goldstone magnons, tend to destroy the long-range
order, i.e. in these systems one has $T_{c}\ll T_{c}^{MF}$. In fact these
fluctuations correspond to the internal Cooper pair fluctuations

In the following we analyze s-wave pairing only, where the solution of Eq. (%
\ref{gap-eq}) gives $T_{c}^{MF}=V_{0}/4$ and $\Delta _{0}=2T_{c}^{MF}$.
(Note, that in the BCS case one has $\Delta _{0}=1.76$ $T_{c}^{MF}$.). The
coherence length is defined by $\xi =v_{F}/\pi \Delta _{0}$. As we said, the
important fact is that in the case of long-ranged superconductors the
Heisenberg-like Hamiltonian in the momentum space is short-ranged giving
rise to low-laying spin-wave spectrum. The latter spectrum are in fact the
low-energy bound states (excitons) which loosely correspond to the
low-energy collective modes (in the true many-body theory based on Eq. (\ref%
{H-long})). This problem was studied in \cite{Yang} for the long-range (but
finite) potential $V(r)=V_{0}\exp \{-r^{2}/2r_{c}^{2}\}$ \ (its Fourier
transform is $V_{k}=(2\pi r_{c}^{2})V_{0}\exp \{-k^{2}r_{c}^{2}/2\}$), and
it was found that a large number $N_{cm}\sim \pi k_{F}r_{c}/6\xi $ (for $%
r_{c}\gg \xi $) of the excitonic like collective modes $\omega _{mn}^{exc}$
exist at zero momentum. These excitonic modes lie between the ground state
and the two-particle continuum, i.e. their energies are $\omega <2\Delta
_{0} $. Note that since we assume that $\Delta _{0}\ll E_{F}$, the system is
far from the Bose-Einstein condensation limit.

The above analysis in terms of the pseudo-spins is useful for physical
understanding, but the full many-body fluctuation problem, which is based on
the Hamiltonian in Eq. (\ref{H-long}), is studied in \cite{Yang} where the
Ginzburg-Landau (G-L) equation is derived for the long-ranged
superconductor. Due to fluctuations of the internal wave-function the G-L
free-energy functional $F\{\Delta (\mathbf{R},\mathbf{k})\}$ for the order
parameter $\Delta (\mathbf{R},\mathbf{k})=\int d\mathbf{r}\Delta (\mathbf{R}-%
\mathbf{r/2},\mathbf{R}+\mathbf{r/2})\exp \{-i\mathbf{kr}\}$ has a more
complicated form

\[
F\{\Delta (\mathbf{R},\mathbf{k})\}=\sum_{k}\int d\mathbf{R}\{A_{\mathbf{k}%
}\mid \Delta (\mathbf{R},\mathbf{k})\mid ^{2}+B_{\mathbf{k}}\mid \Delta (%
\mathbf{R},\mathbf{k})\mid ^{2}
\]
\begin{equation}
+\frac{1}{2M}\mid \partial _{\mathbf{k}}\Delta (\mathbf{R},\mathbf{k})\mid
^{2}+\frac{1}{2m_{\mathbf{k}}}\mid \partial _{\mathbf{R}}\Delta (\mathbf{R},%
\mathbf{k})\mid ^{2}\},  \label{F-GL}
\end{equation}
where $M=r_{c}^{2}V_{0}$ and
\[
A_{\mathbf{k}}=\frac{1}{V_{0}}-\frac{\tanh (\beta \xi _{\mathbf{k}}/2)}{2\xi
_{\mathbf{k}}}
\]
\[
\frac{1}{2m_{\mathbf{k}}}=\frac{\beta ^{2}v_{F}^{2}\sinh (\beta \xi _{%
\mathbf{k}}/2)}{32\xi _{\mathbf{k}}\cosh ^{3}(\beta \xi _{\mathbf{k}}/2)}
\]
\begin{equation}
B_{\mathbf{k}}=\frac{\tanh (\beta \xi _{\mathbf{k}}/2)}{8\xi _{\mathbf{k}%
}^{3}}-\frac{\beta }{16\xi _{\mathbf{k}}^{2}\cosh ^{3}(\beta \xi _{\mathbf{k}%
}/2)}.  \label{Coeff}
\end{equation}

The term with the partial derivative $\partial _{\mathbf{k}}$ is a direct
consequence of the long-ranged pairing potential, and it describes
fluctuations of the internal Cooper pair wave-function. The effect of these
fluctuations, described by the free-energy functional in Eq. (\ref{F-GL}),
was studied in the Hartree-Fock approximation in the limit $r_{c}\gg \xi $,
and a large reduction of the mean-field critical temperature is found
\begin{equation}
T_{c}\sim \frac{T_{c}^{MF}}{(r_{c}/\xi )}.  \label{Tc-fl}
\end{equation}

The latter result means that $T_{c}$ in the long-range superconductors is
\textit{controlled by thermal fluctuations of collective modes} which is in
contrast with the 3D short-range (BCS-like) superconductivity, where phase
fluctuations dominate but only slightly suppress $T_{c}^{MF}$. In the
temperature interval $T_{c}<T<T_{c}^{MF}$, the system with long-range
pairing force is in PG regime where the electrons are paired but there is
\textit{no long-range phase coherence}. The latter sets in only at $T<T_{c}$%
. We shall not further discuss this interesting approach but only stress
that it can be generalized by including repulsive interactions, for instance
due to spin fluctuations and residual Coulomb interaction. This will be
studied elsewhere \cite{Kulic-PG}.

In conclusion, FSP in EPI gives rise to the long-range pairing (in real
space) which produces internal fluctuations of Cooper pairs. These are soft
excitonic modes of the internal Cooper wave function which may reduce $T_{c}$
strongly. In the region $T_{c}<T<T_{c}^{MF}$ PG phase is realized. In this
approach PG possess the same symmetry as the superconducting gap. Since the
cuprates are systems near the Mott-Hubbard insulating state due to strong
correlations, this is responsible for very small phase stiffness and strong
phase fluctuations. Additionally, strong correlations (in conjunction with
the long-range Madelung interaction due to ionic-metallic structure of
cuprates) make EPI long-ranged, which causes strong internal fluctuations of
Cooper pairs. In a realistic theory both type of fluctuations must be taken
into account \cite{Kulic-PG}.

\section{Summary and conclusions}

A number of experiments, such as optics ($IR$ and Raman), transport,
tunnelling, ARPES, neutron scattering, etc. give convincing evidence that
EPI in cuprates is sufficiently strong and contributes to pairing. The ARPES
experiments made a breakthrough in the physics of cuprates, since they
allowed a direct study of quasiparticle properties, such as self-energy
effects. These experiments also give evidence for the presence of strong
correlations which modify EPI not only quantitatively but also
qualitatively. The most spectacular result of the EPI theory in strongly
correlated systems is the appearance of FSP in EPI, as well as in other
charge scattering processes such as the residual Coulomb interaction,
scattering on non-magnetic impurities - the \textit{FSP theory } \cite%
{Kulic1}, \cite{Kulic2}, \cite{Kulic3}, \cite{Kulic-Review}. FSP is
especially pronounced at lower doping $\delta $. This fundamental result
allows us to resolve at least qualitatively a number of experimental facts
which can not be explained by the old theory based on isotropic
Migdal-Eliashberg equations for EPI. The most \textit{important predictions
of the FSP theory of EPI} and other charge scattering processes are: (%
\textbf{1}) the transport coupling constant $\lambda _{tr}$
(entering the resistivity, $\varrho \sim \lambda _{tr}T$) is
\textit{much smaller} than the pairing $\lambda $, i.e. $\lambda
_{tr}<\lambda /3$; (\textbf{2}) the strength of pairing in
cuprates is basically \textit{due to EPI}, while the residual
Coulomb repulsion (including spin--fluctuations) \textit{triggers}
d-wave pairing; (\textbf{3}) d-wave pairing is very
\textit{robust} in presence of non-magnetic impurities with FSP;
(\textbf{4}) the nodal kink in the quasiparticle spectrum is
\textit{not-shifted} in the superconducting state, while the
anti-nodal singularity is shifted by the maximal superconducting
gap; (\textbf{5}) there is a collapse of the elastic impurity
scattering rate in the superconducting state at the anti-nodal
point giving rise to sharpening of the ARPES features;
(\textbf{6}) internal fluctuations of Cooper pairs, due to the
long-ranged scattering potential, can additionally increase the
temperature range of the PG effect.

We point out the following two facts that stem from the theoretical
analysis: (\textbf{i}) FSP in EPI of strongly correlated systems is a
general phenomenon and it \textit{affects electron coupling to all phonons};
(\textbf{ii}) the existence of FSP in EPI is confirmed numerically by Monte
Carlo calculations for the Hubbard-Holstein model with finite U \cite%
{Scalapino-Hanke}, by exact diagonalization \cite{Tohyama}, as well as by
some other methods \cite{Pietro-SB}. Thanks to these numerical results which
confirm the correctness of the FSP theory \cite{Kulic1}, \cite{Kulic2}, \cite%
{Kulic3}, \cite{Kulic-Review}, a number of new results were recently
reported \cite{Gunnar-rediscovery}, which are in favor of strong EPI
interaction and the FSP theory.

We would like to stress that the calculation of the momentum dependent EPI
coupling constant $\lambda _{\mathbf{k}}$ in cuprates is a big challenge for
the theory. The latter must include properly: (\textbf{a}) \textit{%
ionic-metallic coupling} due to the long-range Madelung energy as well as
\textit{covalent coupling,} and (\textbf{b}) \textit{strong electronic
correlations}. From this point of view, one can say that we still do not
have a quantitative microscopic theory of pairing in cuprates.

In the last several years a large number of papers has been devoted to the
study of spin-fluctuation interaction as a mechanism of pairing in cuprates.
In spite of much effort led by the greatest authorities in the field, which
have opened some new research directions in the theory of electron
magnetism, there is no theoretical evidence for the effectiveness of
non-phononic mechanism of pairing. Moreover, recent numerical calculations
\cite{Pryadko} show that probably there is no high-temperature
superconductivity in the t-J model. If it exists at all, its T$_{c}$ is
extremely low. Finally, numerical calculations in the Hubbard model \cite%
{D-Ds} show that the repulsive Hubbard interaction is unfavorable for
high-temperature superconductivity, contrary to the attractive interaction
which favors it.

In conclusion, we would like to stress again: in order to explain the high
critical temperature in cuprates, it is necessary to include EPI which is
renormalized by strong electronic correlations. We hope that in the future
such projects will be supported much more.

\textbf{Acknowledgement}. I express my deep gratitude and respect to Vitalii
Lazarevich Ginzburg for his support over many years, for sharing his deep
understanding of physics generously with us - his students, collaborators
and friends. I am thankful to Oleg V. Dolgov and Evgenii G. Maksimov for
collaboration and numerous elucidating discussions on superconductivity
theory. I am thankful to Ivan Bo\v{z}ovi\'{c} for valuable discussions on
optical properties of oxides, for his scientific support, for careful
reading of the paper and useful comments.


\begin{references}
\bibitem{VL} The members of the Ginzburg group, by including the present author as well as
many others, used to call Vitalii Lazarevich Ginzburg simply VL.
This is a way of expressing our love, devotion and respect to him.
Some of us (probably most of us) by calling him VL have been
expressing our wish to be closer friends of this great and
remarkable man.
\end{references}
\end{document}